\documentclass[aps,prx,twocolumn,showpacs,letterpaper,tighten,float,superscriptaddress,nofootinbib,reprint]{revtex4-2}
\usepackage{amssymb}
\usepackage{amsbsy}
\usepackage{amsmath}
\usepackage{mathrsfs}
\usepackage{epsfig}
\usepackage{graphicx}
\usepackage{array}
\usepackage{textcomp}
\usepackage{color}
\usepackage{braket}
\usepackage{hhline}
\usepackage{comment}
\usepackage{bm,hyperref}
\usepackage[justification=raggedright,font=small]{caption}
\usepackage[titletoc,toc,title]{appendix}
\usepackage[normalem]{ulem}
\usepackage[labelformat=simple]{subcaption}

\definecolor{myblue}{rgb}{.93, .93, 1}

\definecolor{darkgreen}{rgb}{0,0.7,0}

\newcommand{\beq}{\begin{equation}}
\newcommand{\eeq}{\end{equation}}

\newcommand{\bpm}{\begin{pmatrix}}
\newcommand{\epm}{\end{pmatrix}}
\newcommand{\bmm}{\begin{matrix}}
\newcommand{\emm}{\end{matrix}}
\newcommand{\bb}[1]{\boldsymbol{\mathbf{#1}}}

\newcommand{\mE}{\mathcal{E}}

\newcommand{\mC}{\mathcal{C}}
\newcommand{\mA}{\mathcal{A}}
\newcommand{\mS}{\mathcal{S}}
\newcommand{\mZ}{\mathcal{Z}}

\newcommand{\mbZ}{\mathbb{Z}}
\newcommand{\mH}{\mathcal{H}}
\newcommand{\mG}{\mathcal{G}}
\newcommand{\mL}{\mathcal{L}}
\newcommand{\mF}{\mathcal{F}}

\newcommand{\kket}[1]{\ket{#1}\rangle}

\graphicspath{{./Figures/}}

\makeatletter
\def\@fnsymbol#1{\ensuremath{\ifcase#1\or \ket{}\or \bra{}\or \dagger\or \ddagger\or
\mathsection\or \mathparagraph\or \|\or **\or \dagger\dagger
\or \ddagger\ddagger \else\@ctrerr\fi}}
\makeatother

\begin{document}

\author{Ramanjit Sohal}\thanks{rsohal@princeton.edu}
\affiliation{Pritzker School of Molecular Engineering, University of Chicago, Chicago, IL 60637, USA}
\affiliation{Department of Physics, Princeton University, Princeton, New Jersey 08544, USA}
\author{Abhinav Prem}\thanks{aprem@ias.edu\\Author order is symmetric under exchange of bra and ket spaces.}
\affiliation{School of Natural Sciences, Institute for Advanced Study, Princeton, New Jersey 08540, USA}

\date{\today}

\title{A Noisy Approach to Intrinsically Mixed-State Topological Order}
\date{\today}

\begin{abstract}
We propose a general framework for studying two-dimensional (2D) topologically ordered states subject to local correlated errors and show that the resulting mixed-state can display \textit{intrinsically mixed-state topological order} (imTO)---topological order which is not expected to occur in the ground state of 2D local gapped Hamiltonians. Specifically, we show that decoherence, previously interpreted as anyon condensation in a doubled Hilbert space, is more naturally phrased as, and provides a physical mechanism for, ``gauging out" anyons in the original Hilbert space. We find that gauging out anyons generically results in imTO, with the decohered mixed-state strongly symmetric under certain anomalous 1-form symmetries. This framework lays bare a striking connection between the decohered density matrix and \textit{topological subsystem codes}, which can appear as anomalous surface states of 3D topological orders. Through a series of examples, we show that the decohered state can display a classical memory, encode logical qubits (i.e., exhibit a quantum memory), and even host chiral or non-modular topological order. We argue that a partial classification of imTO is given in terms of non-modular braided fusion categories.
\end{abstract}

\maketitle


\section{Introduction}
\label{sec:intro}

Quantum many-body states with non-trivial entanglement serve as resource states for various tasks in  quantum information processing. Quintessential amongst these are states with \textit{topological order}, which support fractionalized excitations (anyons) and may serve as platforms for topological quantum computation~\cite{kitaev2003,nayakreview}. These states, which arise as locally indistinguishable degenerate ground-states of certain gapped Hamiltonians, form the code space for topological quantum error correcting codes (QECC), with the Hamiltonians provably robust to weak, local perturbations~\cite{bravyi2010,bravyi2011short,michalakis2013}: their utility as resource states thus extends to all states within the topologically ordered \textit{phase}. 

Recent years have witnessed remarkable progress in preparing and manipulating such states in programmable quantum simulators~\cite{satzinger2021,semeghini2021,andersen2022,iqbal2023,iqbal2024}. Decoherence is invariably present in these platforms and thus identifying a sharp notion of \textit{mixed-state topological order} is not merely of fundamental interest, but also of immediate practical import. While any finite temperature is known to destroy topological order (TO) in two spatial dimensions (2D)~\cite{hastings2011finiteT,lu2020negativity}, for local decoherence below a certain threshold, the quantum information encoded in a topological QECC is recoverable \cite{Dennis2002}. Indeed, the persistence and eventual breakdown of topological order in a pure state $\ket{\psi}$ subject to a local decoherence channel $\mathcal{E}$ has recently been studied through the lens of the entanglement properties of the ``corrupted" density matrix $\mathcal{E}[\ket{\psi}\bra{\psi}]$~\cite{coser2019class,bao2023mixed,fan2023mixed,lee2023decohere,chen2023separable,chen2023separable2,ma2023avg,sang2023mixed,dai2023ssto,li2024replica,lavasani2024qec}. While na\"ively one expects that local errors destroy quantum correlations (and hence TO), the decohered state is not the Gibbs state and can in principle encode structured entanglement. Indeed, decohered density matrices can display \textit{intrinsically mixed} SPT order~\cite{degroot2022og,ma2023prx,zhang2022strange,lee2022aspt,chirame2024,guo2024} -- that is, SPT order which does not arise in the ground state of a local Hamiltonian. Recently, Ref.~\cite{wang2023mixed} recently observed a non-trivial topological entanglement negativity in a decohered Toric code, which was taken to indicate the presence of a novel mixed-state topological order. However, a  systematic framework for addressing, let alone defining, such \textit{intrinsically mixed-state topological order} (imTO) is lacking. 

\begin{figure}
    \centering
    \includegraphics[width=0.5\textwidth]{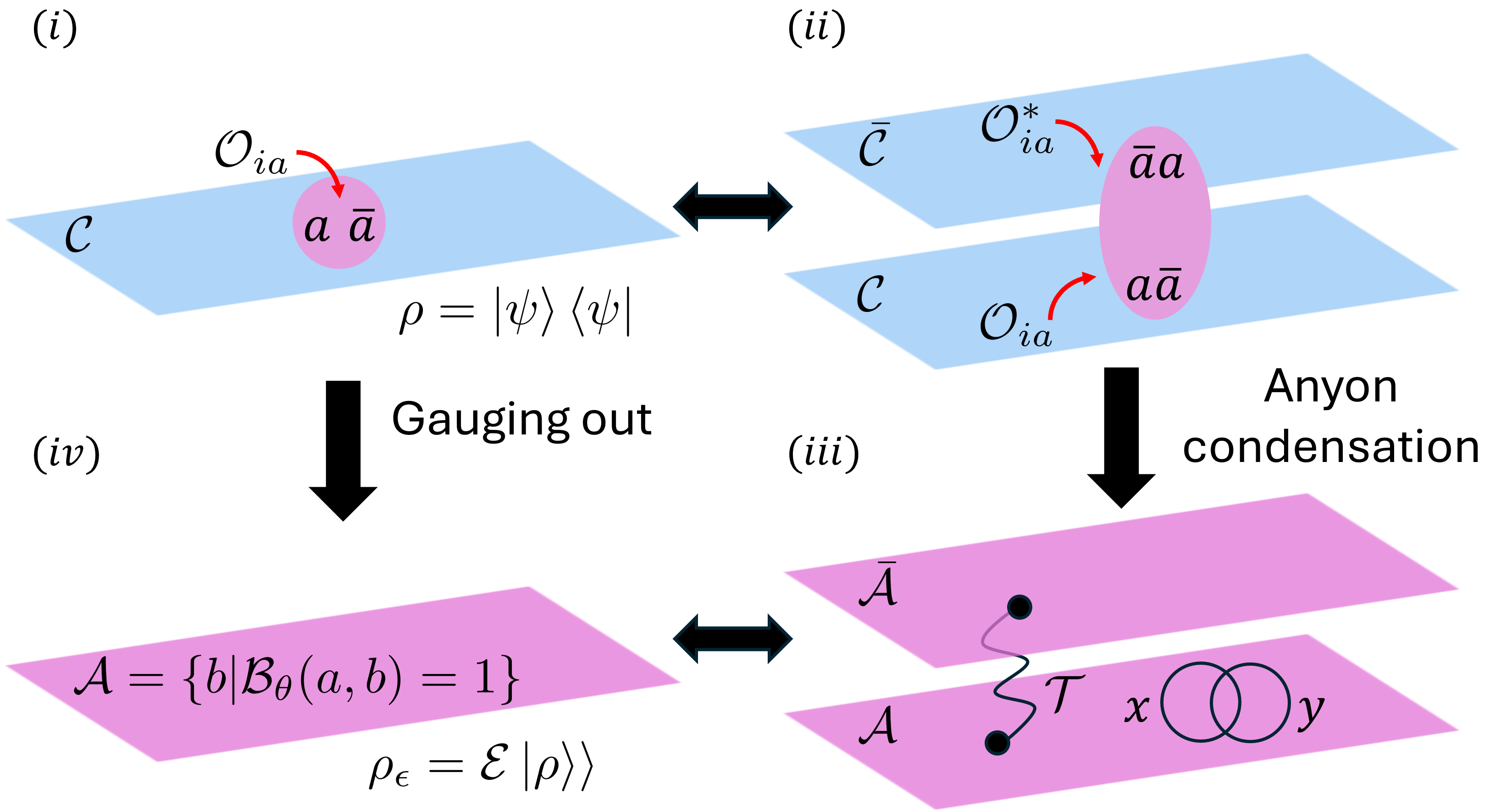}
    \caption{(i) A pure state with TO described by a UMTC $\mathcal{C}$ subject to locally correlated noise can be represented as (ii) a vector in a doubled Hilbert space with $\mC \times \bar{\mC}$ TO undergoing anyon condensation. (iii) The decohered state in the doubled Hilbert space has TO given by $\mathcal{A}$ ($\bar{\mathcal{A}}$) in the bra (ket) space, along with some transparent anyons $\mathcal{T}$. (iv) In the physical Hilbert space, this process corresponds to gauging out the decohered anyon $a$, with the \textit{quantum} TO in the mixed-state described by the anyon theory $\mathcal{A}$ which only includes anyons from $\mC$ which braid trivially with $a$.}
    \label{fig:introfig}
\end{figure}

In this paper, we propose a general framework for characterizing the topological order in mixed-states that are obtained by subjecting arbitrary 2D topologically ordered pure states to local decoherence (as described by finite-depth local quantum channels). For initial pure states with Abelian TO, locally correlated noise can induce imTO in the resulting mixed-state, which we show can be described by \emph{topological subsystem codes}~\cite{bombin2009,bombin2010,bombin2012,bombin2014}. As special cases, we recover previous mixed-state topological orders~\cite{bao2023mixed,wang2023mixed}, which support only classical memories~\footnote{Note that the these states can retain nontrivial quantum entanglement, even though they only encode a classical memory.}. In general however, topological subsystem codes (TSSCs) describe states supporting both classical and \textit{quantum} memories i.e., they can encode logical qubits. We attribute this memory structure in the decohered state to certain anomalies in the 1-form symmetries of the TSSC; these in turn stem from the 1-form symmetries of the parent topological order and of the decoherence channel.

Remarkably, topological subsystem codes can describe \emph{non-modular} (equivalently, pre-modular) topological orders---phases in which certain anyons are ``invisible" to all other anyons---and even \emph{chiral} topological phases, which admit no gapped boundary to vacuum. It is widely believed on physical grounds that non-modular TO \emph{cannot} exist in the ground state of a local gapped Hamiltonian in 2D. Our work nevertheless provides a physical mechanism for realizing such codes and hence, non-modular TO in \emph{mixed} states: indeed, we expect that all TSSCs can be realized by subjecting twisted quantum doubles~\cite{ellison2022} to local noise. Formally, this implies that the classification of imTO -- which we define as any mixed-state topological order that is non-modular -- is \textit{at least} as rich as that of TSSCs. On the practical side, we show that Abelian topological orders (which permit gapped boundaries) are resource states for preparing anomalous topological phases (including non-modular and chiral phases) under locality preserving quantum channels (LPQC)~\cite{piroli2020}, where this would otherwise require sequential quantum circuits~\cite{shirley2022qca,chen2024sequence} or measurements and feedback circuits (which suffer from post-processing bottlenecks). 

As in prior works, an essential step in our analysis is mapping the decohered density matrix to a vector in a ``doubled" Hilbert space. In this doubled Hilbert space, the decoherence channel can be understood as inducing anyon condensation across the two layers~\cite{bao2023mixed,fan2023mixed,lee2023decohere}. Our key insight is that in the original (physical) Hilbert space, this process corresponds to the \textit{``gauging out"} of an anyon in the original topological order~\cite{ellison2023} (see Fig.~\ref{fig:introfig}). While only bosonic anyons are permitted to condense~\cite{burnellreview}, decoherence allows us to gauge out anyons with \textit{any} spin (e.g., fermions or semions). This is intimately related to the fact that TSSCs appear as anomalous surface states of certain topological orders, lattice realizations of which are given by Walker-Wang (WW) models~\cite{walkerwang}: locally correlated errors hence provide a means of exfoliating surface states of 3D WW models, which can be anomalous~\cite{hsin2019oneform,moy2023oblique}. We further extend our framework to incorporate initial pure states with arbitrary topological orders (as described by a unitary modular tensor category), including those with non-Abelian anyons; strictly speaking, the resulting mixed-state is no longer a TSSC (which are characterized by Abelian anyon theories) but nevertheless corresponds to an anomalous WW surface state. This general construction leads us to characterize mixed states supporting imTO as those which are strongly symmetric under a non-modular 1-form symmetry. We will expand on the distinction between strong and weak symmetries, including the role the latter play, as well as the subtleties involved in defining imTO states as \emph{phases} of matter. From this, we conclude that imTO is (partially) classified in terms of non-modular unitary braided fusion categories. We also provide a finer characterization of these imTOs in terms of the set of locally detectable anyons outside the code-space, which are analogous to quasiparticle excitations in conventional pure state TOs. Finally, using the anomalies of their 1-form symmetries, we discuss the sense in which the imTOs we obtain constitute genuine mixed-state phases of matter.

The balance of this paper is organized as follows: in Sec.~\ref{sec:tssc}, we discuss the set of local decoherence channels under consideration and show that, for maximal decoherence, the resulting mixed-state belongs to the code space of a TSSC. In particular, we show that decoherence provides a physical mechanism for ``gauging out" certain anyons, whereby only those anyons which braid trivially with the decohered anyons remain as deconfined excitations in the resulting theory. We illustrate this framework through examples in Sec.~\ref{sec:examples}, where we show that the decohered state can host a quantum memory as well as chiral or even non-modular topological order. In Sec.~\ref{sec:general}, we argue that our framework can naturally be generalized to include parent non-Abelian theories, which leads to our claim that braided tensor categories provide a partial classification for imTO. In Sec.~\ref{sec:excitations}, we introduce the notion of locally detectable anyon types for imTOs. We then state our results in the language of strong 1-form symmetries in Sec.~\ref{sec:higherform}, where we also show an analogy between imTOs and surface states of Walker-Wang models. We conclude in Sec.~\ref{sec:cncls} with a discussion of open questions and future directions. 


\section{Topological subsystem codes via Decoherence}
\label{sec:tssc}

While our framework extends to generic TOs, we first illustrate our construction with Abelian TOs that admit gapped boundaries, in the context of lattice models realizing topological stabilizer codes (which serve as parent codes for topological subsystem codes). Consider a square lattice with periodic boundary conditions and place a $d$-dimensional qudit on each vertex. We define the Pauli operators $X_i$ and $Z_i$ acting on site $i$, which satisfy the $\mbZ_d$ algebra
\begin{align}
	Z_i X_i = \omega X_i Z_i, \qquad \omega = \exp(2\pi i / d).
\end{align}
We consider commuting projector translation-invariant Hamiltonians:
\begin{align}
\label{eq:ham}
	H_{\mC} = \sum_{i,\alpha} \frac{1 - \theta_i^\alpha}{2} + \mathrm{H.c.} \, ,
\end{align}
where $\theta_i^\alpha$ are constructed from finite, local products of Pauli operators acting near site $i$ and are mutually commuting: $[\theta_i^\alpha , \theta_j^\beta] = 0 \, \forall, \alpha,\beta$. The $\theta_i^\alpha$ are \emph{stabilizers} which generate the stabilizer group $\mS = \langle \{ \theta_i^\alpha \} \rangle$. The index $\alpha$ labels different families of stabilizers acting at site $i$. Since the Hamiltonian is positive semi-definite, the ground state manifold, also known as the \emph{code space} $\mathcal{H}_C$, is uniquely specified by the set of all states satisfying $\theta_i^\alpha \ket{\psi} = \ket{\psi}$ $\forall$ $i,\alpha$.

We are interested in topological stabilizer models, whose ground states are topologically ordered. Recall that a TO $\mC$ is described by a braided unitary fusion category: a finite set of anyons $\{a , b, \dots \}$, their fusion rules $a \times b = \sum_c N_{ab}^c c$ (with $N_{ab}^c \in \mathbb{Z}_+$), and their braiding statistics $\mathcal{B}_\theta(a,b)\equiv\theta_{ab}$. It is generally believed that local gapped Hamiltonians in 2D can at most support topological order described by unitary \emph{modular} fusion categories, with the modularity constraint being that the only excitation that braids trivially with itself and all other anyons is the vacuum superselection sector (equivalently, the $S$-matrix is unitary). Here, we will use the term ``anyon theory" to refer generally to unitary braided fusion categories i.e., without the modularity constraint and specify when the anyon theory is modular~\footnote{Although the condition of modularity may seem like a technical constraint, it is expected on physical grounds that TO described by a non-modular anyon theory cannot arise in the ground state of a local 2D Hamiltonian. Remarkably, non-modular anyon theories can arise on the \textit{surfaces} of 3D topological orders, as we will review in Section~\ref{sec:higherform}.}.

Now, given a topological stabilizer model realizing the UMTC $\mC$, each anyon is associated to a ``string-like" operator which violates the stabilizer conditions $\theta_i^\alpha = 1$ only at its endpoints; anyons correspond to \emph{errors} that take one out of the code space. We can thus interpret the $\theta_i^\alpha$ as generating contractible loops of anyons: $\theta_i^\alpha \neq 1$ indicates that the anyon generated by $\theta_i^\alpha$ accrues a nontrivial phase by braiding around the anyon corresponding to the local error. To each anyon, we also associate Wilson-loop operators, $W_{x,y}^a$, which wrap around the $x$ and $y$ cycles of the torus, respectively, and physically correspond to locally creating an anyon $a$ and its conjugate $\bar{a}$, before transporting $a$ around one cycle of the torus and then annihilating it with $\bar{a}$. These Wilson-loop operators commute with the stabilizers and thus preserve the code space, corresponding to a non-trivial ground state degeneracy of $H_{\mathcal{C}}$ on the torus. The nontrivial braiding of the anyons is encoded in the Wilson-loop algebra: $W^a_x W^b_y = e^{i\theta_{ab}} W^b_y W^a_x$. Since topological stabilizer models can only realize \emph{modular} TOs~\cite{ellison2023}, each anyon $a$ braids non-trivially with at least one other anyon $b$. The Wilson-loop operators thus correspond to \emph{logical operators}, in that they provide a representation of the Pauli algebra on the code space; topological orders hence provide quantum memories. The paradigmatic example is provided by the Toric code, for which the Wilson loops associated to the $e$ and $m$ anyon excitations satisfy the Pauli algebra, $\{W_x^e,W_y^m\} = \{W_x^m,W_y^e\} = 0$, such that the code space encodes two logical qubits.

In our framework, we will always take as input a topologically ordered pure state, which hosts TO described by a UMTC $\mC$. To make an explicit connection with topological stabilizer codes and TSSCs, we discuss Abelian theories that admit gapped boundaries here, though we will later relax this restriction. We are now interested in the fate of the TO (equivalently, the code space) under locally correlated noise. Such error processes correspond to quantum channels, where we consider translation-invariant channels of the form
\begin{align}
    \mE_a[\rho] = \prod_i \mE_{a,i}[\rho] , \qquad \mE_{a,i}[\rho] = \sum_{m=0}^{k-1} p_m O_{i,a}^m \rho (O_{i,a}^m)^\dagger \, , \label{eq:error-channel}
\end{align}
where $\sum_m p_m = 1$ and $\rho$ is the density matrix of the system, obtained after tracing out some environment. We will typically take the initial state to be pure: $\rho = \ket{\psi}\bra{\psi}$, where $\ket{\psi}$ is some pure state, but this is easily relaxed. Here, $O_{i,a}$ is a local operator supported near site $i$ corresponding to a ``short" Wilson string creating an anyon $a$ and its conjugate $\bar{a}$ near $i$. We take $O_{i,a}$ to be a product of Pauli operators such that $(O_{i,a})^k = 1$ for some integer $k \leq d$. We restrict ourselves to the case of maximal decoherence $p_m = 1/k$. These error channels thus have the physical interpretation of incoherently proliferating anyons of the type $a^m$, for $m=0,\dots , k-1$. For a general Abelian twisted quantum double model, which can be expressed in the general form Eq.~\eqref{eq:ham}, the set of anyons $\{a^m\}$ ($m=0,\dots,k-1$) can for instance be taken as the set of gauge charges of the model, which generate a Lagrangian subgroup~\cite{ellison2022}. These Pauli stabilizer models realize all Abelian quantum double models (equivalently, all Abelian TOs that admit gapped boundaries~\cite{kapustin2011}), which is the class of systems we now consider.

Let $\rho$ be an arbitrary density matrix in the ground state manifold (code space) of $H_{\mC}$, such that $\theta_i^\alpha \rho = \rho (\theta_i^\alpha)^\dagger = \rho$. Na\"ively, one expects decoherence will wash out any long-range entanglement present in the state, but we will now show that while the TO is indeed reduced (consistent with our error channels being LPQCs~\cite{piroli2020}), it can remain non-trivial and represent a genuine mixed-state quantum phase of matter, which is not expected to be realized as the gapped ground state of any local 2D Hamiltonian but instead does arise as an anomalous surface state of a 3D TO.

To characterize the topological order in the decohered density matrix $\rho_{\mE} \equiv \mE[\rho]$, it will prove conceptually fruitful to represent the density matrix as a vector in a doubled Hilbert space, through the Choi-Jamio{\l}kowski isomorphism~\cite{jamio1972,choi1975}. Explicitly, we map $\rho = \sum_{a,b} \rho_{ab} \ket{a}\bra{b}$ to the pure state $\kket{\rho} = \sum_{a,b} \rho_{ab} \ket{a}\ket{b}^* \in \mH_+ \otimes \mH_-$, where $\ket{b}^* \equiv K \ket{b}$, $K$ is complex conjugation in the computational basis, and $\mH_{\sigma = \pm}$ are the ket and bra spaces, respectively. In the doubled space, the stabilizer conditions become
\begin{align}
    \theta_{i+}^\alpha \kket{\rho} = (\theta_{i-}^\alpha)^* \kket{\rho} = \kket{\rho},
\end{align}
where $\theta_{i \pm}^\alpha$ is the action of $\theta_{i}^\alpha$ on $\mH_\pm$. Hence, $\kket{\rho}$ lies in the ground state manifold of the topological order $\mC \times \overline{\mC}$ in the doubled Hilbert space, with the two factors living on the ket and bra spaces, respectively. Consider the decohered density matrix $\rho_\mE$ in the doubled Hilbert space:
\begin{align}
    \kket{\rho_\mE} = \mE_a \kket{\rho} = \prod_i \left( \sum_{m=0}^{k-1} \frac{1}{k} O_{i,a+}^m (O_{i,a-}^m)^*  \right)\kket{\rho} \, .
\end{align}
For maximal decoherence, the vectorized error channel thus has the effect of projecting $\kket{\rho}$ to the subspace satisfying $O_{i,a+} (O_{i,a-})^* = +1$. As previously noted (see e.g. Ref.~\cite{bao2023mixed}), the effect of local error channels of the form Eq.~\eqref{eq:error-channel}, associated with decohering the set of anyons $\hat{\mathcal{A}} = \{a^m\}$, can be understood as inducing anyon condensation of the anyon pair $a_+ a_-$ in the doubled Hilbert space representation of the initially pure density matrix~\footnote{While we consider only maximal decoherence here, for finite decoherence, the decoherence \textit{transition} can be phrased in terms of anyon condensation on the boundary~\cite{bao2023mixed,fan2023mixed}.}. Since the two anyons $a_+$ and $a_-$ have opposite spins, their composite is a boson and can be condensed.

We now turn to one of the key results of this work by providing a finer characterization of the resulting decohered state $\kket{\rho_\mE}$. In particular, let us understand the effect of decoherence on the code space of the original stabilizer group $\mS$. In the doubled Hilbert space, let $\mS_\pm$ be the groups generated by the original stabilizers on the ket and bra spaces: $\mS_{\mC,\sigma} = \langle \{ \theta_{i\sigma}^\alpha\}_{i,\alpha} \rangle$. By assumption, $O_{i,a+} (O_{i,a-})^*$ do not commute with the original stabilizers (these create anyons and hence take us out of the code space). Thus, defining the group generated by the errors, $\mathcal{F}_{\mE} = \langle \{O_{i,a+} (O_{i,a-})^*\}_{i,a} \rangle$, the state $\kket{\rho_{\mE}}$ is stabilized by the group of mutually commuting elements in the union of the groups $\mathcal{F}_{\mE}$, $\mS_{\mC,+}$, and $ \mS_{\mC,-}$---that is to say, their centralizer $\mS_{\mC\times\bar\mC , \mE} \equiv \mZ( \mathcal{F}_{\mE} \cup \mS_{\mC,+} \cup \mS_{\mC,-} )$. Note that, in the doubled Hilbert space, $\kket{\rho}_\mE$ is still an element of the code space of a topological stabilizer code~\footnote{A quick way to see this is that condensing a boson in a UMTC always results in another UMTC which, for Abelian TOs, can always be described by a topological stabilizer model~\cite{ellison2023}.}.

Let us now restrict our attention to those elements $S_+ \in \mS_{\mC\times\bar\mC , \mE}$ which act non-trivially only on $\mathcal{H}_+$. In the physical Hilbert space, these are precisely those stabilizers which commute with the errors: $[S,O_{i,a}] = 0$. In particular, if $S_+ \in \mS_{\mC\times\bar\mC , \mE}$, then this implies $S_-^* \in \mS_{\mC\times\bar\mC , \mE}$. These operators satisfy $S_+ \kket{\rho_\mE} = S_-^* \kket{\rho_\mE} = \kket{\rho_\mE}$ or, equivalently, 
\begin{equation}
\label{eq:strongsym}
S \rho_{\mE} = \rho_{\mE} S^\dagger  = \rho_{\mE} \, , \forall S: [S,O_{i,a}] = 0 \, .
\end{equation}
Let $\mF = \langle \{ O_{i,a} \}_{i} \rangle$ be the group generated by all local errors and let $\mG = \langle \mS , \mF \rangle $. Since the elements of $\mS$ and $\mF$ do not commute, the group $\mG$ is in general non-Abelian. The set of stabilizers of $\rho_\mE$ is then given by the center of $\mG$, $\mS_\mE = \mZ(\mG)$. In other words, $\rho_\mE$ is an element of the code space of $\mS_\mE$, but this need not be the code space of a topological stabilizer code. 

Remarkably, the group structure given by the stabilizers $\mS_\mE$ and $\mG$, the latter of which is known as a ``gauge group" (not to be confused with the gauge group of a gauge theory), precisely realizes the structure of a \emph{topological subsystem code}, which leads to one of our main results: the set of decohered states $\rho_\mE$ on the torus form the code space for a TSSC. We briefly discuss the structure of TSSCs here, but refer the reader to Ref.~\cite{ellison2023} for a thorough exposition. As in the case of stabilizer codes discussed above, the Hilbert space for a TSSC can be written as a direct sum of the code space and its orthogonal complement $\mH = \mH_C \oplus \mH_C^\perp$. For a subsystem code, the code space further factorizes $\mH_C = \mH_\mL \otimes \mH_\mG$ such that the logical information is only encoded in the logical subsystem $\mH_\mL$, while $\mH_\mG$ is referred to as the gauge subsystem~\cite{bombin2010}. The gauge group $\mG$ comprises a set of Pauli operators which preserve the code space (commute with the stabilizers), but their action within the code space induces the factorization of the code space. For a gauge group $\mG$ that is proportional to the stabilizer group $\mS$, the gauge subsystem $\mH_\mG$ is trivial and one again has a topological stabilizer code. If the gauge group is non-trivial, TSSCs can support non-local stabilizers which cannot be generated by local stabilizers; moreover, TSSCs must satisfy the constraint that there should be no non-local stabilizers or logical operators on an infinite plane.

Recently, Ref.~\cite{ellison2023} discussed a general procedure for generating TSSCs from parent topological stabilizer codes by ``gauging out" appropriate anyons (see also Ref.~\cite{bombin2010}). In brief, given a parent Abelian TO with UMTC $\mC$ that admits a gapped boundary (equivalently a topological stabilizer group $\mS$), gauging out the set of anyons $\hat{\mathcal{A}} = \{a^m\}$ proceeds as follows: denote by $\mF$ the group of short string operators for the set of anyons $\hat{\mathcal{A}}$. Note that $\mF$ is only Abelian if $a$ is a boson. Gauging out then takes the stabilizer group $\mS$ to the gauge group $\mG = \langle \mS, \mF \rangle$. This means that the short string operators for the anyons in $\hat{\mathcal{A}}$ get appended to the original Abelian gauge group ($\propto \mS$). Physically, this means that any anyon $c \in \mC$ that braids non-trivially with the anyons in $\hat{\mathcal{A}}$ is confined, since the short string operators for the gauged out anyons do not commute with the Wilson loop operators for $c$. Further, the Wilson loop operators for anyons in $\hat{\mathcal{A}}$ are now given by products of gauge operators and if an anyon $x \in \hat{\mathcal{A}}$ is transparent in $\hat{\mathcal{A}}$, it becomes a transparent anyon in the TSSC. This procedure is distinct from anyon condensation in that the gauged out anyons are not necessarily identified with the vacuum, and so the excitations that only differ up to fusion with anyons in $\hat{\mathcal{A}}$ are not identified. 

With that brief review, let us return to the decohered mixed-state $\rho_\mE$. Following our discussion, it is clear that locally correlated errors induced by the short string operators for the set $\hat{\mathcal{A}}$ have the effect of gauging out these anyons, since the decohered density matrix $\rho_\mE$ (which is an element of the code space) is stabilized by precisely those stabilizers which commute with $O_{i,a}$ (see Eq.~\ref{eq:strongsym}). As violations of these stabilizers correspond to those anyons from the parent TO $\mC$ that braid trivially with anyons in $\hat{\mathcal{A}}$, we see that the decohered state has TO defined by a proper subset of anyons
\begin{equation}
\mathcal{A} \equiv \{b \in \mC | \mathcal{B}_{\theta}(a,b) = 1  \}
\end{equation}
where $\mathcal{B}_\theta(x,y)$ denotes the braiding statistics between anyons $x$ and $y$ (encoded in the parent UMTC $\mC$). By definition, $\mG$ contains the short string operators for the set of anyons in $\hat{\mathcal{A}}$: thus, their Wilson loop operators are given by products of gauge operators. If an anyon $x \in \hat{\mathcal{A}}$ is invisible to all other anyons in that set, its logical operator corresponds to a non-local stabilizer~\cite{ellison2023}, such that it becomes a transparent anyon in the decohered theory. Crucially, here $\mathcal{A}$ is an Abelian anyon theory which can be non-modular and corresponds to a TSSC. If $\mathcal{A}$ has no opaque (i.e., detectable via braiding with anyons in $\mathcal{A}$) anyons but still has transparent anyons, then the resulting mixed-state is a \textit{classical} self-correcting memory~\cite{yoshida2011,poulin2019}. In contrast, on the torus Wilson loops for opaque anyons in the (generally non-modular) Abelian anyon theory $\mathcal{A}$ correspond to logical operators for the TSSC, and $\rho_\mE$ can hence encode a \textit{quantum} memory. 

The identification of the decohered mixed-state $\rho_\mE$ with a TSSC provides a powerful framework within which to study mixed-state topological order, since we can leverage several known results about the former to characterize the latter. First, we have shown that decohering anyons in topological stabilizer model provides a physical mechanism for gauging out anyons. Since decoherence $\cong$ gauging out, the results of Ref.~\cite{ellison2023}, which established that every Abelian anyon theory (not necessarily modular) can be obtained by gauging out anyons from Abelian twisted quantum doubles, immediately imply that the classification of mixed-state topological order is at least as rich as that of Abelian anyon theories. This also suggests that the decohered states we have obtained should be viewed as being \textit{intrinsically mixed}: $\rho_\mE$ belongs to the code space of a TSSC which, unlike topological stabilizer codes, can realize \textit{non-modular} and even \textit{chiral} topological order, the latter of which is believed to not occur in the ground state of  a locally commuting parent Hamiltonian in 2D~\cite{kitaev2006}. It is also widely accepted that local gapped Hamiltonians in 2D only support modular anyon theories, implying that mixed states supporting \emph{non-modular} TO lie outside the classification of pure state phases of matter.

Indeed, motivated by the fact that non-modular TO cannot arise in the ground state of a local gapped 2D Hamiltonian, we say that a density matrix describes an \emph{intrinsically mixed-state topological order} (imTO) if the set of anyons $\mathcal{A}$ -- precisely, the associated set of Wilson loop operators -- describe a non-modular anyon theory. Although we have restricted ourselves to Abelian anyon theories thus far, in Section~\ref{sec:general} we will argue that the same characterization of imTO can be made for general (i.e. non-Abelian) anyon theories $\mathcal{A}$.  Let us also emphasize that the anyon theory $\mathcal{A}$ \emph{does not} describe the set of anyon ``excitations" (or errors) above the code space. We will elaborate on this point in Section~\ref{sec:excitations}. 
In Section~\ref{sec:higherform}, we will discuss how this characterization of imTO can be re-framed in the language of generalized symmetries -- namely, a density matrix has imTO if its set of \emph{strong} 1-form symmetries (i.e. the set $\mathcal{A}$) is non-modular and thus cannot be consistently realized in the ground state of a 2D local Hamiltonian. 

At this point, it is natural to ask whether these imTO states characterize genuine \emph{phases} of matter. Indeed, here and throughout, we focus on ``fixed-point" models obtained by subjecting fixed-point models of pure state topological order to maximal decoherence channels. The practical defining feature of these fixed-point states is that the Wilson loops and hence the corresponding anyon theory $\mathcal{A}$ can be constructed explicitly, as it is a subset of the parent pure state. The question is then whether perturbed versions of these fixed-point states still lie in the same phase, as characterized by the Wilson loop algebra of $\mathcal{A}$ (i.e. the set of strong 1-form symmetries to be discussed in Section~\ref{sec:higherform}). This is a subtle question, as there does not yet exist a consensus on how to define a phase of matter for mixed-states. It has been shown in the Toric code (and claimed for more general topological orders) that the loss of logical information due to decoherence should coincide with a phase transition, according to one definition of mixed-state phases~\cite{sang2023mixed}. As imTO states can be distinguished based on their logical spaces, we expect there to be a suitable definition of mixed-state phases that identifies these states as genuine phases of matter. We will offer additional speculation on these points in Section~\ref{sec:higherform}, but at no point will we make rigorous claims about the stability of imTO.

Subjecting a 2D pure state to a locally correlated noise channel induces topological order that would otherwise require a sequential quantum circuit~\cite{chen2024sequence} or measurement with feedback, and we take this to be a defining feature of imTO. An appealing perspective is then that ground states of (non-chiral) topological stabilizer codes furnish resource states for the dissipative-preparation of chiral (or non-modular) topological order under locally correlated noise. Given that a large class of topologically ordered gapped ground states can in principle be realized in quantum simulators using single-shot measurement and feedback~\cite{feedforward2023}, our results suggest that engineered dissipation can play a crucial role in the preparation of quantum states with imTO. As noted above, the decohered state $\rho_\mE$ can in principle encode a quantum memory (see examples below), and the lifetime of this encoded quantum information will remain infinite in the presence of any noise that respects the stabilizer symmetry of the TSCC (see Eq.~\eqref{eq:strongsym}). Indeed, an appealing interpretation of the decohered code space is that of a \textit{noiseless} or \textit{decoherence free} subspace~\cite{zanardi1997,zanardi1998,lidar1998,zanardi2003}, where the noise only acts within the gauge subsystem, leaving the logical subspace intact.


\section{Examples: Parent Abelian TOs}
\label{sec:examples}

With the general framework for imTO established, we now analyze several concrete examples that illustrate our results. In the process, we also discuss how anyon condensation in the doubled Hilbert space is equivalent to gauging out in the physical Hilbert space. As mentioned earlier, in principle one can straightforwardly obtain any (non-modular) Abelian anyon theory by decohering the gauge charges of the twisted quantum double models presented in Ref.~\cite{ellison2023} (which also furnishes the appropriate short-string operators and verifies that these satisfy the required braiding and fusion properties). 

\subsection{$\mbZ_2^{(0)}$ and $\mbZ_2^{(1)}$ TSSC from $\mathbb{Z}_2$ Toric code}
\label{sec:2DTC}

As the paradigmatic example of a topological stabilizer code, the stability of the $\mathbb{Z}_2$ Toric code to decoherence has been extensively investigated~\cite{Dennis2002,bao2023mixed,lee2023decohere,chen2023separable2,sang2023mixed}. We revisit this problem here in light of our interpretation of the decohered state as a TSSC. Consider a system of qubits placed on the edges of a square lattice with periodic boundary conditions, with the Hamiltonian given by
\begin{align}
    H_{\mathbb{Z}_2} = \sum_s \frac{1-A_s}{2} + \sum_p \frac{1-B_p}{2} , \, A_s = \prod_{i \in s} X_i , \, B_p = \prod_{i \in p} Z_i, \label{eq:Z2-TC-Hamiltonian}
\end{align}
where $s$ and $p$ denote stars and plaquettes, as usual. This Hamiltonian exhibits $\mbZ_2$ topological order ($\mC = \mbZ_2 \times \mbZ_2$), with anyons given by the electric charge $e$, the magnetic charge $m$, and their fermionic composite $f = e \times m$. As a quantum memory, the Toric code supports two logical qubits, with logical operators given by the Wilson loops of the $e$ and $m$ anyons: $W^e_{x,y} = \prod_{i \in \Gamma_{x,y}} Z_i$ and $ W^m_{x,y} = \prod_{i \in \hat{\Gamma}_{x,y}} X_i$, which satisfy $\{W^e_x, W^m_y\} = \{W^e_y, W^m_x\} = 0$. Here, $\Gamma_{x,y}$ and $\hat{\Gamma}_{x,y}$ are the corresponding non-contractible paths on the direct and dual lattices, respectively. 

We now consider two distinct error channels of the form Eq.~\eqref{eq:error-channel}, which proliferate errors associated with the $e$ and $f$ anyons, respectively:
\begin{align}
    \mE_{i,e}[\rho] = \frac{\rho + Z_i \rho Z_i}{2}, \, \mE_{i,f}[\rho] = \frac{\rho + Z_i X_{i+\bb{\delta}} \rho Z_i X_{i+\bb{\delta}}}{2},
\end{align}
where $\bb{\delta} = (\frac{1}{2},-\frac{1}{2})$~\footnote{As noted in Ref.~\cite{wang2023mixed}, while $Y$-errors locally create $f$ anyons, arbitrary configurations of such errors can create unbalanced numbers of $e$ and $m$ anyons. It is hence crucial to consider ``framed" short-string operators to generate strictly $f$ type errors.}. Here the short string operators $O$ are given by the operators $Z_i$ and $Z_i X_{i+\delta}$ for $e$ and $f$ respectively.

Given an arbitrary state $\rho$ in the ground state manifold of Eq.~\eqref{eq:Z2-TC-Hamiltonian}, we wish to characterize the decohered states $\rho_{e,f} \equiv \mE_{e,f}[\rho]$. Clearly, $\mE_e[W^e_{x,y}] = W^e_{x,y}$ while $\mE_e[W^m_{x,y}] = 0$, and so $\rho_e$ only forms a classical memory with a single bit of information encoded in each of $W_{x,y}^e$. Likewise, one also finds that $\rho_f$ forms a classical memory, with classical bits stored in the $f$ Wilson loops, defined as $W_{x,y}^f = \prod_{i \in \Gamma_{x,y}} X_i Z_{i+\bb{\delta}}$. While superficially it appears that errors have rendered the state ``trivial," we now show that $\rho_{e,f}$ exhibit richer structure.

Recall that since $A_s \rho = \rho A_s = \rho$ and $B_p \rho = \rho B_p = \rho$, ground states of Eq.~\eqref{eq:Z2-TC-Hamiltonian} can be interpreted as closed loop condensates of the $e$, $m$, and $f$ anyons. After maximal decoherence, we instead only have $B_p \rho_e = \rho_e B_p = \rho_e$ and $A_s \rho_e A_s = \rho_e$. Physically, the $e$-noise has the effect of ``freezing" the $m$-loops (and hence also the $f$ loops) into a classical ensemble, while leaving the ``quantum" condensate of $e$-loops untouched. More precisely, $e$-errors break the strong 1-form magnetic symmetry of the original pure state down to a weak 1-form symmetry, while leaving the strong 1-form electric symmetry intact. We will later provide a general discussion of the role 1-form symmetries play in characterizing generic imTOs (see Sec.~\ref{sec:higherform}).

One might thus be inclined to view $\rho_e$ as describing a topologically ordered state in which the only deconfined anyon excitation is the bosonic $e$ anyon of the parent Toric code. Indeed, in the notation of Ref.~\cite{bonderson2012thesis}, a phase with anyon content given by the vacuum and a single $e$ anyon corresponds to the $\mathbb{Z}_2^{(0)}$ topological order~\footnote{In general, $\mathbb{Z}_N^{(p)}$ topological order is generated by a single anyon $a$ such that $a^N = 1$ and $a$ has spin $\theta_a = \exp(2\pi i p / N)$.}. Notably, this topological order is \emph{non-modular}: since $e$ is the only non-trivial anyon in the theory, it cannot be detected by braiding with any other anyons i.e., it is transparent. While non-modular topological orders cannot be realized by topological stabilizer models, they do arise in the aforementioned topological stabilizer codes. We can in fact make the correspondence with TSSCs precise, following the preceding general analysis in Sec.~\ref{sec:tssc}. The gauge group for $\rho_e$ is given by~\footnote{$\mG$ must include appropriate roots of unity to ensure that it generates a representation of the Pauli group.}
\begin{align}
    \mG_e = \langle i, Z_i , A_s \rangle,
\end{align}
such that the stabilizer group is given by $\mZ(\mG_e) = \langle B_p, W_{x,y}^e \rangle$. This precisely describes the $\mathbb{Z}_2^{(0)}$ topological subsystem code, which is shown to exhibit the $\mathbb{Z}_2^{(0)}$ topological order in Ref.~\cite{ellison2023}. Since $e$ is transparent in this theory, there are no logical operators and $\rho_e$ does not encode any qubits.

Similar considerations hold for $\rho_f$. We have that $A_s B_{s-\mathbf{y}} \rho_f = \rho_f A_s B_{s-\mathbf{y}} = \rho_f$, where $A_s B_{s-\mathbf{y}}$ generates a closed $f$ loop and $s-\mathbf{y}$ denotes the plaquette to the south-east of vertex $s$, while we only have that $A_s \rho_f A_s = \rho_f$. Thus, $f$-noise freezes both the $e$ and $m$ loops but leaves the $f$ loops untouched, such that $\rho_f$ describes a quantum condensate of \emph{fermionic excitations} (stated otherwise, $\rho_f$ retains a strong 1-form symmetry): this is not expected to occur in the ground state of a gapped, local 2D Hamiltonian. Remarkably, despite starting with a \textit{bosonic} topological order, decoherence has resulted in a state effectively described as a condensate of \emph{fermions}.
In a rough sense, local decoherence allows one to ``peel off" half of the original state. 

Again, this heuristic interpretation can be formalized by following our general analysis in Sec.~\ref{sec:tssc}---the gauge group of $\rho_f$ is given by
\begin{align}
    \mG_f = \langle i, Z_i X_{i+\delta}, A_s \rangle 
\end{align}
which yields the stabilizer group $\mS_f = \langle A_s B_{s-\mathbf{y}}, W_{x,y}^f \rangle$. This precisely describes a topological subsystem code describing the $\mathbb{Z}_2^{(1)}$ topological order which, again, is  \emph{non-modular} \cite{ellison2023}. Like the previous case, $f$ is transparent (it braids trivially with itself) and hence the decohered state encodes no quantum memory, consistent with Ref.~\cite{wang2023mixed}.

It will be instructive to study these mixed-states through the complementary perspective of the doubled Hilbert space. As discussed earlier, the vectorized initial density matrix $\kket{\rho}$ lies in the ground state manifold of a bilayer Toric code, with anyon content
\begin{align}
	\mathcal{C} \times \overline{\mathcal{C}} = \{1_+ , e_+ , m_+ , f_+ \} \times \{ 1_- , e_- , m_-, f_- \},
\end{align}
where $\pm$ subscripts denote the ket and bra spaces respectively. In this picture, the $e$ and $f$ noise channels have the effect of condensing the anyons $e_+ e_-$ and $f_+ f_-$ respectively, which for maximal decoherence lead to the resulting daughter topological orders
\begin{align}
    \mC_e &= \{ 1_+ 1_-  , e_+ , m_+ m_- , f_+ m_-  \}, \\
    \mC_f &= \{ 1_+ 1_- , e_+ e_- , e_+ m_-  , f_+  \}.
\end{align}
It is readily apparent that the resulting TO in either case is that of a single $\mbZ_2$ Toric code, with the fusion group of the Abelian anyons given by $\mbZ_2 \times \mbZ_2$. This can also be directly verified with the explicit forms of $\kket{\rho_{e,f}}$ in the lattice model. In light of our above stabilizer analysis however, we note a key distinction between the mixed-states $\mC_e$ and $\mC_f$. Restricting attention to anyons with support on only the ket or bra space, we see that both orders support a single such anyon. For $\mC_e$, this is the \emph{boson} $e_+ \sim e_-$, while for $\mC_f$, this is the \emph{fermion} $f_+ \sim f_-$, where the equivalences are up to fusion with the condensed anyon. This is consistent with our observation in the stabilizer analysis that under $e$ and $f$ noise, the sole remaining coherent closed loops are simply those corresponding to the original $e$ and $f$ anyons, respectively.

We now show that this anyon condensation across the ket and bra spaces, at the level of the density matrix in the original Hilbert space, corresponds to gauging out an anyon. Recall that anyon condensation proceeds in two steps (in Abelian theories): to condense an anyon $a$, one first (i) projects out from the theory those anyons which braid non-trivially with $a$ (i.e. they become confined) and then (ii) identifies those anyon types which differ by fusion with $a$. For instance, in the Toric code, condensing $e$ confines the $m$ and $f$ anyons while $e$ becomes identified with the vacuum: the resulting state has no remaining anyon excitations and is trivial. Gauging out an anyon however corresponds to only performing step (i) of this process. For instance, gauging out $e$ still confines $m$ and $f$, but leaves $e$ distinct from the vacuum, such that one is left with the anyon content $\{1 , e \}$---precisely that of the $\mathbb{Z}_2^{(0)}$ non-modular TO realized via decoherence of $e$. 

Crucially, one can also gauge out anyons that \emph{cannot} be condensed (i.e. non-bosonic anyons); analogously, one can decohere non-bosonic anyons $a$ since this corresponds to the conventional condensation of the bosonic pair $(a_+ a_-)$ in the doubled Hilbert space. For instance, one may gauge out $f$ from the $\mbZ_2$ Toric code to obtain the $\mathbb{Z}_2^{(1)}$ TO, precisely replicating the effect of $f$ errors. Surprisingly, as this simple example illustrates, locally correlated errors (which correspond to anyon condensation under the Choi map) provide a \emph{physical} implementation of the gauging out procedure, which thus far remained a conceptual device for generating topological subsystem codes from topological stabilizer codes \cite{bombin2009,bombin2010,bombin2012,bombin2014,ellison2023}.

Let us pause to recapitulate our observations in the context of the Toric code. We found that anyonic decoherence, previously shown to correspond to anyon condensation across the ket and bra spaces, implements the gauging out of anyons in the original Hilbert space, including those which are forbidden from condensing under purely unitary evolution. This process led to mixed-states supporting non-modular topological order (corresponding to topological subsystem codes) which is believed to be forbidden in the ground state of a locally gapped Hamiltonian in 2D. Thus, we claim that both the $e$- and $f$-decohered Toric codes represent \textit{intrinsically mixed} topological states of matter. However, there is an important distinction between the two cases: namely, the strong 1-form symmetry of the $e$-decohered Toric code is non-anomalous, while that of the $f$-decohered Toric code is anomalous. In Sec.~\ref{sec:higherform}, we will discuss the implications of these anomalies and the sense in which we expect them to characterize mixed-state \emph{phases} of matter. Note that for the specific case of the $\mbZ_2$ Toric code subject to $f$ errors, Ref.~\cite{wang2023mixed} numerically verified the robustness of the resulting imTO against finite noise channels that explicitly break the strong 1-form symmetry.

\subsection{$\mbZ_4^{(1)}$ TSSC from $\mathbb{Z}_4$ Toric code}
\label{sec:2DTCZ4}

\begin{figure}
    \centering
    \includegraphics[width=0.3\textwidth]{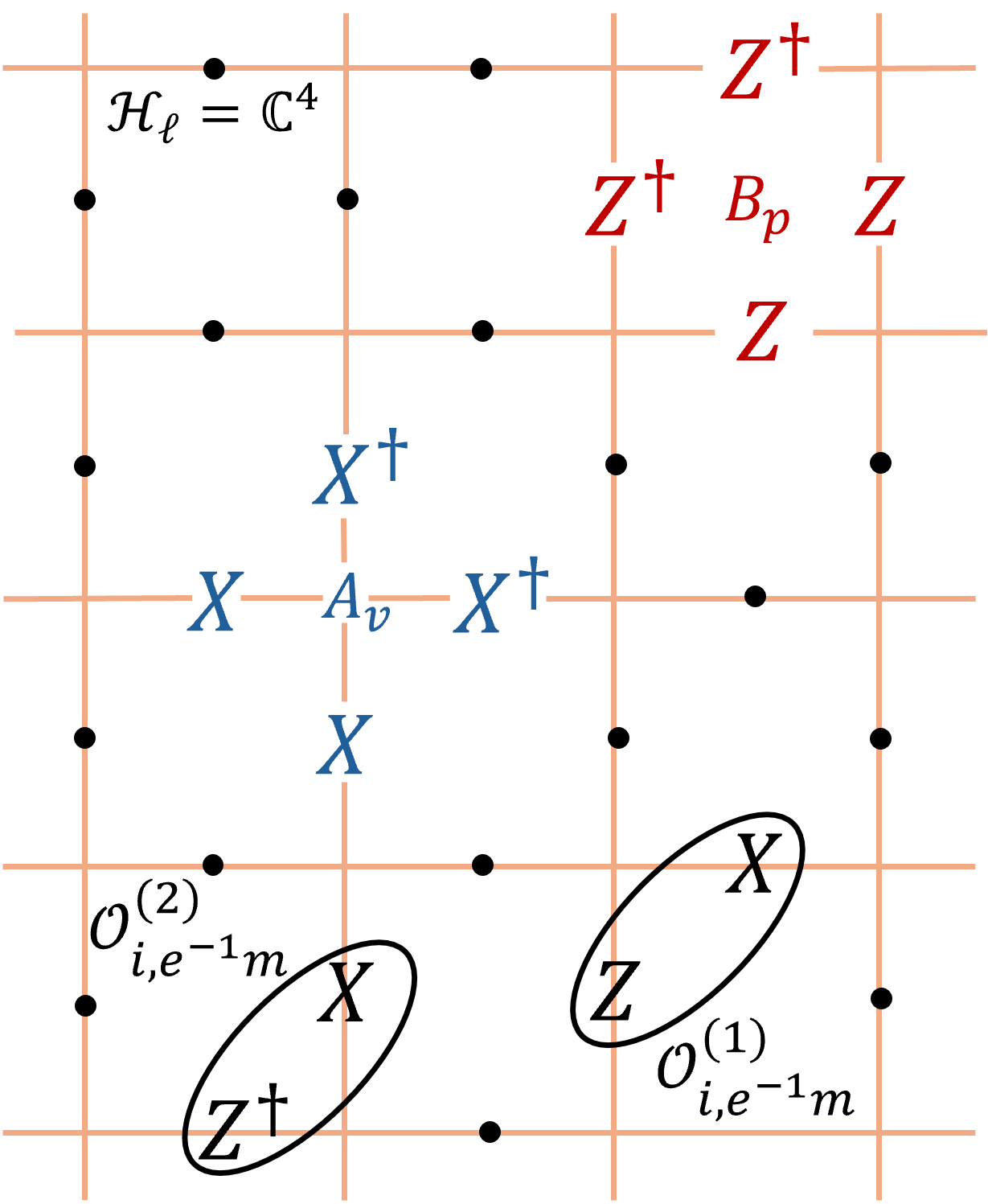}
    \caption{The $\mbZ_4$ Toric code is defined on the 2D square lattice with a $d=4$ qudit (black) on each link. The star (plaquette) stabilizers are shown here in blue (red). Short string operators for the $e^{-1}m$ anyon are also shown.}
    \label{fig:mainfig}
\end{figure}

We next consider a square lattice with $d=4$ qudits on each edge. A Hamiltonian realizing $\mathbb{Z}_4$ TO is given by
\begin{align}
    H_{\mathbb{Z}_4} = -\sum_v ( A_v + A_v^\dagger) - \sum_p (B_p + B_p^\dagger) ,
    \label{eq:Z4-TC-Hamiltonian}
\end{align}
with the star and plaquette operators defined in Fig. \ref{fig:mainfig}. The ground state manifold is determined by the constraints $A_v = B_p = 1$, violations of which indicate the presence of electric $e$ and magnetic $m$ excitations, respectively. Explicitly, $Z_i$ applied to the ground state excites an $e$ and an $e^{-1}$ anyon at vertices connected by the edge $i$. Likwise, applying $X_i$ creates an $m$ and $m^{-1}$ on plaquettes separated by the edge $i$. These anyons satisfy $\mbZ_4 \times \mbZ_4$ fusion rules $e^4 = m^4 = 1$ and the braiding statistics between two composite objects $e^a m^b$ and $e^c m^d$ is given by $\mathcal{B}_\theta(ab,cd) = i^{ad+bc}$. On the torus, the non-contractible Wilson loops $W^e_{x,y} = \prod_{i \in \Gamma_{x,y}} Z_i$ and $ W^m_{x,y} = \prod_{i \in \hat{\Gamma}_{x,y}} X_i$ serve as the logical operators and satisfy the algebra $W_{x/y}^e W_{y/x}^m = i W_{y/x}^m W_{x/y}^e$, such that the code space stores two $d=4$ qudits.

Here, we consider local errors for the set of anyons generated by the $e^{-1} m$ anyon, $\hat{\mathcal{A}}=\{1, e^{-1}m, e^2 m^2, e m^3\}$. The corresponding decoherence channel is given by Eq.~\eqref{eq:error-channel} with the generating short string operators $O_{i,e^{-1}m}^{(1)}$, $O_{i,e^{-1}m}^{(2)}$ for $e^{-1}m$ shown in Fig.~\ref{fig:mainfig}. Here, the group of local errors $\mF$ is precisely the group generated by the short string operators of $e^{-1}m$ (see Fig.~\ref{fig:mainfig}).

Now, for an arbitrary state $\rho$ in the ground state manifold of the $\mbZ_4$ Toric code, we wish to characterize $\rho_{e^{-1}m} \equiv \mE_{e^{-1}m}[\rho]$. Let us proceed formally first in this case. Following the general prescription in Sec.~\ref{sec:tssc}, the gauge group $\mG_{e^{-1}m} = \langle \mS, \mF \rangle$ is given by
\begin{equation}
\mG_{e^{-1}m} = \langle e^{i\pi/2}, A_v, B_p, O_{i,e^{-1}m}^{(1)}, O_{i,e^{-1}m}^{(2)}  \rangle \, .
\end{equation}
The stabilizer group for the decohered density matrix is then $\mS_{e^{-1}m} = \langle A_v B_{v+\bb{y}}, W_{x,y}^{e^2 m^2}\rangle$, where $v+\bb{y}$ denotes the plaquette to the north-east of vertex $v$, $W_{x,y}^{e^2 m^2} = \prod_{i \in \Gamma_{x,y}} X_i^2 Z_i^2$ is the Wilson loop operator for the $e^2 m^2$ anyons, and $A_v B_{v + \bb{y}}$ generates a closed loop of $em$ anyons. This is precisely the topological subsystem code corresponding to the $\mbZ_4^{(1)}$ topological order, which is given by the Abelian anyon theory $\mathcal{A} = \{1, em, e^2 m^2, e^{-1}m^{-1}\}$, in which both $em$ and $e^3 m^3$ are semions and $e^2 m^2$ is a transparent boson (it braids trivially with all other anyons in $\mathcal{A}$). This stems from the fact that the open Wilson line operator for the $e^2 m^2$ is built out of gauge operators and commutes with all of the stabilizers in $\mS_{e^{-1}m}$ at its endpoints. The code space, stabilized by $\mS_{e^{-1}m}$ has two logical operators on the torus, which are the Wilson loop operators of the two semions, and encodes a single logical qubit in its logical subystem. Thus, $\rho_{e^{-1},m}$ realizes imTO as it is a \textit{non-modular} Abelian anyon theory that cannot be the ground state of a gapped local Hamiltonian in 2D and also realizes a quantum memory. 

Recall that the original pure state $\rho$ satisfies $A_s \rho = \rho A_s = \rho$ and $B_p \rho = \rho B_p = \rho$ and can be thought of as a closed loop condensate of all non-trivial anyons $e^a m^b$. Clearly, $\mE[W_{x,y}^{\alpha}] = 0$ for any anyons $\alpha$ that braid non-trivially with anyons in $\hat{\mathcal{A}}$. Since only the anyons in $\mathcal{A} = \{1, em, e^2 m^2, e^{-1}m^{-1}\}$ braid trivially with those in $\hat{\mathcal{A}}$, decoherence does not affect their Wilson loops: $\mE[W_{x,y}^{b \in \mathcal{A}}] = W_{x,y}^{b \in \mathcal{A}}$. Intuitively, decoherence has thus frozen out the loops for any anyons $\notin \mathcal{A}$ into a classical ensemble, while the quantum condensate of anyons in $\mathcal{A}$ is left untouched.

Said more formally, in the language of higher-form symmetries to be discussed in Section~\ref{sec:higherform}, $\hat{\mathcal{A}}$ errors break most of the strong 1-form symmetries of $\rho$ down to weak 1-form symmetries, while leaving the strong 1-form symmetries corresponding to $\mathcal{A}$ anyons intact. This is encoded in Eq.~\eqref{eq:strongsym} and the fact that the stabilizer group $\mS_{e^{-1}m}$ for the TSSC is generated by small loops for the $em$ anyon (which generates $\mathcal{A}$). Finally, since $e^2 m^2$ is transparent in $\hat{\mathcal{A}}$, it remains transparent in $\mathcal{A}$ by definition. We thus obtain the same result as above: the set of decohered density matrices on the torus form the code space for a TSSC which describes a non-modular Abelian anyon theory $\mathcal{A}$, whose non-trivial anyons are two semions and a transparent boson. The mutual statistics of the semions result in this mixed-state encoding a logical qubit in its logical subsystem. Since this mixed-state encodes non-trivial logical information, we expect on physical grounds that it is robust (up to a finite noise threshold) against finite-depth local quantum channels~\cite{brownreview,sang2023mixed} and so represents a genuine \emph{imTO phase} of matter although, as emphasized above and in Section~\ref{sec:higherform}, this requires careful verification. 

It is instructive to once again consider the gauging out procedure from the perspective of the doubled Hilbert space. In this picture, decoherence of the anyons in $\hat{\mathcal{A}}$ corresponds to condensing $\{[1]_+[1]_-,$ $ [e^{-1}m]_+ [e^{-1}m]_-, [e^2m^2]_+ [e^2m^2]_-,[em^3]_+[em^3]_-\}$, which form a Lagrangian subgroup of the TO in the doubled space $\mC \times \bar{\mC}$, where $\mC = \mbZ_4 \times \mbZ_4$. Anyon condensation proceeds in the usual way: each anyon from $\hat{\mathcal{A}}$ is identified with the vacuum. Next, any excitation which braids non-trivially with any condensed anyon becomes confined and, of the remaining deconfined excitations, any that differ only up to fusion by anyons in $\hat{\mathcal{A}}$ are identified. A simple calculation shows that the resulting topological order is that of a $\mbZ_4$ gauge theory, with only the following anyons supported solely on the ket space: $[em]_+$, and $[e^{-1}m^{-1}]_+ $, while $[e^2 m^2]_+ \sim [e^2 m^2]_-$ can move freely between the ket and bra spaces, and is a transparent anyon. As expected, these correspond precisely to those anyons in $\mathcal{A}$, obtained by applying the gauging out procedure in the original Hilbert space.

This example already displays much of the rich structure that emerges when anyonic errors are introduced into a pure topologically ordered state, with the most striking features being the presence of a robust quantum memory alongside a non-modular Abelian anyon theory that is generally believed to not occur in the ground state of a locally gapped Hamiltonian.

Equipped with the preceding understanding of the correspondence between decoherence, gauging out, and anyon condensation in the doubled Hilbert space, we now briefly discuss two other examples which illustrate the breadth of Abelian anyon theories that can be ``peeled off" via decoherence. Moreover, we have established a mapping from the space of imTOs that result from decohering a set of anyons $\hat{\mathcal{A}}$ when starting from a parent topological stabilizer code to the space of TSSCs that results from gauging out $\hat{\mathcal{A}}$ from the same parent topological stabilizer code. Thus, we can directly use results from Ref.~\cite{ellison2023}, which provides a thorough exploration of TSSCs. In particular, once we specify the parent TO and the set of decohered anyons $\hat{\mathcal{A}}$, we can immediately read off the gauge group and the structure of the code space from the results contained in Ref.~\cite{ellison2023}. 

\subsection{Chiral Semion from Double Semion}
\label{sec:semion}

As an instance of this mapping, let us take the doubled semion anyon theory as our parent TO. This theory can be realized as a Pauli stabilizer Hamiltonian~\cite{ellison2022} and its anyons form a $\mbZ_2 \times \mbZ_2$ group under fusion, with elements $\{1,s,\bar{s}, s \bar{s}\}$. Here, $s$ is a semion (it has self-statistics $\theta(s) = i$), $\bar{s}$ is an anti-semion ($\theta(\bar{s})=-i$), and $s \bar{s}$ is a boson. Now, we subject a ground state of this system to an error channel that incoherently proliferates the semion $s$ i.e., $\hat{\mathcal{A}} = \{1,s\}$. This corresponds to gauging out $s$, which braids trivially with $\bar{s}$. The resulting anyon theory for the decohered mixed-state is given by $\mathcal{A} = \{1,\bar{s}\}$ i.e., it is the \textit{chiral} (anti)-semion Abelian anyon theory. Since $\bar{s}$ has non-trivial self-statistics, the decohered code space encodes exactly one logical qubit; this example represents the minimal model in which one obtains a chiral anyon theory with an encoded logical qubit. Again, the presence of a logical qubit confers stability to this state against finite-depth local channels and it represents a mixed-state phase of matter. While this mixed state does \emph{not} correspond to an imTO, as it is described by a \emph{modular} anyon theory, chiral UMTCs cannot arise in the ground states of locally commuting Hamiltonians~\cite{kitaev2006} and thus cannot be realized in fixed point wave-functions (with finite-dimensional local Hilbert spaces). Nonetheless, here we have shown that a chiral UMTC can, in fact, arise in a \textit{mixed state}.

\subsection{Three-Fermion from $\mbZ_2 \times \mbZ_2$ Toric code}
\label{sec:3F}

Take the initial pure state to be a ground state of the $\mbZ_2 \times \mbZ_2$ Toric code. The anyons in this theory form a $\mbZ_2^4$ group under fusion, with elements $\{1,e_1,m_1,f_1 \} \times \{1,e_2,m_2,f_2\}$. As noted in Ref.~\cite{bombin2012}, the anyon types can be relabeled $\{1, f_1, e_1 f_2, m_1 f_2 \} \times \{1, f_2, f_1 e_2, f_1 m_2 \}$ which is equivalent to two copies of the \textit{Three-Fermion} (3F) anyon theory ($f_1 = e_1 m_1$ and $f_2 = e_2 m_2$ are fermions). The 3F anyon theory is a chiral Abelian UMTC which contains the anyons $\{1,\psi_1,\psi_2,\psi_3 \}$ where $\theta(\psi_i) = -1$ for $i=1,2,3$, and with the braiding between the fermions given by $\mathcal{B}_\theta(\psi_i,\psi_j) = -1$ for any $i,j = 1,2,3$. 

We now wish to ``peel off" a single 3F theory (which is a chiral Abelian UMTC) from the parent topological order. For this, we need to identify a set of anyons $\hat{\mathcal{A}}$ that braid trivially only with three fermions in the parent TO. One can choose the set $\hat{\mathcal{A}} = \{1, f_1, f_2 e_1, f_1 f_2 e_1 \}$. Maximally decohering the initial density matrix with respect to these error channels amounts to gauging these anyons out, with the resulting anyon theory $\mathcal{A} = \{1, f_2, e_2 f_1, e_2 f_1 f_2 \}$ identical to a single 3F anyon theory. Due to the self and mutual braiding statistics of this theory, its logical subspace encodes 2 logical qubits.

As a final remark, finite temperature mixed-states also provide simple instances of our general framework. For example, consider the $D=2,3,4$ Toric code at finite temperature~\cite{castelnovo2007classical,lu2020negativity}. For $D=2$, any finite-T state corresponds to both the $e,m$ anyons being incoherently proliferated: the resulting state hosts no deconfined anyons and is hence trivial. Now, in $D=3$, the $e$ charges proliferate at any non-zero temperature, but below a critical $T_c$, the flux-loops of the 3D Toric code remain deconfined but are now transparent. This corresponds to a TSSC that does not encode any logical qubits in its logical subsystem but still has a non-trivial classical memory due to the transparent loops. Finally, for the 4D Toric code which has only loop-like excitations, there exist two critical temperatures: below the first, none of the excitations proliferate and the finite-T mixed-state is a TSSC that is equivalent to a topological stabilizer code (i.e., its gauge group is proportional to the stabilizer group). Hence, it represents finite-T topological order. Above the first, but below the second critical temperature, only one of the loop excitations proliferate and one obtains a TSSC with a classical memory. Above the second critical temperature, all anyons are condensed and the mixed-state is topologically trivial. Thus, prior results on topological order at finite temperature are straightforwardly incorporated into our general framework. We note that an infinite temperature state with quantum memory based on a subsystem code was previously proposed in Ref.~\cite{domtom}, and in our framework, constitutes an imTO.

In general, the map from decohered density matrices to TSSCs conveniently allows one to use results regarding the latter to obtain a partial classification of the former. In particular, since Ref.~\cite{ellison2023} showed that any (non-modular) Abelian anyon theory can be realized by a TSSC, it immediately provides a partial classification of Abelian imTOs in terms of non-modular Abelian anyon theories. 


\section{Decoherence as gauging out in general UMTCs}
\label{sec:general}

While the precise relation between imTOs and TSSCs can only be made in the context of Abelian anyon theories with gappable boundaries, we expect that the general relation between decoherence induced imTO, anyon condensation in a doubled Hilbert space, and ``gauging out" anyons should hold more generally. Indeed, gauging out a proper subset of anyons $\hat{\mathcal{A}}$ from a parent UMTC $\mC$ is nothing but anyon condensation in a doubled Hilbert space\footnote{This is distinct from anyon condensation in a \textit{physical} bilayer.} with the resulting deconfined anyons $\mathcal{A}$ given by those that have support purely in the ket or bra space. We will show through examples that gauging out Abelian anyons in an otherwise non-Abelian theory is conceptually straightforward. Similarly, Abelian anyons can also be gauged out from parent chiral UMTCs. This suggests the intriguing possibility of realizing non-modular anyon theories by appropriately gauging out anyons from a UMTC. We schematically describe this below, leaving a complete algebraic decription for future work.

Let us assume that we always begin with a pure state that is the ground state of some local, gapped Hamiltonian in 2D. That is, our parent theory has TO characterized by a UMTC $\mC$ with a finite set of anyons $\{a\}$. As is well-established by now, in the doubled Hilbert space this corresponds to the doubled TO $\mC \times \bar{\mC}$, with anyons labeled by the ordered pair $a \bar{b} = (a_+b_-)$. Note that the TO in the doubled space is nothing but the Drinfeld centre of $\mC$: $\mathcal{Z}(\mC) = \mC \times \bar{\mC}$. The theory in the doubled space is then equivalent to that of a string-net model~\cite{levinwen}, for which the input theory is the UMTC $\mC$. In such a theory, it is always possible to condense excitations of the form $(a_+ a_-)$, which are obviously bosonic~\cite{bonderson2012thesis, burnell2011,burnell2012,cagenet}. In the physical Hilbert space, this corresponds to subjecting the initial pure state to local error channels, which can be written in terms of short string operators for $a \in \mC$. 

In the doubled Hilbert space, maximal decoherence corresponds to conventional anyon condensation~\cite{burnellreview}, whereby any anyons $(r_+s_-)$ that braid non-trivially with $(a_+a_-)$ are confined and, of the resulting anyons, those that differ only up to fusion by $(a_+a_-)$ are identified. For any non-Abelian anyons that remain deconfined, one must also check their fusion rules: if the the vacuum superselection sector appears more than once, then the non-Abelian anyon splits into other deconfined anyons. From our preceding discussion, we know that the resulting mixed-state TO is encoded in the set of anyons with support only on the ket (or bra) space (the Wilson loops of the remainder are frozen into classical ensembles). These are given by the set $\mathcal{A} = \{r\in \mC | \mathcal{B}_{\theta}(a,r) = 1\}$, of which some may be transparent anyons i.e., the resulting anyon theory may be non-modular or even chiral, both of which we have already encountered. 

Thus, we can now define gauging out anyons in the same way as before, but in a more general context: starting with a UMTC $\mC$ and a proper subset of anyons $\hat{\mathcal{A}}$ to be gauged out, the resulting anyon theory (the code space of the decohered theory) is given by those anyons in $\mC$ which braid trivially with those in $\hat{\mathcal{A}}$. Moreover, if any anyons in $\hat{\mathcal{A}}$ are transparent in $\hat{\mathcal{A}}$, they remain transparent in $\mathcal{A}$, which will generically be a braided fusion category (without the modularity restriction). Formally, given a UMTC $\mC$ and a proper subset of objects (anyons) $\hat{\mathcal{A}}$ (i.e., a full subcategory of $\mC$), the anyon theory $\mathcal{A}$ that results upon gauging out $\hat{\mathcal{A}}$ is given by the centralizer $C_C(\hat{\mathcal{A}})$ of $\hat{\mathcal{A}}$ in $\mC$:
\beq
\mathcal{A} \equiv \{ x \in \mC | \mathcal{B}_\theta(x,y) = 1 \, \forall y \in \hat{\mathcal{A}} \}, 
\eeq
which is a braided fusion category (see Ref.~\cite{mueger2002}). One could in principle then generate another braided fusion category by gauging out anyons from $\mathcal{A}$ and generate a cascade of imTOs by iteratively gauging out anyons. As discussed in Section~\ref{sec:tssc}, we define a density matrix with imTO as one where $\mathcal{A}$ is non-modular, even in the non-Abelian case.

We believe that this picture for obtaining braided fusion categories from parent UMTCs falls squarely within the general class of \textit{mixed topological quantum field theories (TQFTs)} proposed by Zini and Wang in Ref.~\cite{wangmixedtqft}, but where the input to the parent 2+1D Turaev-Viro (TV) type TQFT is always modular. In our context, this restriction is physically motivated since we take as input the ground state of a local Hamiltonian (so the anyon theory is a UMTC) and then subject it to \textit{local} noise. 
In fact, note that the doubled semion example we previously considered, in which the resulting mixed state supports the chiral semion TO, is presented as an example of a mixed TQFT in Ref.~\cite{wangmixedtqft}. Specifically, in that case the input was the doubled semion UMTC and the output mixed TQFT was the chiral semion UMTC, where we can view the latter as the Reshetikhin-Turaev (RT) TQFT of the former. Ref.~\cite{wangmixedtqft} described this as ``tracing" out the anti-chiral degrees of freedom, which we believe corresponds to the process of gauging out presented above. This supports our claim that (unitary) non-modular braided fusion categories provide a partial classification of imTOs. More generally, if the parent theory is some doubled Chern-Simons topological quantum field theory (which admits a gapped boundary to vacuum), one might expect that local error channels will lead to the underlying chiral Chern-Simons TQFT in the decohered mixed-state---we hope to investigate this general procedure in the context of a continuum field theory description in a forthcoming work. For now, we consider some simple examples that go beyond Abelian anyon theories to show the generality of our framework.

\subsection{$\mbZ_2^{(1)}$ TSSC from Chiral Ising UMTC}
\label{sec:isingchiral}

The chiral Ising anyon theory consists of the anyons $\{1, \sigma, \psi \}$ which satisfy the fusion algebra $\psi \times \psi = 1$, $\sigma \times \psi = \psi \times \sigma = \sigma$, and $\sigma \times \sigma = 1 + \psi$. Here, $\psi$ is a fermion and $\sigma$ is an Ising anyon, whose non-integer quantum dimension $\sqrt{2}$ reflects its non-Abelian nature. The topological spin (self-statistics) of the theory are $\theta(\psi) = -1$ and $\theta(\sigma) = e^{i \pi/8}$ from which, combined with the fusion rules, one can derive the non-trivial braiding between $\psi$ and $\sigma$: $\mathcal{B}_{\theta}(\sigma,\psi) = -1$.

A physical Hamiltonian that supports a phase with chiral Ising TO is furnished by the Kitaev honeycomb model~\cite{kitaev2006}. We consider gauging out the $\psi$ fermion: since $\sigma$ braids non-trivially with $\psi$, the resulting anyon theory describing the decohered state is simply given by $\{1,\psi\}$, which does not encode any quantum memory but still yields a classical memory and retains a well-defined fermionic excitation. 

We may also consider the corresponding analysis in the doubled Hilbert space. Here, we are condensing $(\psi_+ \psi_-)$ in the doubled Ising Chern-Simons theory $\mC \times \bar{\mC}$. It is well-known that the condensed phase has the following deconfined excitations: $\{1, \psi_+,\psi_-,\sigma_+ \sigma_-\}$ where $\sigma_+ \sigma_-$ splits since the vacuum sector appears twice in its fusion rules: $\sigma_+ \sigma_- \times \sigma_+ \sigma_- = 1 + \psi_+ + \psi_- + \psi_+ \psi_-$ (where $\psi_+ \psi_- \sim 1$) which is identical to the fusion $(e + m) \times (e + m)$ in the $\mbZ_2$ Toric code. Thus, the TO in the doubled Hilbert space is a $\mbZ_2$ gauge theory, but back in the physical Hilbert space, this corresponds to the freezing of $\sigma$ loops into a classical ensemble while $\psi$ remains a well-defined excitation.

\subsection{Non-modular imTO from Doubled Ising UMTC}
\label{sec:isingdouble}

Building on the previous example, let us now consider a pure state which belongs to the ground state manifold of the doubled Ising string-net~\cite{levinwen}. The anyons in this theory are $\{1,\psi,\sigma\} \times \{1,\bar{\psi},\bar{\sigma}\}$ with fusion rules that can be inferred from those of the chiral Ising UMTC. Now, suppose we wish to consider $\psi \bar{\psi}$ errors: these are induced by local short-string operators that are explicitly provided in e.g. Ref.~\cite{cagenet}. As above, we will not delve into details of the specific lattice Hamiltonian or the short-string operators here as we can directly infer the imTO of the decohered density matrix.

Maximal decoherence of the $\psi \bar{\psi}$ errors is equivalent to gauging out this bosonic anyon. As before, only those excitations that braid trivially with $\psi \bar{\psi}$ remain as deconfined anyons in the resulting decohered state. Thus, the resulting mixed-state TO is given by the set $\mathcal{A} = \{1, \psi, \bar{\psi}, \sigma \bar{\sigma},\psi \bar{\psi} \}$. Notably, this is distinct from typical anyon condensation of $\psi \bar{\psi}$ in the doubled Ising string-net, where $\psi \bar{\psi}$ disappears into the condensate, $\psi$ and $\bar{\psi}$ are identified, and $\sigma \bar{\sigma}$ splits into Abelian anyons. Decohering $\psi \bar{\psi}$ instead results in a \textit{non-modular} imTO, characterized by the anyons $\mathcal{A}$, amongst which $\psi \bar{\psi}$ is transparent~\footnote{For the cognoscenti, we note that maximally decohering $\psi \bar{\psi}$ in the doubled Ising string-net results in precisely the same anyon content as in each layer of the Ising cage-net~\cite{cagenet}; this suggests that ``p-string" condensation~\cite{ma2017pstring} may have an interpretation in terms of gauging out certain anyons.}. We can infer the presence of a quantum memory in the logical subsystem of the decohered code space from the presence of non-trivial braiding between the remaining opaque anyons in $\mathcal{A}$.

\subsection{Non-modular imTOs from Doubled $SU(2)_k$ UMTC}
\label{sec:SU2}

As a final example, we can consider doubled SU(2)$_k$ string-net models, whose lattice models and short-string operators are given in Ref.~\cite{burnell2011}. Anyons in this theory are labeled by pairs $(j_1,j_2)$ where $j=0,1/2,1,\dots,k/2$. Let us now subject a ground state of this model to local errors that incoherently proliferate the anyon $(k/2,k/2)$ (which is a boson). In order to read off the resulting imTO in the decohered density matrix, after $(k/2,k/2)$ has been gauged out, we require the braiding relations of this theory. In particular, the braiding between an anyon $(j_1,j_2)$ and $(k/2,k/2)$ is given by $\mathcal{B}_\theta((j_1,j_2),(k_1,k_2)) = (-1)^{2(j_1 + j_2)}$. Thus, the resulting imTO is characterized by the anyon theory $\mathcal{A} = \{ (j_1,j_2)| j_1 + j_2 \in \mbZ \}$ with $j_1,j_2 = 0,1/2,1,\dots,k/2.$ Of these, $(k/2,k/2)$ is a transparent boson, which is sufficient to conclude that the decohered theory is a \textit{non-modular} anyon theory.

Thus, we can obtain a large family of imTOs by exposing the ground states of string-net models to local error channels, where the decohered code space generically retains logical information i.e., it is a decoherence-free subspace. The presence of non-trivial logical information (or a quantum memory) is encoded in the Wilson-loop algebra (equivalently, the $S$-matrix of $\mathcal{A}$). Since we have shown that the resulting imTOs can host transparent anyons (corresponding to a non-modular anyon theory), we obtain a partial classification of imTO in terms of non-modular unitary braided fusion categories. 


\section{Locally Detectable Anyons}
\label{sec:excitations}

Thus far, we have characterized the topological order exhibited by mixed states in terms of their anyon data $\mathcal{A}$ i.e., the set of anyons whose closed Wilson loops remain coherent after subjecting the original pure state to decoherence. In the Abelian case, these correspond to the stabilizer group and, as we will discuss below, in general these correspond to the set of strong 1-form symmetries respected by the mixed state. However, one could also characterize topological order in terms of the distinct anyon types which remain as locally detectable \emph{excitations} outside the code space~\footnote{We thank Michael Levin for a discussion prompting us to include this perspective.}. Indeed, in pure state topological order, the detectable anyon types are generated by open Wilson lines, which are in one-to-one correspondence with the Wilson loops generating the ground state degeneracy on the torus. In contrast, while in the present context there is no Hamiltonian and hence no notion of an excitation gap (see however Ref.~\cite{sang2024stable}), one may still identify states (outside the code space) with local errors as ``excited states" of the mixed state. As we now show, if the TO characterizing the mixed state $\mathcal{A}$ is modular, then $\mathcal{A}$ also describes the set of locally detectable anyon types. However, if $\mathcal{A}$ is non-modular, then $\mathcal{A}$ is no longer in one-to-one correspondence with the set of locally detectable anyons. The physical meaning of this will be made clearer in the following section, when we identify imTO as corresponding to the surface of a Walker-Wang model.

Let us consider a mixed state $\rho$ obtained via decoherence of an anyon $a$ in a parent TO $\mathcal{C}$. Then, from the preceding discussion, $\rho$ is characterized by the (potentially non-modular) TO $\mathcal{A}$ obtained by gauging out $\hat{\mathcal{A}} = \{ a^m \}$ in $\mathcal{C}$. Now, let $c$ be an anyon in the \emph{parent} TO $\mathcal{C}$ and $W_{x_1,x_2}^c$ the Wilson line operator creating $c$ and its conjugate $\overline{c}$ at well-separated points $x_1$ and $x_2$. Likewise, let $W_\Gamma^d$ be the operator creating a Wilson loop of $d \in \mathcal{C}$ along the contractible cycle $\Gamma$. Then we will call 
\begin{align}
    \rho_c = W_{x_1,x_2}^c \rho (W_{x_1,x_2}^c)^\dagger
\end{align}
an ``excited state" relative to $\rho$ if we can detect the presence of $c$ via braiding with some anyon $d$. That is, we wish to compute
\begin{align}
    \mathrm{Tr}[W_\Gamma^d \rho_c] = \mathrm{Tr}[W_\Gamma^d W_{x_1,x_2}^c \rho (W_{x_1,x_2}^c)^\dagger ] = \mathcal{B}_\theta(d,c)  \mathrm{Tr}[W_\Gamma^d \rho] \, ,
\end{align}
where $\Gamma$ encloses only, say, the point $x_1$ and we have used the braiding between anyons $c$ and $d$. In order for this expression to be non-zero, we require $W_\Gamma^d$ to be a stabilizer of $\rho$ and hence $d$ must be in $\mathcal{A}$. Note that there is no such restriction on $c$. However, $c$ is an excitation if and only if $\mathcal{B}_\theta(d,c)\neq 1$. Note that this means that \emph{transparent} anyons in $\mathcal{A}$ are not locally detectable anyon types while all of the opaque anyons in $\mathcal{A}$ represent genuine quantum excitations. In particular, any two anyons in the parent theory $\mathcal{C}$ which remain detectable after decoherence and differ by fusion with a transparent anyon become indistinguishable as excitations. Thus, given an imTO $\mathcal{A}$ that results from incoherently proliferating anyons in a UMTC $\mathcal{C}$, the set of locally detectable anyons is given by
\begin{align}
    \mathcal{L} = \{c \in \mathcal{C} | \mathcal{B}_\theta(c,d) \neq 1 \text{ for any } d \in \mathcal{A} \} \, ,
\end{align}
with the identification that $r \sim r \otimes t$, where $r \in \mathcal{L}$ and $t \in \mathcal{T}$, the set of transparent anyons. The mathematical structure underlying $\mathcal{L}$ remains mysterious; note for instance that it need not even be closed under fusion!

In the doubled space picture, we thus see that the locally detectable anyon types in the original Hilbert space correspond to anyons of the form $(c_+,c_-)$, while the observables (namely the Wilson loops which detect other anyons via braiding and hence the stabilizers) are in correspondence with anyons of the form $(d_+,1)$. 

Let us understand the structure of the set of locally detectable anyons by way of a few examples. First consider a parent TO that can be written as a product of two modular TOs, $\mathcal{C} = \mathcal{A} \boxtimes \hat{\mathcal{A}}$, and subject it to decoherence that gauges out $\hat{\mathcal{A}}$ -- the DS theory discussed above is one such example. Since, by definition, every $d \in \mathcal{A}$ braids trivially with every $c \in \hat{\mathcal{A}}$, the only locally detectable and hence genuine quantum excitations, are labelled by the anyons in $\mathcal{A}$. Thus, for a modular theory, the set of locally detectable anyons types is in one-to-one correspondence with the braided fusion category $\mathcal{A}$ characterizing the mixed state TO.

This correspondence does not hold for non-modular imTO. Indeed, let us consider the simple example of the Toric code subjected to $e$ deocherence, yielding the $\mathbb{Z}_2^{(0)}$ mixed state discussed above. While the $\mathbb{Z}_2^{(0)}$ is characterized by the anyon content $\{1,e\}$, the $e$ anyon does \emph{not} label a genuine quantum excitation, as it is a transparent anyon. This is trivially seen from the fact that $Z_i \rho Z_i = \rho$. In contrast, the original $m$ and $f$ anyons of the parent Toric code do exist as genuine quantum excitations, as they may be detected via expectation values of $W^e_\Gamma$ Wilson loops~\footnote{Note that while the $e$ anyons do not exist as genuine quantum excitations, we can nevertheless use braiding of $e$ to detect other anyons.}. Moreover, the $m$ and $f$ anyons are identified as excitations in the decohered theory, as there are no stabilizers that can distinguish their spin.

While for a generic imTO, the anyon theory (equivalently, the braided fusion category) $\mathcal{A}$ describing the mixed state $\rho$ does not label the set of locally detectable anyon excitations, the latter is still fully determined by the former. As such, the set $\mathcal{L}$ provides a finer characterization of the imTO, which should be a feature of all states within the same imTO phase and not simply fixed point states. We elaborate on this in the following section in the language of anomalies between strong and/or weak symmetries. This framework also clarifies how the set of locally detectable anyon types can be determined simply by knowing the symmetries of $\rho$ and their associated anomalies.


\section{Higher-Form Symmetry and non-unitary exfoliation of Walker-Wang models}
\label{sec:higherform}

We now place our results in a broader context by characterizing imTO states via their \emph{higher-form} symmetry structure \cite{gaiotto2015generalized,mcgreevy2023generalized}, which we have already alluded to above in specific examples, and by relating them to anomalous surface states of 3D pure state TO. First, we recall that $q$-form symmetries are generated by operators acting on a closed, codimension $q-1$ manifold of spacetime. In the $2+1$-dimensional case, 1-form symmetries thus are both generated by, and act on, one-dimensional loop-like objects. Indeed, in a 2D TO, the Wilson loops associated with (Abelian) anyons may be understood as being generators of 1-form symmetries. 
In this language, the non-trivial ground state degeneracy---and hence the non-trivial code space---of a TO on the torus is often understood in terms of the spontaneous breaking of these 1-form symmetries. For instance, the Toric code possesses a $\mathbb{Z}_2^e \times \mathbb{Z}_2^m$ 1-form symmetry, generated by the $e$ and $m$ Wilson loops. Like conventional symmetries, 1-form symmetries can be gauged which, in the context of 2D TO, amounts to condensing the corresponding anyon~\footnote{The relation between gauging 1-form symmetries and anyon condensation only holds for Abelian anyons; more generally, condensing non-Abelian anyons can be understood in terms of gauging \textit{non-invertible} 1-form symmetries.}. Thus, a 1-form symmetry is anomalous if the corresponding anyon has non-trivial self-statistics (i.e., is not bosonic). In the $\mbZ_2$ Toric code, the $\mathbb{Z}_2^e$ and $\mathbb{Z}_2^m$ 1-form symmetries are hence not individually anomalous, as we may gauge either to obtain a trivial state; correspondingly, we may condense either of these anyons. Instead, the 1-form symmetries for $e$ and $m$ have a mixed anomaly, reflecting the non-trivial braiding between $e$ and $m$, and that we cannot condense $f$. 

In order to extend this analysis to mixed-state order, 
we must distinguish between \emph{strong} and \emph{weak} symmetries of a density matrix~\cite{buca2012,albert2014sym,degroot2022og}. Given a unitary representation $U_g$ of a symmetry $g$ in some symmetry group $G$, we say that the density matrix $\rho$ is strongly symmetric under $G$ if for every $g \in G$, $U_g \rho = \rho U_g^\dagger = \rho$. In the doubled space picture, this constraint translates to $U_{g+}\kket{\rho} = U_{g-}^* \kket{\rho} = \kket{\rho}$. Conversely, $\rho$ is weakly symmetric if we only have $U_g \rho U_g^\dagger = \rho$ or, equivalently, $U_{g+}U_{g-}^*\kket{\rho} = \kket{\rho}$. 

Let us focus on the Toric code first for concreteness. The initial pure state TO density matrix trivially has a \emph{strong} $\mbZ_2^e \times \mbZ_2^m$ 1-form symmetry. Working in the doubled space picture, we then see that the $\mbZ_2^{(0)}$ imTO resulting from $e$-decoherence still has a strong $\mbZ_2^e$ 1-form symmetry generated by Wilson loops associated to the anyon $e_+$, but only a \emph{weak} $\mbZ_2^m$ 1-form symmetry, generated by $m_+ m_-$. Indeed, in each of the examples we have studied, we see that decoherence has a non-trivial effect on the underlying strong 1-form symmetries of the parent TO. We can thus rephrase our results for generic Abelian TOs in the language of 1-form symmetry: when gauging out an anyon $a$ via decoherence, the resulting imTO density matrix retains \emph{strong} 1-form symmetries for those symmetries generated by anyons which braid trivially with $a$, while the remaining 1-form symmetries are reduced to \emph{weak} symmetries. In other words, only those 1-form symmetries which do not have a mixed anomaly with the 1-form symmetry generated by Wilson loops of $a$ remain as strong 1-form symmetries, while the remainder are reduced to weak symmetries. As noted previously, in the TSSC framework, the strong 1-form symmetries of the decohered state are manifest in Eq.~\eqref{eq:strongsym}, where the stabilizers may be viewed as closed Wilson loops for the deconfined anyons. This observation further reinforces the idea that the logical subsystem forms a decoherence free subspace under local noise that incoherently proliferates certain anyons, namely those affecting only degrees of freedom in the gauge subsystem.

In this language, we can hence rephrase our characterization of imTO as follows: a density matrix $\rho$ exhibits imTO if its set of strong 1-form symmetries form a non-modular anyon theory -- that is, its set of strong 1-form symmetries cannot be consistently realized in the ground state of a gapped, local 2D Hamiltonian. The 1-form symmetry structure of imTOs provides a useful language for characterizing the utility of these states as quantum memories. Note that each strong 1-form symmetry implies the existence of non-local operators commuting with the stabilizer group---the corresponding anyon Wilson loop along the non-contractible cycles of the torus. We may then employ the anomalies of the 1-form symmetries to characterize the structure of the code space. Specifically, if two strong 1-form symmetries have a mixed anomaly, they give rise to a pair of logical operators and hence a quantum memory, as in the $\mbZ_4^{(1)}$ TSSC. If a strong 1-form symmetry has no mixed anomalies but has a $\mbZ_N$ anomaly with $N>2$, its corresponding Wilson loops along the two cycles of the torus also yield logical operators and a quantum memory, as in the chiral semion TSSC. Finally, if a strong 1-form symmetry has no mixed anomalies and at most a $\mbZ_2$ anomaly (i.e. it is either a boson or fermion), its corresponding non-contractible Wilson loops only yield non-local stabilizers, thus yielding a classical memory. Higher form symmetries thus provide a convenient language with which to characterize the TSSC structure of Abelian imTOs.

Indeed, the mixed anomalies between strong and weak 1-form symmetries play a central role in determining the set of locally detectable anyons $\mathcal{L}$ in the imTO. In particular, any strong 1-form symmetry of a state $\rho$ that has a $\mathbb{Z_N}$ self-anomaly (with $N>2$) is associated with a locally detectable anyon. If any strong 1-form symmetries have a mixed anomaly, they too correspond to anyons in the imTO. One can think of such anomalous strong 1-form symmetries as corresponding to $\mathcal{A} \backslash \mathcal{T}$ i.e., the anyons in the (generically non-modular) braided fusion category $\mathcal{A}$ minus the set of transparent anyons $\mathcal{T}$ in $\mathcal{A}$. Finally, weak 1-form symmetries which have a mixed anomaly with a strong 1-form symmetry also correspond to locally detectable anyons (see Sec.~\ref{sec:excitations}); while the transparent anyons do not have any self anomalies (with $\mathbb{Z}_{N>2}$) or mixed anomalies with other strong symmetries, they can have mixed anomalies with weak 1-form symmetries and therefore influence the structure of $\mathcal{L}$. Hence, the anomaly structure of the strong and weak symmetries of a short-range correlated 2D mixed state $\rho$ provides a detailed characterization of the corresponding imTO.

There is also a striking analogy between imTO and anomalous surface states of certain pure state 3D TOs, which suggests potential generalizations of our scheme to other intrinsically mixed-states. This also allows us to generalize the discussion from the preceding paragraph to the non-Abelian case. As we have emphasized throughout, imTO generally supports chiral and non-modular TO in a purely 2D system. In the context of local gapped Hamiltonians, such states naturally arise at the 2D surfaces of 3D topological orders, specifically those realized in the Walker-Wang (WW) models~\cite{walkerwang}. These are 3D exactly solvable lattice models which, given a potentially non-modular TO $\mA$, realize $\mA$ as its surface theory. In particular, if we consider a slab geometry with open boundary conditions in, say, the $z$ direction, one obtains $\mA$ on the top surface and $\bar{\mA}$ on the bottom surface. If $\mA$ is modular, then the bulk has trivial topological order. Conversely, if $\mA$ is non-modular, then the bulk is topologically ordered and supports both point-like and loop-like excitations, generated at the ends of Wilson lines and edges of Wilson surfaces, respectively, which braid non-trivially with each other. These loop-like excitations can be absorbed by the surfaces. Importantly, the transparent anyons in $\mA$ also correspond to deconfined point-like excitations in the bulk, and so can freely move from the top $\mA$ surface, into the bulk, and onto the bottom $\bar{\mA}$ surface. 
Additionally, a ``tube-like" Wilson surface stretching between a loop on the top surface and a loop on the bottom surface serves as a symmetry of the ground state, as the loop-like excitations are condensed on the surfaces. 

Remarkably, this structure \textit{exactly} parallels that of the vectorized density matrix for an imTO in the doubled Hilbert space, with the ket and bra spaces identified with the top and bottom surfaces of a WW model. Much like the surface states of WW models, the deconfined excitations with support solely on the ket (or bra) space can realize non-modular or chiral TO. The aforementioned weak 1-form symmetries (which act simultaneously on the ket and bra spaces) mirror the effect of the tube-like Wilson surfaces in the WW model, when they terminate on the top and bottom surfaces. Moreover, in the doubled Hilbert space representation of the imTO, the transparent anyons can move freely between the ket and bra spaces, just as the transparent anyons in the WW model can move between the top and bottom surfaces. Indeed, in the doubled Hilbert space, if we condense $(a_+ a_-)$, all anyons of the form $(a_+^m 1_-)$ (for integer $m$) are transparent and are equivalent to anyons of the form $(1_+ a_-^m)$ via fusion with the condensate.

This picture also provides a convenient way of understanding the set of detectable anyons discussed in the preceding section. Indeed, focusing on a single surface of a WW model in a slab geometry, there are two classes of surface excitations which can be detected via braiding with anyons in the surface anyon theory $\mA$ -- that is to say, detectable via computing the expectation value of a closed Wilson loop of an anyon $d \in \mA$. The first are the opaque anyons in $\mA$, generated by open Wilson lines on the surface; by definition, the transparent anyons are undetectable via braiding with anyons in $\mA$. The second class is generated by open Wilson surfaces in the bulk, the boundaries of which intersect the surface on an open line -- the endpoints correspond to point-like excitations on the surface~\footnote{While these loop-like excitations are strictly speaking confined as there is an energy cost proportional to the length of the boundary of the Wilson surface, they still nominally exist as excitations in the boundary of the WW model.}. These are in one-to-one correspondence with the tube-like Wilson surfaces mentioned above. These two sets of surface excitations exactly correspond to the detectable anyons of the imTO characterized by $\mA$ discussed previously. Indeed, in the analogy with the double space picture, the two sets of excitations in the WW model correspond to anyons of the form $(c_+ , c_-)$ such that $c$ braids non-trivially with some $d \in \mA$; the genuine surface Wilson lines in the WW model map to those anyons in the double space such that $c$ braids trivially with the set of gauged out anyons $\hat{\mA}$, while the bulk Wilson surfaces generating surface excitations map to those anyons in the double space such that $c$ braids non-trivially with $\hat{\mA}$.

Thus, at least at the level of analogy, decoherence induced imTO provides a physical means of realizing anomalous surface states of 3D pure state TO as realized by WW models. This lends further credence to our claim that a partial classification of imTOs is provided by non-modular unitary braided tensor categories, since the classification of WW models includes these. One may view decoherence as a means of ``exfoliating" surface states of a 3D TO into a purely 2D mixed-state via finite-depth LPQCs. Indeed, we can also imagine applying such non-unitary exfoliation to isolate anomalous surface states of other exotic pure states, an avenue we intend to pursue in future work.

\subsection*{imTO Phases}
Finally, let us comment on the extent to which the imTOs we have discussed thus far constitute genuine mixed-state \textit{phases} and not simply fixed-point states. We note that this is a subtle question, as there is as of yet no consensus on what constitutes a mixed-state phase of matter. This requires a sharp notion of an equivalence relation that determines when two density matrices lie within the same phase, and there exist several proposals in the literature~\cite{coser2019class,sang2023mixed,rakovszky2023stable,sang2024stable,ma2024symmetryprotectedtopologicalphases}. For simplicity, we restrict to the Abelian case here. As we have emphasized, the 1-form symmetries (strong and weak) and their associated anomalies encode both the braided fusion category data $\mathcal{A}$ as well as the locally detectable anyon types $\mathcal{L}$. As such, these should constitute invariants of a given mixed-state phase which, in the case of a non-modular $\mathcal{A}$ provides an instance of an imTO phase.

Consider the ``two-way connectivity" relation described in Ref.~\cite{sang2023mixed}: namely, two short-range correlated mixed-states $\rho_1$ and $\rho_2$ are in the same mixed-state phase if there exist quasi-local quantum channels $\Sigma_{12}$ and $\Sigma_{21}$ such that $\rho_1 = \Sigma_{12} \rho_2$ and $\rho_2 = \Sigma_{21} \rho_1$ (see~\cite{sang2023mixed} for a precise definition of a quasi-local channel). Physically, this relation encodes the fact that while such channels can trivialize long-range correlations, they cannot create them in an arbitrarily short amount of time. Now, note that the $\mathbb{Z}_2^{(0)}$ imTO (obtained from $Z$-decoherence of the $\mathbb{Z}_2$ Toric code) can be considered to be a purely classical state. Indeed, it is completely separable. Further, when defined on a spatial manifold with trivial genus -- the plane or the sphere -- one can show this state is two-way connected to a trivial product state and hence, under the equivalence relation of Ref.~\cite{sang2023mixed},  belongs to the trivial phase. However, as discussed in Section~\ref{sec:excitations}, there exist local, detectable excitations (i.e. errors) above this state, which reflect the mixed anomaly between the strong electric and weak 1-form symmetries. Moreover, when placed on a torus, the corresponding code space encodes two classical bits of information; while we do not have an explicit proof, this fact suggests that certain states within the code-space cannot be two-way connected to the trivial state via a finite-depth local quantum channel on the torus. 

More generally, on physical grounds we expect that logical information (encoded in anomalies between the strong 1-form symmetries) will remain robust under quasi-local quantum channels, since these satisfy a Lieb-Robinson bound and cannot generate arbitrary long-range correlations that destroy this information at infinitesimally small noise rates~\cite{brownreview}. On the other hand, classical information (encoded in the mixed strong-weak 1-form anomalies) can be smoothly erased as already seen in the example above. Equivalently, the transparent anyons need not be preserved 
and, since these are required to capture the complete set of locally detectable anyon types, the set $\mathcal{L}$ is not an invariant under this definition of a mixed-state phase. Note that this subtlety does not arise when $\mathcal{A}$ is modular (since $\mathcal{L} = \mathcal{A}$) and only exists for intrinsically mixed states. Therefore, we contend that a finer equivalence relation is required to accurately describe the invariant data of imTO phases and to distinguish between ``classical" imTO phases (such as the $\mathbb{Z}_2^{(0)}$ Toric code) from purely trivial states. For instance, an equivalence relation based on the Markov length under local Lindbladian evolution could provide an alternative definition for mixed state phases~\cite{sang2024stable}. We leave a detailed investigation of this matter to future work and speculate no further.


\section{Discussion}
\label{sec:cncls}

In this work, we have proposed a framework for classifying a large family of intrinsically mixed-state topological orders, obtained via local decoherence of parent pure state topological order. We demonstrated that local decoherence, previously shown to correspond to anyon condensation in the vectorized density matrix obtained via the Choi-Jamio{\l}kowski isomorphism, in fact provides a physical mechanism for the gauging out of anyons. As a consequence, for parent Abelian topological order, the resulting imTO is naturally characterized as a topological subsystem code and thus classified in terms of (degenerate) braided tensor category theory. Hence, 2D pure state TOs provide resource states, under \textit{local decoherence}, for the preparation of non-modular and even chiral states. We also illustrated that this procedure naturally extends to non-Abelian states, though the resulting imTOs are no longer identified as TSSCs. Finally, we characterized the family of imTOs under consideration by their strong and weak 1-form symmetries, and demonstrated that they correspond to the anomalous surface states of 3D pure state topological orders, to wit, Walker-Wang models. This provides a natural interpretation of decoherence as a means of non-unitarily exfoliating surface states of topological states in one higher dimension, a perspective which may find use in generating other classes of intrinsically mixed phases of matter. Our general framework provides many exciting avenues for further exploration, some of which we address in forthcoming work.

As we have discussed in Sec.~\ref{sec:higherform}, perhaps the most outstanding question is establishing an equivalence relation on the space of short-range correlated mixed-states in 2D that provides a clear notion of imTO \textit{phases}. Based on our work, we expect that all states within the same phase should share the same 1-form anomaly structure, with imTO phases distinguished by the presence of non-modular strong 1-form symmetries. Importantly, this data encodes not only the (non-modular) braided fusion category $\mathcal{A}$ but also the set of locally detectable anyons $\mathcal{L}$. Since an equivalence relation that requires two-way channel connectivity~\cite{sang2023mixed} is only sensitive to the logical information (i.e., anomalies between strong 1-form symmetries), a finer diagnostic is required to also capture the classical information (i.e., mixed strong-weak 1-form anomalies). It is unclear whether requiring two-way channel connectivity on any closed manifold will suffice, or if the equivalence relation based on the Markov length along local Lindblad evolution~\cite{sang2024stable} is more suitable.

A pressing issue is to characterize our family of imTOs via their entanglement structure. While the entanglement entropy has previously been studied in mixed-state TO~\cite{castelnovo2007classical,hermanns2014classical}, a more natural probe of entanglement in mixed-states is the entanglement negativity which, unlike the entanglement entropy, is a good measure of quantum correlations in a mixed state~\cite{peres1996,zyczkowski1998,eisert1999,vidal2002,plenio2005}.  
In pure state TO, the negativity receives universal contributions which are sensitive to the modular data of the TO (namely, the total quantum dimension)~\cite{kitaev2006tee,levin2006tee,lee2013negativity,castelnovo2013negativity,wen2016edge,wen2016surgery,lu2020negativity}. This topological entanglement negativity (TEN) has also been shown to be sensitive to the breakdown of TO in thermal states~\cite{hart2018negativity,lu2020negativity}, which the entanglement entropy does not accurately reflect. Since we have shown that imTO is generally characterized by \emph{non-modular} anyon theories, it is an intriguing question as to what universal data the TEN captures in these states. In one specific instance, Ref. \cite{wang2023mixed} distinguished between the $\mbZ_2^{(0)}$ and $\mbZ_2^{(1)}$ imTOs (obtained via decoherence of the $\mbZ_2$ Toric code) by the respective absence and presence of topological contributions to the negativity. Recalling that these two states correspond to quantum condensates of bosonic and fermionic loops, it is tempting to conjecture that the TEN remains sensitive to the spins of the underlying deconfined anyon excitations. In the future, we intend to address more comprehensively the connection between TEN and the braided tensor category structure of imTO. It would likewise be interesting to understand novel decoherence induced negativity transitions~\cite{fan2023mixed,wang2023mixed,eckstein2024,chen2024unconventional} that may result from (competition between) the different decoherence channels discussed in this work.

While the entanglement negativity is a good measure of bipartite entanglement, it has recently been understood that pure state TO can be more finely characterized by its \emph{tripartite} entanglement structure~\cite{siva2021,liu2021,zou2022modular,kim2022modulara,kim2022modularb,fan2022modular,sohal2023tripartite,liu2024tripartite}. Specifically, it has been argued that chiral TO supports tripartite entanglement beyond that of the Greenberger-Horne-Zeilinger (GHZ) type~\cite{siva2021,liu2021,liu2024tripartite}. As we have shown, decoherence of the double semion state can induce a mixed state characterized by the chiral semion TO. Intriguingly, this suggests that decoherence has transmuted one form of many-body tripartite entanglement (i.e. GHZ-like entanglement) into another. It is conceivable that non-unitary processes may stabilize patterns of multipartite entanglement in many-body systems which do not arise naturally in the ground states of local gapped Hamiltonians. Understanding in more detail the multipartite entanglement of imTOs and the transmutation between different classes of entanglement via non-unitary processes promises to be a fruitful direction for further research. In a similar vein, Ref.~\cite{lessa2024anomaly} recently argued that a state that is strongly symmetric with respect to an anomalous 0-form symmetry in 2D must be 4-partite non-separable. It is an intriguing question whether similar constraints exist for systems respecting an anomalous strong 1-form symmetry (see Ref.~\cite{li2024entanglementneededemergentanyons} for a recent discussion).

In the spirit of fleshing out the structure of the family of imTOs we have obtained, an important avenue for further development is a more thorough classification of non-Abelian imTO. While we have demonstrated that the process of gauging out via decoherence extends naturally to the non-Abelian case, we do not yet have a comprehensive understanding of the algebraic structure of the resulting imTO, although we have provided compelling evidence that the appropriate mathematical framework is that of non-modular braided fusion categories. To that end, it would be prudent to understand more fully the connections with the mixed TQFTs proposed by Zini and Wang~\cite{wangmixedtqft}. In particular, it remains to be understood whether the class of mixed TQFTs proposed in that work can be realized in a physical setting i.e., by exposing some parent state to \textit{local} noise. On a related note, it would be interesting to understand whether there exists a non-equilibrium, continuum field theory formulation for describing generic imTO, most likely in the language of the Schwinger-Keldysh path integral.

The general framework we have developed also has potential exciting applications beyond the context of 2D mixed-state topological order. An obvious extension is to incorporate the ground states of 3D local gapped Hamiltonians into our framework and study the resulting decohered mixed-states. Our picture of decoherence in $d$-dimensions as non-unitary exfoliation of anomalous surface states of $d+1$-dimensional systems suggests a route towards realizing anomalous 3D topological orders~\cite{fidkowski2022-3d,chen2023-4d,chen2023-4dlattice} via local noise channels, where these states generically host transparent loop-like excitations (in analogy with the transparent anyons in our imTOs). We also expect that 3D fracton orders can be prepared by subjecting a 3D stack of 2D TO layers to an appropriate noise channel. Secondly, it is natural to consider the possibility of replacing correlated decoherence with correlated \emph{disorder}; as in the context of intrinsically average SPTs stabilized by disorder~\cite{ma2023prx,ma2023avg}, one may expect intrinsically average TO, the classification of which would likely be similar to, but distinct from, that of imTO.

We conclude by commenting on practical implications of our work for imTO in open quantum systems. Currently, preparing states with chiral TO requires sequential quantum circuits~\cite{chen2024sequence} and, although unproven, it is widely believed that no finite-depth quantum circuit can disentangle such states from the surface of a 3D WW model; for instance, Ref.~\cite{shirley2022qca} proved that either there exists a commuting projector Hamiltonian for the 2D chiral semion TO (which Ref.~\cite{kitaev2003} argued should not exist) or that the circuit that disentangles this TO from the surface of a 3D WW is not finite-depth. On the other hand, single shot measurement and feedback protocols for preparing ground states with Abelian TO have recently been proposed~\cite{feedforward2023}. Our results thus open the door towards the dissipative preparation of chiral TOs using finite-depth LPQCs: given a parent TO that can be prepared using a single-shot measurement and feedback circuit, we have shown that appropriately engineering a locally correlated noise channel can lead to chiral imTO. Surprisingly, since the doubled state in our construction can always be represented as a fixed point projected entangled pair state (PEPS) with finite bond dimension as we have a topological stabilizer code in the doubled space, our work also indicates the existence of a fixed point projected entangled pair operator (PEPO) representation for density matrices exhibiting chiral topological order (with finite bond dimension). This is an intriguing implication, as it is widely believed---though not proven---that there do not exist exact PEPS representations for (interacting) chiral topological pure states. 

More generally, we can imagine beginning from a topologically ordered pure state that can be efficiently prepared using existing protocols. Exposing such a state to noise channels will generically decrease its encoded logical information (as in each of our examples), such that the resulting decohered state represents a genuinely distinct phase of matter~\cite{coser2019class,sang2023mixed}. Heuristically, this is clear since no quasi-local recovery map can reconstruct the logical information stored in the parent state. We can then imagine a cascade of descendant TOs that be prepared from a parent state by carefully selecting error channels that gauge out anyons in a prescribed manner. This suggests a classification of mixed-state phases of matter in terms of the complexity of their code space, whereby no state in the sequence can recover the information of a precursor via LPQCs. Understanding the appropriate equivalence relation on the space of mixed-states is a question we intend to address in a future work. 


\begin{acknowledgements}
We are grateful to Meng Cheng, Tyler Ellison, Dominic Else, Tarun Grover, Tim Hsieh, Kohei Kawabata, Michael Levin, Ruochen Ma, Shinsei Ryu, Shengqi Sang, Nati Seiberg, Wilbur Shirley, Nikita Sopenko, Xiao-Qi Sun, Ruben Verresen, Chong Wang, and Carolyn Zhang for stimulating discussions. R.S. is especially grateful to Shinsei Ryu for pointing out, and discussions regarding, Ref. \cite{wangmixedtqft}. A. P. thanks Fiona Burnell, Sanket Chirame, and Sarang Gopalakrishnan for collaboration on related projects. This material is based upon work supported by a Simons Investigator Grant from the Simons Foundation (Grant No. 566116) awarded to Shinsei Ryu (R. S.), the Sivian Fund and the Paul Dirac Fund at the Institute for Advanced Study (A. P.), and the U.S. Department of Energy, Office of Science, Office of High Energy Physics under Award Number DE-SC0009988 (A. P.).
\end{acknowledgements}

\emph{Note.---}During the completion of this work, we were informed about Ref.~\cite{ellison2024} which also addresses mixed-state topological order. 


\newpage 

\bibliography{library}

\begin{thebibliography}{102}%
\makeatletter
\providecommand \@ifxundefined [1]{%
 \@ifx{#1\undefined}
}%
\providecommand \@ifnum [1]{%
 \ifnum #1\expandafter \@firstoftwo
 \else \expandafter \@secondoftwo
 \fi
}%
\providecommand \@ifx [1]{%
 \ifx #1\expandafter \@firstoftwo
 \else \expandafter \@secondoftwo
 \fi
}%
\providecommand \natexlab [1]{#1}%
\providecommand \enquote  [1]{``#1''}%
\providecommand \bibnamefont  [1]{#1}%
\providecommand \bibfnamefont [1]{#1}%
\providecommand \citenamefont [1]{#1}%
\providecommand \href@noop [0]{\@secondoftwo}%
\providecommand \href [0]{\begingroup \@sanitize@url \@href}%
\providecommand \@href[1]{\@@startlink{#1}\@@href}%
\providecommand \@@href[1]{\endgroup#1\@@endlink}%
\providecommand \@sanitize@url [0]{\catcode `\\12\catcode `\$12\catcode
  `\&12\catcode `\#12\catcode `\^12\catcode `\_12\catcode `\%12\relax}%
\providecommand \@@startlink[1]{}%
\providecommand \@@endlink[0]{}%
\providecommand \url  [0]{\begingroup\@sanitize@url \@url }%
\providecommand \@url [1]{\endgroup\@href {#1}{\urlprefix }}%
\providecommand \urlprefix  [0]{URL }%
\providecommand \Eprint [0]{\href }%
\providecommand \doibase [0]{https://doi.org/}%
\providecommand \selectlanguage [0]{\@gobble}%
\providecommand \bibinfo  [0]{\@secondoftwo}%
\providecommand \bibfield  [0]{\@secondoftwo}%
\providecommand \translation [1]{[#1]}%
\providecommand \BibitemOpen [0]{}%
\providecommand \bibitemStop [0]{}%
\providecommand \bibitemNoStop [0]{.\EOS\space}%
\providecommand \EOS [0]{\spacefactor3000\relax}%
\providecommand \BibitemShut  [1]{\csname bibitem#1\endcsname}%
\let\auto@bib@innerbib\@empty
\bibitem [{\citenamefont {{Kitaev}}(2003)}]{kitaev2003}%
  \BibitemOpen
  \bibfield  {author} {\bibinfo {author} {\bibfnamefont {A.~Y.}\ \bibnamefont
  {{Kitaev}}},\ }\bibfield  {title} {\bibinfo {title} {{Fault-tolerant quantum
  computation by anyons}},\ }\href
  {https://doi.org/10.1016/S0003-4916(02)00018-0} {\bibfield  {journal}
  {\bibinfo  {journal} {Annals of Physics}\ }\textbf {\bibinfo {volume}
  {303}},\ \bibinfo {pages} {2} (\bibinfo {year} {2003})}\BibitemShut {NoStop}%
\bibitem [{\citenamefont {Nayak}\ \emph {et~al.}(2008)\citenamefont {Nayak},
  \citenamefont {Simon}, \citenamefont {Stern}, \citenamefont {Freedman},\ and\
  \citenamefont {Das~Sarma}}]{nayakreview}%
  \BibitemOpen
  \bibfield  {author} {\bibinfo {author} {\bibfnamefont {C.}~\bibnamefont
  {Nayak}}, \bibinfo {author} {\bibfnamefont {S.~H.}\ \bibnamefont {Simon}},
  \bibinfo {author} {\bibfnamefont {A.}~\bibnamefont {Stern}}, \bibinfo
  {author} {\bibfnamefont {M.}~\bibnamefont {Freedman}},\ and\ \bibinfo
  {author} {\bibfnamefont {S.}~\bibnamefont {Das~Sarma}},\ }\bibfield  {title}
  {\bibinfo {title} {Non-abelian anyons and topological quantum computation},\
  }\href {https://doi.org/10.1103/RevModPhys.80.1083} {\bibfield  {journal}
  {\bibinfo  {journal} {Rev. Mod. Phys.}\ }\textbf {\bibinfo {volume} {80}},\
  \bibinfo {pages} {1083} (\bibinfo {year} {2008})}\BibitemShut {NoStop}%
\bibitem [{\citenamefont {{Bravyi}}\ \emph {et~al.}(2010)\citenamefont
  {{Bravyi}}, \citenamefont {{Hastings}},\ and\ \citenamefont
  {{Michalakis}}}]{bravyi2010}%
  \BibitemOpen
  \bibfield  {author} {\bibinfo {author} {\bibfnamefont {S.}~\bibnamefont
  {{Bravyi}}}, \bibinfo {author} {\bibfnamefont {M.~B.}\ \bibnamefont
  {{Hastings}}},\ and\ \bibinfo {author} {\bibfnamefont {S.}~\bibnamefont
  {{Michalakis}}},\ }\bibfield  {title} {\bibinfo {title} {{Topological quantum
  order: Stability under local perturbations}},\ }\href
  {https://doi.org/10.1063/1.3490195} {\bibfield  {journal} {\bibinfo
  {journal} {Journal of Mathematical Physics}\ }\textbf {\bibinfo {volume}
  {51}},\ \bibinfo {pages} {093512} (\bibinfo {year} {2010})}\BibitemShut
  {NoStop}%
\bibitem [{\citenamefont {{Bravyi}}\ and\ \citenamefont
  {{Hastings}}(2011)}]{bravyi2011short}%
  \BibitemOpen
  \bibfield  {author} {\bibinfo {author} {\bibfnamefont {S.}~\bibnamefont
  {{Bravyi}}}\ and\ \bibinfo {author} {\bibfnamefont {M.~B.}\ \bibnamefont
  {{Hastings}}},\ }\bibfield  {title} {\bibinfo {title} {{A Short Proof of
  Stability of Topological Order under Local Perturbations}},\ }\href
  {https://doi.org/10.1007/s00220-011-1346-2} {\bibfield  {journal} {\bibinfo
  {journal} {Communications in Mathematical Physics}\ }\textbf {\bibinfo
  {volume} {307}},\ \bibinfo {pages} {609} (\bibinfo {year}
  {2011})}\BibitemShut {NoStop}%
\bibitem [{\citenamefont {{Michalakis}}\ and\ \citenamefont
  {{Zwolak}}(2013)}]{michalakis2013}%
  \BibitemOpen
  \bibfield  {author} {\bibinfo {author} {\bibfnamefont {S.}~\bibnamefont
  {{Michalakis}}}\ and\ \bibinfo {author} {\bibfnamefont {J.~P.}\ \bibnamefont
  {{Zwolak}}},\ }\bibfield  {title} {\bibinfo {title} {{Stability of
  Frustration-Free Hamiltonians}},\ }\href
  {https://doi.org/10.1007/s00220-013-1762-6} {\bibfield  {journal} {\bibinfo
  {journal} {Communications in Mathematical Physics}\ }\textbf {\bibinfo
  {volume} {322}},\ \bibinfo {pages} {277} (\bibinfo {year}
  {2013})}\BibitemShut {NoStop}%
\bibitem [{\citenamefont {{Satzinger}}\ \emph {et~al.}(2021)\citenamefont
  {{Satzinger}}, \citenamefont {{Liu}}, \citenamefont {{Smith}}, \citenamefont
  {{Knapp}}, \citenamefont {{Newman}}, \citenamefont {{Jones}}, \citenamefont
  {{Chen}}, \citenamefont {{Quintana}}, \citenamefont {{Mi}}, \citenamefont
  {{Dunsworth}}, \citenamefont {{Gidney}}, \citenamefont {{Aleiner}},
  \citenamefont {{Arute}}, \citenamefont {{Arya}}, \citenamefont {{Atalaya}},
  \citenamefont {{Babbush}}, \citenamefont {{Bardin}}, \citenamefont
  {{Barends}}, \citenamefont {{Basso}}, \citenamefont {{Bengtsson}},
  \citenamefont {{Bilmes}}, \citenamefont {{Broughton}}, \citenamefont
  {{Buckley}}, \citenamefont {{Buell}}, \citenamefont {{Burkett}},
  \citenamefont {{Bushnell}}, \citenamefont {{Chiaro}}, \citenamefont
  {{Collins}}, \citenamefont {{Courtney}}, \citenamefont {{Demura}},
  \citenamefont {{Derk}}, \citenamefont {{Eppens}}, \citenamefont {{Erickson}},
  \citenamefont {{Faoro}}, \citenamefont {{Farhi}}, \citenamefont {{Fowler}},
  \citenamefont {{Foxen}}, \citenamefont {{Giustina}}, \citenamefont
  {{Greene}}, \citenamefont {{Gross}}, \citenamefont {{Harrigan}},
  \citenamefont {{Harrington}}, \citenamefont {{Hilton}}, \citenamefont
  {{Hong}}, \citenamefont {{Huang}}, \citenamefont {{Huggins}}, \citenamefont
  {{Ioffe}}, \citenamefont {{Isakov}}, \citenamefont {{Jeffrey}}, \citenamefont
  {{Jiang}}, \citenamefont {{Kafri}}, \citenamefont {{Kechedzhi}},
  \citenamefont {{Khattar}}, \citenamefont {{Kim}}, \citenamefont {{Klimov}},
  \citenamefont {{Korotkov}}, \citenamefont {{Kostritsa}}, \citenamefont
  {{Landhuis}}, \citenamefont {{Laptev}}, \citenamefont {{Locharla}},
  \citenamefont {{Lucero}}, \citenamefont {{Martin}}, \citenamefont
  {{McClean}}, \citenamefont {{McEwen}}, \citenamefont {{Miao}}, \citenamefont
  {{Mohseni}}, \citenamefont {{Montazeri}}, \citenamefont {{Mruczkiewicz}},
  \citenamefont {{Mutus}}, \citenamefont {{Naaman}}, \citenamefont {{Neeley}},
  \citenamefont {{Neill}}, \citenamefont {{Niu}}, \citenamefont {{O'Brien}},
  \citenamefont {{Opremcak}}, \citenamefont {{Pat{\'o}}}, \citenamefont
  {{Petukhov}}, \citenamefont {{Rubin}}, \citenamefont {{Sank}}, \citenamefont
  {{Shvarts}}, \citenamefont {{Strain}}, \citenamefont {{Szalay}},
  \citenamefont {{Villalonga}}, \citenamefont {{White}}, \citenamefont {{Yao}},
  \citenamefont {{Yeh}}, \citenamefont {{Yoo}}, \citenamefont {{Zalcman}},
  \citenamefont {{Neven}}, \citenamefont {{Boixo}}, \citenamefont {{Megrant}},
  \citenamefont {{Chen}}, \citenamefont {{Kelly}}, \citenamefont
  {{Smelyanskiy}}, \citenamefont {{Kitaev}}, \citenamefont {{Knap}},
  \citenamefont {{Pollmann}},\ and\ \citenamefont {{Roushan}}}]{satzinger2021}%
  \BibitemOpen
  \bibfield  {author} {\bibinfo {author} {\bibfnamefont {K.~J.}\ \bibnamefont
  {{Satzinger}}}, \bibinfo {author} {\bibfnamefont {Y.~J.}\ \bibnamefont
  {{Liu}}}, \bibinfo {author} {\bibfnamefont {A.}~\bibnamefont {{Smith}}},
  \bibinfo {author} {\bibfnamefont {C.}~\bibnamefont {{Knapp}}}, \bibinfo
  {author} {\bibfnamefont {M.}~\bibnamefont {{Newman}}}, \bibinfo {author}
  {\bibfnamefont {C.}~\bibnamefont {{Jones}}}, \bibinfo {author} {\bibfnamefont
  {Z.}~\bibnamefont {{Chen}}}, \bibinfo {author} {\bibfnamefont
  {C.}~\bibnamefont {{Quintana}}}, \bibinfo {author} {\bibfnamefont
  {X.}~\bibnamefont {{Mi}}}, \bibinfo {author} {\bibfnamefont {A.}~\bibnamefont
  {{Dunsworth}}}, \bibinfo {author} {\bibfnamefont {C.}~\bibnamefont
  {{Gidney}}}, \bibinfo {author} {\bibfnamefont {I.}~\bibnamefont {{Aleiner}}},
  \bibinfo {author} {\bibfnamefont {F.}~\bibnamefont {{Arute}}}, \bibinfo
  {author} {\bibfnamefont {K.}~\bibnamefont {{Arya}}}, \bibinfo {author}
  {\bibfnamefont {J.}~\bibnamefont {{Atalaya}}}, \bibinfo {author}
  {\bibfnamefont {R.}~\bibnamefont {{Babbush}}}, \bibinfo {author}
  {\bibfnamefont {J.~C.}\ \bibnamefont {{Bardin}}}, \bibinfo {author}
  {\bibfnamefont {R.}~\bibnamefont {{Barends}}}, \bibinfo {author}
  {\bibfnamefont {J.}~\bibnamefont {{Basso}}}, \bibinfo {author} {\bibfnamefont
  {A.}~\bibnamefont {{Bengtsson}}}, \bibinfo {author} {\bibfnamefont
  {A.}~\bibnamefont {{Bilmes}}}, \bibinfo {author} {\bibfnamefont
  {M.}~\bibnamefont {{Broughton}}}, \bibinfo {author} {\bibfnamefont {B.~B.}\
  \bibnamefont {{Buckley}}}, \bibinfo {author} {\bibfnamefont {D.~A.}\
  \bibnamefont {{Buell}}}, \bibinfo {author} {\bibfnamefont {B.}~\bibnamefont
  {{Burkett}}}, \bibinfo {author} {\bibfnamefont {N.}~\bibnamefont
  {{Bushnell}}}, \bibinfo {author} {\bibfnamefont {B.}~\bibnamefont
  {{Chiaro}}}, \bibinfo {author} {\bibfnamefont {R.}~\bibnamefont {{Collins}}},
  \bibinfo {author} {\bibfnamefont {W.}~\bibnamefont {{Courtney}}}, \bibinfo
  {author} {\bibfnamefont {S.}~\bibnamefont {{Demura}}}, \bibinfo {author}
  {\bibfnamefont {A.~R.}\ \bibnamefont {{Derk}}}, \bibinfo {author}
  {\bibfnamefont {D.}~\bibnamefont {{Eppens}}}, \bibinfo {author}
  {\bibfnamefont {C.}~\bibnamefont {{Erickson}}}, \bibinfo {author}
  {\bibfnamefont {L.}~\bibnamefont {{Faoro}}}, \bibinfo {author} {\bibfnamefont
  {E.}~\bibnamefont {{Farhi}}}, \bibinfo {author} {\bibfnamefont {A.~G.}\
  \bibnamefont {{Fowler}}}, \bibinfo {author} {\bibfnamefont {B.}~\bibnamefont
  {{Foxen}}}, \bibinfo {author} {\bibfnamefont {M.}~\bibnamefont {{Giustina}}},
  \bibinfo {author} {\bibfnamefont {A.}~\bibnamefont {{Greene}}}, \bibinfo
  {author} {\bibfnamefont {J.~A.}\ \bibnamefont {{Gross}}}, \bibinfo {author}
  {\bibfnamefont {M.~P.}\ \bibnamefont {{Harrigan}}}, \bibinfo {author}
  {\bibfnamefont {S.~D.}\ \bibnamefont {{Harrington}}}, \bibinfo {author}
  {\bibfnamefont {J.}~\bibnamefont {{Hilton}}}, \bibinfo {author}
  {\bibfnamefont {S.}~\bibnamefont {{Hong}}}, \bibinfo {author} {\bibfnamefont
  {T.}~\bibnamefont {{Huang}}}, \bibinfo {author} {\bibfnamefont {W.~J.}\
  \bibnamefont {{Huggins}}}, \bibinfo {author} {\bibfnamefont {L.~B.}\
  \bibnamefont {{Ioffe}}}, \bibinfo {author} {\bibfnamefont {S.~V.}\
  \bibnamefont {{Isakov}}}, \bibinfo {author} {\bibfnamefont {E.}~\bibnamefont
  {{Jeffrey}}}, \bibinfo {author} {\bibfnamefont {Z.}~\bibnamefont {{Jiang}}},
  \bibinfo {author} {\bibfnamefont {D.}~\bibnamefont {{Kafri}}}, \bibinfo
  {author} {\bibfnamefont {K.}~\bibnamefont {{Kechedzhi}}}, \bibinfo {author}
  {\bibfnamefont {T.}~\bibnamefont {{Khattar}}}, \bibinfo {author}
  {\bibfnamefont {S.}~\bibnamefont {{Kim}}}, \bibinfo {author} {\bibfnamefont
  {P.~V.}\ \bibnamefont {{Klimov}}}, \bibinfo {author} {\bibfnamefont {A.~N.}\
  \bibnamefont {{Korotkov}}}, \bibinfo {author} {\bibfnamefont
  {F.}~\bibnamefont {{Kostritsa}}}, \bibinfo {author} {\bibfnamefont
  {D.}~\bibnamefont {{Landhuis}}}, \bibinfo {author} {\bibfnamefont
  {P.}~\bibnamefont {{Laptev}}}, \bibinfo {author} {\bibfnamefont
  {A.}~\bibnamefont {{Locharla}}}, \bibinfo {author} {\bibfnamefont
  {E.}~\bibnamefont {{Lucero}}}, \bibinfo {author} {\bibfnamefont
  {O.}~\bibnamefont {{Martin}}}, \bibinfo {author} {\bibfnamefont {J.~R.}\
  \bibnamefont {{McClean}}}, \bibinfo {author} {\bibfnamefont {M.}~\bibnamefont
  {{McEwen}}}, \bibinfo {author} {\bibfnamefont {K.~C.}\ \bibnamefont
  {{Miao}}}, \bibinfo {author} {\bibfnamefont {M.}~\bibnamefont {{Mohseni}}},
  \bibinfo {author} {\bibfnamefont {S.}~\bibnamefont {{Montazeri}}}, \bibinfo
  {author} {\bibfnamefont {W.}~\bibnamefont {{Mruczkiewicz}}}, \bibinfo
  {author} {\bibfnamefont {J.}~\bibnamefont {{Mutus}}}, \bibinfo {author}
  {\bibfnamefont {O.}~\bibnamefont {{Naaman}}}, \bibinfo {author}
  {\bibfnamefont {M.}~\bibnamefont {{Neeley}}}, \bibinfo {author}
  {\bibfnamefont {C.}~\bibnamefont {{Neill}}}, \bibinfo {author} {\bibfnamefont
  {M.~Y.}\ \bibnamefont {{Niu}}}, \bibinfo {author} {\bibfnamefont {T.~E.}\
  \bibnamefont {{O'Brien}}}, \bibinfo {author} {\bibfnamefont {A.}~\bibnamefont
  {{Opremcak}}}, \bibinfo {author} {\bibfnamefont {B.}~\bibnamefont
  {{Pat{\'o}}}}, \bibinfo {author} {\bibfnamefont {A.}~\bibnamefont
  {{Petukhov}}}, \bibinfo {author} {\bibfnamefont {N.~C.}\ \bibnamefont
  {{Rubin}}}, \bibinfo {author} {\bibfnamefont {D.}~\bibnamefont {{Sank}}},
  \bibinfo {author} {\bibfnamefont {V.}~\bibnamefont {{Shvarts}}}, \bibinfo
  {author} {\bibfnamefont {D.}~\bibnamefont {{Strain}}}, \bibinfo {author}
  {\bibfnamefont {M.}~\bibnamefont {{Szalay}}}, \bibinfo {author}
  {\bibfnamefont {B.}~\bibnamefont {{Villalonga}}}, \bibinfo {author}
  {\bibfnamefont {T.~C.}\ \bibnamefont {{White}}}, \bibinfo {author}
  {\bibfnamefont {Z.}~\bibnamefont {{Yao}}}, \bibinfo {author} {\bibfnamefont
  {P.}~\bibnamefont {{Yeh}}}, \bibinfo {author} {\bibfnamefont
  {J.}~\bibnamefont {{Yoo}}}, \bibinfo {author} {\bibfnamefont
  {A.}~\bibnamefont {{Zalcman}}}, \bibinfo {author} {\bibfnamefont
  {H.}~\bibnamefont {{Neven}}}, \bibinfo {author} {\bibfnamefont
  {S.}~\bibnamefont {{Boixo}}}, \bibinfo {author} {\bibfnamefont
  {A.}~\bibnamefont {{Megrant}}}, \bibinfo {author} {\bibfnamefont
  {Y.}~\bibnamefont {{Chen}}}, \bibinfo {author} {\bibfnamefont
  {J.}~\bibnamefont {{Kelly}}}, \bibinfo {author} {\bibfnamefont
  {V.}~\bibnamefont {{Smelyanskiy}}}, \bibinfo {author} {\bibfnamefont
  {A.}~\bibnamefont {{Kitaev}}}, \bibinfo {author} {\bibfnamefont
  {M.}~\bibnamefont {{Knap}}}, \bibinfo {author} {\bibfnamefont
  {F.}~\bibnamefont {{Pollmann}}},\ and\ \bibinfo {author} {\bibfnamefont
  {P.}~\bibnamefont {{Roushan}}},\ }\bibfield  {title} {\bibinfo {title}
  {{Realizing topologically ordered states on a quantum processor}},\ }\href
  {https://doi.org/10.1126/science.abi8378} {\bibfield  {journal} {\bibinfo
  {journal} {Science}\ }\textbf {\bibinfo {volume} {374}},\ \bibinfo {pages}
  {1237} (\bibinfo {year} {2021})}\BibitemShut {NoStop}%
\bibitem [{\citenamefont {{Semeghini}}\ \emph {et~al.}(2021)\citenamefont
  {{Semeghini}}, \citenamefont {{Levine}}, \citenamefont {{Keesling}},
  \citenamefont {{Ebadi}}, \citenamefont {{Wang}}, \citenamefont {{Bluvstein}},
  \citenamefont {{Verresen}}, \citenamefont {{Pichler}}, \citenamefont
  {{Kalinowski}}, \citenamefont {{Samajdar}}, \citenamefont {{Omran}},
  \citenamefont {{Sachdev}}, \citenamefont {{Vishwanath}}, \citenamefont
  {{Greiner}}, \citenamefont {{Vuleti{\'c}}},\ and\ \citenamefont
  {{Lukin}}}]{semeghini2021}%
  \BibitemOpen
  \bibfield  {author} {\bibinfo {author} {\bibfnamefont {G.}~\bibnamefont
  {{Semeghini}}}, \bibinfo {author} {\bibfnamefont {H.}~\bibnamefont
  {{Levine}}}, \bibinfo {author} {\bibfnamefont {A.}~\bibnamefont
  {{Keesling}}}, \bibinfo {author} {\bibfnamefont {S.}~\bibnamefont {{Ebadi}}},
  \bibinfo {author} {\bibfnamefont {T.~T.}\ \bibnamefont {{Wang}}}, \bibinfo
  {author} {\bibfnamefont {D.}~\bibnamefont {{Bluvstein}}}, \bibinfo {author}
  {\bibfnamefont {R.}~\bibnamefont {{Verresen}}}, \bibinfo {author}
  {\bibfnamefont {H.}~\bibnamefont {{Pichler}}}, \bibinfo {author}
  {\bibfnamefont {M.}~\bibnamefont {{Kalinowski}}}, \bibinfo {author}
  {\bibfnamefont {R.}~\bibnamefont {{Samajdar}}}, \bibinfo {author}
  {\bibfnamefont {A.}~\bibnamefont {{Omran}}}, \bibinfo {author} {\bibfnamefont
  {S.}~\bibnamefont {{Sachdev}}}, \bibinfo {author} {\bibfnamefont
  {A.}~\bibnamefont {{Vishwanath}}}, \bibinfo {author} {\bibfnamefont
  {M.}~\bibnamefont {{Greiner}}}, \bibinfo {author} {\bibfnamefont
  {V.}~\bibnamefont {{Vuleti{\'c}}}},\ and\ \bibinfo {author} {\bibfnamefont
  {M.~D.}\ \bibnamefont {{Lukin}}},\ }\bibfield  {title} {\bibinfo {title}
  {{Probing topological spin liquids on a programmable quantum simulator}},\
  }\href {https://doi.org/10.1126/science.abi8794} {\bibfield  {journal}
  {\bibinfo  {journal} {Science}\ }\textbf {\bibinfo {volume} {374}},\ \bibinfo
  {pages} {1242} (\bibinfo {year} {2021})}\BibitemShut {NoStop}%
\bibitem [{\citenamefont {{Andersen}}\ \emph {et~al.}(2022)\citenamefont
  {{Andersen}}, \citenamefont {{Lensky}}, \citenamefont {{Kechedzhi}},
  \citenamefont {{Drozdov}}, \citenamefont {{Bengtsson}}, \citenamefont
  {{Hong}}, \citenamefont {{Morvan}}, \citenamefont {{Mi}}, \citenamefont
  {{Opremcak}}, \citenamefont {{Acharya}}, \citenamefont {{Allen}},
  \citenamefont {{Ansmann}}, \citenamefont {{Arute}}, \citenamefont {{Arya}},
  \citenamefont {{Asfaw}}, \citenamefont {{Atalaya}}, \citenamefont
  {{Babbush}}, \citenamefont {{Bacon}}, \citenamefont {{Bardin}}, \citenamefont
  {{Bortoli}}, \citenamefont {{Bourassa}}, \citenamefont {{Bovaird}},
  \citenamefont {{Brill}}, \citenamefont {{Broughton}}, \citenamefont
  {{Buckley}}, \citenamefont {{Buell}}, \citenamefont {{Burger}}, \citenamefont
  {{Burkett}}, \citenamefont {{Bushnell}}, \citenamefont {{Chen}},
  \citenamefont {{Chiaro}}, \citenamefont {{Chik}}, \citenamefont {{Chou}},
  \citenamefont {{Cogan}}, \citenamefont {{Collins}}, \citenamefont {{Conner}},
  \citenamefont {{Courtney}}, \citenamefont {{Crook}}, \citenamefont
  {{Curtin}}, \citenamefont {{Debroy}}, \citenamefont {{Del Toro Barba}},
  \citenamefont {{Demura}}, \citenamefont {{Dunsworth}}, \citenamefont
  {{Eppens}}, \citenamefont {{Erickson}}, \citenamefont {{Faoro}},
  \citenamefont {{Farhi}}, \citenamefont {{Fatemi}}, \citenamefont
  {{Ferreira}}, \citenamefont {{Flores Burgos}}, \citenamefont {{Forati}},
  \citenamefont {{Fowler}}, \citenamefont {{Foxen}}, \citenamefont {{Giang}},
  \citenamefont {{Gidney}}, \citenamefont {{Gilboa}}, \citenamefont
  {{Giustina}}, \citenamefont {{Gosula}}, \citenamefont {{Grajales Dau}},
  \citenamefont {{Gross}}, \citenamefont {{Habegger}}, \citenamefont
  {{Hamilton}}, \citenamefont {{Hansen}}, \citenamefont {{Harrigan}},
  \citenamefont {{Harrington}}, \citenamefont {{Heu}}, \citenamefont
  {{Hilton}}, \citenamefont {{Hoffmann}}, \citenamefont {{Huang}},
  \citenamefont {{Huff}}, \citenamefont {{Huggins}}, \citenamefont {{Ioffe}},
  \citenamefont {{Isakov}}, \citenamefont {{Iveland}}, \citenamefont
  {{Jeffrey}}, \citenamefont {{Jiang}}, \citenamefont {{Jones}}, \citenamefont
  {{Juhas}}, \citenamefont {{Kafri}}, \citenamefont {{Khattar}}, \citenamefont
  {{Khezri}}, \citenamefont {{Kieferov{\'a}}}, \citenamefont {{Kim}},
  \citenamefont {{Kitaev}}, \citenamefont {{Klimov}}, \citenamefont {{Klots}},
  \citenamefont {{Korotkov}}, \citenamefont {{Kostritsa}}, \citenamefont
  {{Kreikebaum}}, \citenamefont {{Landhuis}}, \citenamefont {{Laptev}},
  \citenamefont {{Lau}}, \citenamefont {{Laws}}, \citenamefont {{Lee}},
  \citenamefont {{Lee}}, \citenamefont {{Lester}}, \citenamefont {{Lill}},
  \citenamefont {{Liu}}, \citenamefont {{Locharla}}, \citenamefont {{Lucero}},
  \citenamefont {{Malone}}, \citenamefont {{Martin}}, \citenamefont
  {{McClean}}, \citenamefont {{McCourt}}, \citenamefont {{McEwen}},
  \citenamefont {{Miao}}, \citenamefont {{Mieszala}}, \citenamefont
  {{Mohseni}}, \citenamefont {{Montazeri}}, \citenamefont {{Mount}},
  \citenamefont {{Movassagh}}, \citenamefont {{Mruczkiewicz}}, \citenamefont
  {{Naaman}}, \citenamefont {{Neeley}}, \citenamefont {{Neill}}, \citenamefont
  {{Nersisyan}}, \citenamefont {{Newman}}, \citenamefont {{How Ng}},
  \citenamefont {{Nguyen}}, \citenamefont {{Nguyen}}, \citenamefont {{Yuezhen
  Niu}}, \citenamefont {{O'Brien}}, \citenamefont {{Omonije}}, \citenamefont
  {{Petukhov}}, \citenamefont {{Potter}}, \citenamefont {{Pryadko}},
  \citenamefont {{Quintana}}, \citenamefont {{Rocque}}, \citenamefont
  {{Rubin}}, \citenamefont {{Saei}}, \citenamefont {{Sank}}, \citenamefont
  {{Sankaragomathi}}, \citenamefont {{Satzinger}}, \citenamefont {{Schurkus}},
  \citenamefont {{Schuster}}, \citenamefont {{Shearn}}, \citenamefont
  {{Shorter}}, \citenamefont {{Shutty}}, \citenamefont {{Shvarts}},
  \citenamefont {{Skruzny}}, \citenamefont {{Clarke Smith}}, \citenamefont
  {{Somma}}, \citenamefont {{Sterling}}, \citenamefont {{Strain}},
  \citenamefont {{Szalay}}, \citenamefont {{Torres}}, \citenamefont {{Vidal}},
  \citenamefont {{Villalonga}}, \citenamefont {{Vollgraff Heidweiller}},
  \citenamefont {{White}}, \citenamefont {{Woo}}, \citenamefont {{Xing}},
  \citenamefont {{Yao}}, \citenamefont {{Yeh}}, \citenamefont {{Yoo}},
  \citenamefont {{Young}}, \citenamefont {{Zalcman}}, \citenamefont {{Zhang}},
  \citenamefont {{Zhu}}, \citenamefont {{Zobrist}}, \citenamefont {{Neven}},
  \citenamefont {{Boixo}}, \citenamefont {{Megrant}}, \citenamefont {{Kelly}},
  \citenamefont {{Chen}}, \citenamefont {{Smelyanskiy}}, \citenamefont {{Kim}},
  \citenamefont {{Aleiner}},\ and\ \citenamefont {{Roushan}}}]{andersen2022}%
  \BibitemOpen
  \bibfield  {author} {\bibinfo {author} {\bibfnamefont {T.~I.}\ \bibnamefont
  {{Andersen}}}, \bibinfo {author} {\bibfnamefont {Y.~D.}\ \bibnamefont
  {{Lensky}}}, \bibinfo {author} {\bibfnamefont {K.}~\bibnamefont
  {{Kechedzhi}}}, \bibinfo {author} {\bibfnamefont {I.}~\bibnamefont
  {{Drozdov}}}, \bibinfo {author} {\bibfnamefont {A.}~\bibnamefont
  {{Bengtsson}}}, \bibinfo {author} {\bibfnamefont {S.}~\bibnamefont {{Hong}}},
  \bibinfo {author} {\bibfnamefont {A.}~\bibnamefont {{Morvan}}}, \bibinfo
  {author} {\bibfnamefont {X.}~\bibnamefont {{Mi}}}, \bibinfo {author}
  {\bibfnamefont {A.}~\bibnamefont {{Opremcak}}}, \bibinfo {author}
  {\bibfnamefont {R.}~\bibnamefont {{Acharya}}}, \bibinfo {author}
  {\bibfnamefont {R.}~\bibnamefont {{Allen}}}, \bibinfo {author} {\bibfnamefont
  {M.}~\bibnamefont {{Ansmann}}}, \bibinfo {author} {\bibfnamefont
  {F.}~\bibnamefont {{Arute}}}, \bibinfo {author} {\bibfnamefont
  {K.}~\bibnamefont {{Arya}}}, \bibinfo {author} {\bibfnamefont
  {A.}~\bibnamefont {{Asfaw}}}, \bibinfo {author} {\bibfnamefont
  {J.}~\bibnamefont {{Atalaya}}}, \bibinfo {author} {\bibfnamefont
  {R.}~\bibnamefont {{Babbush}}}, \bibinfo {author} {\bibfnamefont
  {D.}~\bibnamefont {{Bacon}}}, \bibinfo {author} {\bibfnamefont {J.~C.}\
  \bibnamefont {{Bardin}}}, \bibinfo {author} {\bibfnamefont {G.}~\bibnamefont
  {{Bortoli}}}, \bibinfo {author} {\bibfnamefont {A.}~\bibnamefont
  {{Bourassa}}}, \bibinfo {author} {\bibfnamefont {J.}~\bibnamefont
  {{Bovaird}}}, \bibinfo {author} {\bibfnamefont {L.}~\bibnamefont {{Brill}}},
  \bibinfo {author} {\bibfnamefont {M.}~\bibnamefont {{Broughton}}}, \bibinfo
  {author} {\bibfnamefont {B.~B.}\ \bibnamefont {{Buckley}}}, \bibinfo {author}
  {\bibfnamefont {D.~A.}\ \bibnamefont {{Buell}}}, \bibinfo {author}
  {\bibfnamefont {T.}~\bibnamefont {{Burger}}}, \bibinfo {author}
  {\bibfnamefont {B.}~\bibnamefont {{Burkett}}}, \bibinfo {author}
  {\bibfnamefont {N.}~\bibnamefont {{Bushnell}}}, \bibinfo {author}
  {\bibfnamefont {Z.}~\bibnamefont {{Chen}}}, \bibinfo {author} {\bibfnamefont
  {B.}~\bibnamefont {{Chiaro}}}, \bibinfo {author} {\bibfnamefont
  {D.}~\bibnamefont {{Chik}}}, \bibinfo {author} {\bibfnamefont
  {C.}~\bibnamefont {{Chou}}}, \bibinfo {author} {\bibfnamefont
  {J.}~\bibnamefont {{Cogan}}}, \bibinfo {author} {\bibfnamefont
  {R.}~\bibnamefont {{Collins}}}, \bibinfo {author} {\bibfnamefont
  {P.}~\bibnamefont {{Conner}}}, \bibinfo {author} {\bibfnamefont
  {W.}~\bibnamefont {{Courtney}}}, \bibinfo {author} {\bibfnamefont {A.~L.}\
  \bibnamefont {{Crook}}}, \bibinfo {author} {\bibfnamefont {B.}~\bibnamefont
  {{Curtin}}}, \bibinfo {author} {\bibfnamefont {D.~M.}\ \bibnamefont
  {{Debroy}}}, \bibinfo {author} {\bibfnamefont {A.}~\bibnamefont {{Del Toro
  Barba}}}, \bibinfo {author} {\bibfnamefont {S.}~\bibnamefont {{Demura}}},
  \bibinfo {author} {\bibfnamefont {A.}~\bibnamefont {{Dunsworth}}}, \bibinfo
  {author} {\bibfnamefont {D.}~\bibnamefont {{Eppens}}}, \bibinfo {author}
  {\bibfnamefont {C.}~\bibnamefont {{Erickson}}}, \bibinfo {author}
  {\bibfnamefont {L.}~\bibnamefont {{Faoro}}}, \bibinfo {author} {\bibfnamefont
  {E.}~\bibnamefont {{Farhi}}}, \bibinfo {author} {\bibfnamefont
  {R.}~\bibnamefont {{Fatemi}}}, \bibinfo {author} {\bibfnamefont {V.~S.}\
  \bibnamefont {{Ferreira}}}, \bibinfo {author} {\bibfnamefont
  {L.}~\bibnamefont {{Flores Burgos}}}, \bibinfo {author} {\bibfnamefont
  {E.}~\bibnamefont {{Forati}}}, \bibinfo {author} {\bibfnamefont {A.~G.}\
  \bibnamefont {{Fowler}}}, \bibinfo {author} {\bibfnamefont {B.}~\bibnamefont
  {{Foxen}}}, \bibinfo {author} {\bibfnamefont {W.}~\bibnamefont {{Giang}}},
  \bibinfo {author} {\bibfnamefont {C.}~\bibnamefont {{Gidney}}}, \bibinfo
  {author} {\bibfnamefont {D.}~\bibnamefont {{Gilboa}}}, \bibinfo {author}
  {\bibfnamefont {M.}~\bibnamefont {{Giustina}}}, \bibinfo {author}
  {\bibfnamefont {R.}~\bibnamefont {{Gosula}}}, \bibinfo {author}
  {\bibfnamefont {A.}~\bibnamefont {{Grajales Dau}}}, \bibinfo {author}
  {\bibfnamefont {J.~A.}\ \bibnamefont {{Gross}}}, \bibinfo {author}
  {\bibfnamefont {S.}~\bibnamefont {{Habegger}}}, \bibinfo {author}
  {\bibfnamefont {M.~C.}\ \bibnamefont {{Hamilton}}}, \bibinfo {author}
  {\bibfnamefont {M.}~\bibnamefont {{Hansen}}}, \bibinfo {author}
  {\bibfnamefont {M.~P.}\ \bibnamefont {{Harrigan}}}, \bibinfo {author}
  {\bibfnamefont {S.~D.}\ \bibnamefont {{Harrington}}}, \bibinfo {author}
  {\bibfnamefont {P.}~\bibnamefont {{Heu}}}, \bibinfo {author} {\bibfnamefont
  {J.}~\bibnamefont {{Hilton}}}, \bibinfo {author} {\bibfnamefont {M.~R.}\
  \bibnamefont {{Hoffmann}}}, \bibinfo {author} {\bibfnamefont
  {T.}~\bibnamefont {{Huang}}}, \bibinfo {author} {\bibfnamefont
  {A.}~\bibnamefont {{Huff}}}, \bibinfo {author} {\bibfnamefont {W.~J.}\
  \bibnamefont {{Huggins}}}, \bibinfo {author} {\bibfnamefont {L.~B.}\
  \bibnamefont {{Ioffe}}}, \bibinfo {author} {\bibfnamefont {S.~V.}\
  \bibnamefont {{Isakov}}}, \bibinfo {author} {\bibfnamefont {J.}~\bibnamefont
  {{Iveland}}}, \bibinfo {author} {\bibfnamefont {E.}~\bibnamefont
  {{Jeffrey}}}, \bibinfo {author} {\bibfnamefont {Z.}~\bibnamefont {{Jiang}}},
  \bibinfo {author} {\bibfnamefont {C.}~\bibnamefont {{Jones}}}, \bibinfo
  {author} {\bibfnamefont {P.}~\bibnamefont {{Juhas}}}, \bibinfo {author}
  {\bibfnamefont {D.}~\bibnamefont {{Kafri}}}, \bibinfo {author} {\bibfnamefont
  {T.}~\bibnamefont {{Khattar}}}, \bibinfo {author} {\bibfnamefont
  {M.}~\bibnamefont {{Khezri}}}, \bibinfo {author} {\bibfnamefont
  {M.}~\bibnamefont {{Kieferov{\'a}}}}, \bibinfo {author} {\bibfnamefont
  {S.}~\bibnamefont {{Kim}}}, \bibinfo {author} {\bibfnamefont
  {A.}~\bibnamefont {{Kitaev}}}, \bibinfo {author} {\bibfnamefont {P.~V.}\
  \bibnamefont {{Klimov}}}, \bibinfo {author} {\bibfnamefont {A.~R.}\
  \bibnamefont {{Klots}}}, \bibinfo {author} {\bibfnamefont {A.~N.}\
  \bibnamefont {{Korotkov}}}, \bibinfo {author} {\bibfnamefont
  {F.}~\bibnamefont {{Kostritsa}}}, \bibinfo {author} {\bibfnamefont {J.~M.}\
  \bibnamefont {{Kreikebaum}}}, \bibinfo {author} {\bibfnamefont
  {D.}~\bibnamefont {{Landhuis}}}, \bibinfo {author} {\bibfnamefont
  {P.}~\bibnamefont {{Laptev}}}, \bibinfo {author} {\bibfnamefont {K.-M.}\
  \bibnamefont {{Lau}}}, \bibinfo {author} {\bibfnamefont {L.}~\bibnamefont
  {{Laws}}}, \bibinfo {author} {\bibfnamefont {J.}~\bibnamefont {{Lee}}},
  \bibinfo {author} {\bibfnamefont {K.}~\bibnamefont {{Lee}}}, \bibinfo
  {author} {\bibfnamefont {B.~J.}\ \bibnamefont {{Lester}}}, \bibinfo {author}
  {\bibfnamefont {A.}~\bibnamefont {{Lill}}}, \bibinfo {author} {\bibfnamefont
  {W.}~\bibnamefont {{Liu}}}, \bibinfo {author} {\bibfnamefont
  {A.}~\bibnamefont {{Locharla}}}, \bibinfo {author} {\bibfnamefont
  {E.}~\bibnamefont {{Lucero}}}, \bibinfo {author} {\bibfnamefont {F.~D.}\
  \bibnamefont {{Malone}}}, \bibinfo {author} {\bibfnamefont {O.}~\bibnamefont
  {{Martin}}}, \bibinfo {author} {\bibfnamefont {J.~R.}\ \bibnamefont
  {{McClean}}}, \bibinfo {author} {\bibfnamefont {T.}~\bibnamefont
  {{McCourt}}}, \bibinfo {author} {\bibfnamefont {M.}~\bibnamefont {{McEwen}}},
  \bibinfo {author} {\bibfnamefont {K.~C.}\ \bibnamefont {{Miao}}}, \bibinfo
  {author} {\bibfnamefont {A.}~\bibnamefont {{Mieszala}}}, \bibinfo {author}
  {\bibfnamefont {M.}~\bibnamefont {{Mohseni}}}, \bibinfo {author}
  {\bibfnamefont {S.}~\bibnamefont {{Montazeri}}}, \bibinfo {author}
  {\bibfnamefont {E.}~\bibnamefont {{Mount}}}, \bibinfo {author} {\bibfnamefont
  {R.}~\bibnamefont {{Movassagh}}}, \bibinfo {author} {\bibfnamefont
  {W.}~\bibnamefont {{Mruczkiewicz}}}, \bibinfo {author} {\bibfnamefont
  {O.}~\bibnamefont {{Naaman}}}, \bibinfo {author} {\bibfnamefont
  {M.}~\bibnamefont {{Neeley}}}, \bibinfo {author} {\bibfnamefont
  {C.}~\bibnamefont {{Neill}}}, \bibinfo {author} {\bibfnamefont
  {A.}~\bibnamefont {{Nersisyan}}}, \bibinfo {author} {\bibfnamefont
  {M.}~\bibnamefont {{Newman}}}, \bibinfo {author} {\bibfnamefont
  {J.}~\bibnamefont {{How Ng}}}, \bibinfo {author} {\bibfnamefont
  {A.}~\bibnamefont {{Nguyen}}}, \bibinfo {author} {\bibfnamefont
  {M.}~\bibnamefont {{Nguyen}}}, \bibinfo {author} {\bibfnamefont
  {M.}~\bibnamefont {{Yuezhen Niu}}}, \bibinfo {author} {\bibfnamefont {T.~E.}\
  \bibnamefont {{O'Brien}}}, \bibinfo {author} {\bibfnamefont {S.}~\bibnamefont
  {{Omonije}}}, \bibinfo {author} {\bibfnamefont {A.}~\bibnamefont
  {{Petukhov}}}, \bibinfo {author} {\bibfnamefont {R.}~\bibnamefont
  {{Potter}}}, \bibinfo {author} {\bibfnamefont {L.~P.}\ \bibnamefont
  {{Pryadko}}}, \bibinfo {author} {\bibfnamefont {C.}~\bibnamefont
  {{Quintana}}}, \bibinfo {author} {\bibfnamefont {C.}~\bibnamefont
  {{Rocque}}}, \bibinfo {author} {\bibfnamefont {N.~C.}\ \bibnamefont
  {{Rubin}}}, \bibinfo {author} {\bibfnamefont {N.}~\bibnamefont {{Saei}}},
  \bibinfo {author} {\bibfnamefont {D.}~\bibnamefont {{Sank}}}, \bibinfo
  {author} {\bibfnamefont {K.}~\bibnamefont {{Sankaragomathi}}}, \bibinfo
  {author} {\bibfnamefont {K.~J.}\ \bibnamefont {{Satzinger}}}, \bibinfo
  {author} {\bibfnamefont {H.~F.}\ \bibnamefont {{Schurkus}}}, \bibinfo
  {author} {\bibfnamefont {C.}~\bibnamefont {{Schuster}}}, \bibinfo {author}
  {\bibfnamefont {M.~J.}\ \bibnamefont {{Shearn}}}, \bibinfo {author}
  {\bibfnamefont {A.}~\bibnamefont {{Shorter}}}, \bibinfo {author}
  {\bibfnamefont {N.}~\bibnamefont {{Shutty}}}, \bibinfo {author}
  {\bibfnamefont {V.}~\bibnamefont {{Shvarts}}}, \bibinfo {author}
  {\bibfnamefont {J.}~\bibnamefont {{Skruzny}}}, \bibinfo {author}
  {\bibfnamefont {W.}~\bibnamefont {{Clarke Smith}}}, \bibinfo {author}
  {\bibfnamefont {R.}~\bibnamefont {{Somma}}}, \bibinfo {author} {\bibfnamefont
  {G.}~\bibnamefont {{Sterling}}}, \bibinfo {author} {\bibfnamefont
  {D.}~\bibnamefont {{Strain}}}, \bibinfo {author} {\bibfnamefont
  {M.}~\bibnamefont {{Szalay}}}, \bibinfo {author} {\bibfnamefont
  {A.}~\bibnamefont {{Torres}}}, \bibinfo {author} {\bibfnamefont
  {G.}~\bibnamefont {{Vidal}}}, \bibinfo {author} {\bibfnamefont
  {B.}~\bibnamefont {{Villalonga}}}, \bibinfo {author} {\bibfnamefont
  {C.}~\bibnamefont {{Vollgraff Heidweiller}}}, \bibinfo {author}
  {\bibfnamefont {T.}~\bibnamefont {{White}}}, \bibinfo {author} {\bibfnamefont
  {B.~W.~K.}\ \bibnamefont {{Woo}}}, \bibinfo {author} {\bibfnamefont
  {C.}~\bibnamefont {{Xing}}}, \bibinfo {author} {\bibfnamefont {Z.~J.}\
  \bibnamefont {{Yao}}}, \bibinfo {author} {\bibfnamefont {P.}~\bibnamefont
  {{Yeh}}}, \bibinfo {author} {\bibfnamefont {J.}~\bibnamefont {{Yoo}}},
  \bibinfo {author} {\bibfnamefont {G.}~\bibnamefont {{Young}}}, \bibinfo
  {author} {\bibfnamefont {A.}~\bibnamefont {{Zalcman}}}, \bibinfo {author}
  {\bibfnamefont {Y.}~\bibnamefont {{Zhang}}}, \bibinfo {author} {\bibfnamefont
  {N.}~\bibnamefont {{Zhu}}}, \bibinfo {author} {\bibfnamefont
  {N.}~\bibnamefont {{Zobrist}}}, \bibinfo {author} {\bibfnamefont
  {H.}~\bibnamefont {{Neven}}}, \bibinfo {author} {\bibfnamefont
  {S.}~\bibnamefont {{Boixo}}}, \bibinfo {author} {\bibfnamefont
  {A.}~\bibnamefont {{Megrant}}}, \bibinfo {author} {\bibfnamefont
  {J.}~\bibnamefont {{Kelly}}}, \bibinfo {author} {\bibfnamefont
  {Y.}~\bibnamefont {{Chen}}}, \bibinfo {author} {\bibfnamefont
  {V.}~\bibnamefont {{Smelyanskiy}}}, \bibinfo {author} {\bibfnamefont {E.-A.}\
  \bibnamefont {{Kim}}}, \bibinfo {author} {\bibfnamefont {I.}~\bibnamefont
  {{Aleiner}}},\ and\ \bibinfo {author} {\bibfnamefont {P.}~\bibnamefont
  {{Roushan}}},\ }\bibfield  {title} {\bibinfo {title} {{Non-Abelian braiding
  of graph vertices in a superconducting processor}},\ }\href
  {https://doi.org/10.48550/arXiv.2210.10255} {\bibfield  {journal} {\bibinfo
  {journal} {arXiv e-prints}\ ,\ \bibinfo {eid} {arXiv:2210.10255}} (\bibinfo
  {year} {2022})}\BibitemShut {NoStop}%
\bibitem [{\citenamefont {{Iqbal}}\ \emph {et~al.}(2023)\citenamefont
  {{Iqbal}}, \citenamefont {{Tantivasadakarn}}, \citenamefont {{Gatterman}},
  \citenamefont {{Gerber}}, \citenamefont {{Gilmore}}, \citenamefont {{Gresh}},
  \citenamefont {{Hankin}}, \citenamefont {{Hewitt}}, \citenamefont {{Horst}},
  \citenamefont {{Matheny}}, \citenamefont {{Mengle}}, \citenamefont
  {{Neyenhuis}}, \citenamefont {{Vishwanath}}, \citenamefont {{Foss-Feig}},
  \citenamefont {{Verresen}},\ and\ \citenamefont {{Dreyer}}}]{iqbal2023}%
  \BibitemOpen
  \bibfield  {author} {\bibinfo {author} {\bibfnamefont {M.}~\bibnamefont
  {{Iqbal}}}, \bibinfo {author} {\bibfnamefont {N.}~\bibnamefont
  {{Tantivasadakarn}}}, \bibinfo {author} {\bibfnamefont {T.~M.}\ \bibnamefont
  {{Gatterman}}}, \bibinfo {author} {\bibfnamefont {J.~A.}\ \bibnamefont
  {{Gerber}}}, \bibinfo {author} {\bibfnamefont {K.}~\bibnamefont {{Gilmore}}},
  \bibinfo {author} {\bibfnamefont {D.}~\bibnamefont {{Gresh}}}, \bibinfo
  {author} {\bibfnamefont {A.}~\bibnamefont {{Hankin}}}, \bibinfo {author}
  {\bibfnamefont {N.}~\bibnamefont {{Hewitt}}}, \bibinfo {author}
  {\bibfnamefont {C.~V.}\ \bibnamefont {{Horst}}}, \bibinfo {author}
  {\bibfnamefont {M.}~\bibnamefont {{Matheny}}}, \bibinfo {author}
  {\bibfnamefont {T.}~\bibnamefont {{Mengle}}}, \bibinfo {author}
  {\bibfnamefont {B.}~\bibnamefont {{Neyenhuis}}}, \bibinfo {author}
  {\bibfnamefont {A.}~\bibnamefont {{Vishwanath}}}, \bibinfo {author}
  {\bibfnamefont {M.}~\bibnamefont {{Foss-Feig}}}, \bibinfo {author}
  {\bibfnamefont {R.}~\bibnamefont {{Verresen}}},\ and\ \bibinfo {author}
  {\bibfnamefont {H.}~\bibnamefont {{Dreyer}}},\ }\bibfield  {title} {\bibinfo
  {title} {{Topological Order from Measurements and Feed-Forward on a Trapped
  Ion Quantum Computer}},\ }\href {https://doi.org/10.48550/arXiv.2302.01917}
  {\bibfield  {journal} {\bibinfo  {journal} {arXiv e-prints}\ ,\ \bibinfo
  {eid} {arXiv:2302.01917}} (\bibinfo {year} {2023})}\BibitemShut {NoStop}%
\bibitem [{\citenamefont {{Iqbal}}\ \emph {et~al.}(2024)\citenamefont
  {{Iqbal}}, \citenamefont {{Tantivasadakarn}}, \citenamefont {{Verresen}},
  \citenamefont {{Campbell}}, \citenamefont {{Dreiling}}, \citenamefont
  {{Figgatt}}, \citenamefont {{Gaebler}}, \citenamefont {{Johansen}},
  \citenamefont {{Mills}}, \citenamefont {{Moses}}, \citenamefont {{Pino}},
  \citenamefont {{Ransford}}, \citenamefont {{Rowe}}, \citenamefont
  {{Siegfried}}, \citenamefont {{Stutz}}, \citenamefont {{Foss-Feig}},
  \citenamefont {{Vishwanath}},\ and\ \citenamefont {{Dreyer}}}]{iqbal2024}%
  \BibitemOpen
  \bibfield  {author} {\bibinfo {author} {\bibfnamefont {M.}~\bibnamefont
  {{Iqbal}}}, \bibinfo {author} {\bibfnamefont {N.}~\bibnamefont
  {{Tantivasadakarn}}}, \bibinfo {author} {\bibfnamefont {R.}~\bibnamefont
  {{Verresen}}}, \bibinfo {author} {\bibfnamefont {S.~L.}\ \bibnamefont
  {{Campbell}}}, \bibinfo {author} {\bibfnamefont {J.~M.}\ \bibnamefont
  {{Dreiling}}}, \bibinfo {author} {\bibfnamefont {C.}~\bibnamefont
  {{Figgatt}}}, \bibinfo {author} {\bibfnamefont {J.~P.}\ \bibnamefont
  {{Gaebler}}}, \bibinfo {author} {\bibfnamefont {J.}~\bibnamefont
  {{Johansen}}}, \bibinfo {author} {\bibfnamefont {M.}~\bibnamefont {{Mills}}},
  \bibinfo {author} {\bibfnamefont {S.~A.}\ \bibnamefont {{Moses}}}, \bibinfo
  {author} {\bibfnamefont {J.~M.}\ \bibnamefont {{Pino}}}, \bibinfo {author}
  {\bibfnamefont {A.}~\bibnamefont {{Ransford}}}, \bibinfo {author}
  {\bibfnamefont {M.}~\bibnamefont {{Rowe}}}, \bibinfo {author} {\bibfnamefont
  {P.}~\bibnamefont {{Siegfried}}}, \bibinfo {author} {\bibfnamefont {R.~P.}\
  \bibnamefont {{Stutz}}}, \bibinfo {author} {\bibfnamefont {M.}~\bibnamefont
  {{Foss-Feig}}}, \bibinfo {author} {\bibfnamefont {A.}~\bibnamefont
  {{Vishwanath}}},\ and\ \bibinfo {author} {\bibfnamefont {H.}~\bibnamefont
  {{Dreyer}}},\ }\bibfield  {title} {\bibinfo {title} {{Non-Abelian topological
  order and anyons on a trapped-ion processor}},\ }\href
  {https://doi.org/10.1038/s41586-023-06934-4} {\bibfield  {journal} {\bibinfo
  {journal} {\nat}\ }\textbf {\bibinfo {volume} {626}},\ \bibinfo {pages} {505}
  (\bibinfo {year} {2024})}\BibitemShut {NoStop}%
\bibitem [{\citenamefont {Hastings}(2011)}]{hastings2011finiteT}%
  \BibitemOpen
  \bibfield  {author} {\bibinfo {author} {\bibfnamefont {M.~B.}\ \bibnamefont
  {Hastings}},\ }\bibfield  {title} {\bibinfo {title} {Topological order at
  nonzero temperature},\ }\href
  {https://doi.org/10.1103/PhysRevLett.107.210501} {\bibfield  {journal}
  {\bibinfo  {journal} {Phys. Rev. Lett.}\ }\textbf {\bibinfo {volume} {107}},\
  \bibinfo {pages} {210501} (\bibinfo {year} {2011})}\BibitemShut {NoStop}%
\bibitem [{\citenamefont {Lu}\ \emph {et~al.}(2020)\citenamefont {Lu},
  \citenamefont {Hsieh},\ and\ \citenamefont {Grover}}]{lu2020negativity}%
  \BibitemOpen
  \bibfield  {author} {\bibinfo {author} {\bibfnamefont {T.-C.}\ \bibnamefont
  {Lu}}, \bibinfo {author} {\bibfnamefont {T.~H.}\ \bibnamefont {Hsieh}},\ and\
  \bibinfo {author} {\bibfnamefont {T.}~\bibnamefont {Grover}},\ }\bibfield
  {title} {\bibinfo {title} {Detecting topological order at finite temperature
  using entanglement negativity},\ }\href
  {https://doi.org/10.1103/PhysRevLett.125.116801} {\bibfield  {journal}
  {\bibinfo  {journal} {Phys. Rev. Lett.}\ }\textbf {\bibinfo {volume} {125}},\
  \bibinfo {pages} {116801} (\bibinfo {year} {2020})}\BibitemShut {NoStop}%
\bibitem [{\citenamefont {Dennis}\ \emph {et~al.}(2002)\citenamefont {Dennis},
  \citenamefont {Kitaev}, \citenamefont {Landahl},\ and\ \citenamefont
  {Preskill}}]{Dennis2002}%
  \BibitemOpen
  \bibfield  {author} {\bibinfo {author} {\bibfnamefont {E.}~\bibnamefont
  {Dennis}}, \bibinfo {author} {\bibfnamefont {A.}~\bibnamefont {Kitaev}},
  \bibinfo {author} {\bibfnamefont {A.}~\bibnamefont {Landahl}},\ and\ \bibinfo
  {author} {\bibfnamefont {J.}~\bibnamefont {Preskill}},\ }\bibfield  {title}
  {\bibinfo {title} {Topological quantum memory},\ }\href
  {https://doi.org/10.1063/1.1499754} {\bibfield  {journal} {\bibinfo
  {journal} {Journal of Mathematical Physics}\ }\textbf {\bibinfo {volume}
  {43}},\ \bibinfo {pages} {4452–4505} (\bibinfo {year} {2002})}\BibitemShut
  {NoStop}%
\bibitem [{\citenamefont {{Coser}}\ and\ \citenamefont
  {{P{\'e}rez-Garc{\'\i}a}}(2019)}]{coser2019class}%
  \BibitemOpen
  \bibfield  {author} {\bibinfo {author} {\bibfnamefont {A.}~\bibnamefont
  {{Coser}}}\ and\ \bibinfo {author} {\bibfnamefont {D.}~\bibnamefont
  {{P{\'e}rez-Garc{\'\i}a}}},\ }\bibfield  {title} {\bibinfo {title}
  {{Classification of phases for mixed states via fast dissipative
  evolution}},\ }\href {https://doi.org/10.22331/q-2019-08-12-174} {\bibfield
  {journal} {\bibinfo  {journal} {Quantum}\ }\textbf {\bibinfo {volume} {3}},\
  \bibinfo {pages} {174} (\bibinfo {year} {2019})}\BibitemShut {NoStop}%
\bibitem [{\citenamefont {{Bao}}\ \emph {et~al.}(2023)\citenamefont {{Bao}},
  \citenamefont {{Fan}}, \citenamefont {{Vishwanath}},\ and\ \citenamefont
  {{Altman}}}]{bao2023mixed}%
  \BibitemOpen
  \bibfield  {author} {\bibinfo {author} {\bibfnamefont {Y.}~\bibnamefont
  {{Bao}}}, \bibinfo {author} {\bibfnamefont {R.}~\bibnamefont {{Fan}}},
  \bibinfo {author} {\bibfnamefont {A.}~\bibnamefont {{Vishwanath}}},\ and\
  \bibinfo {author} {\bibfnamefont {E.}~\bibnamefont {{Altman}}},\ }\bibfield
  {title} {\bibinfo {title} {{Mixed-state topological order and the errorfield
  double formulation of decoherence-induced transitions}},\ }\bibfield
  {journal} {\bibinfo  {journal} {arXiv e-prints}\ }\href
  {https://doi.org/10.48550/arXiv.2301.05687} {10.48550/arXiv.2301.05687}
  (\bibinfo {year} {2023}),\ \Eprint {https://arxiv.org/abs/2301.05687}
  {arXiv:2301.05687 [quant-ph]} \BibitemShut {NoStop}%
\bibitem [{\citenamefont {{Fan}}\ \emph {et~al.}(2023)\citenamefont {{Fan}},
  \citenamefont {{Bao}}, \citenamefont {{Altman}},\ and\ \citenamefont
  {{Vishwanath}}}]{fan2023mixed}%
  \BibitemOpen
  \bibfield  {author} {\bibinfo {author} {\bibfnamefont {R.}~\bibnamefont
  {{Fan}}}, \bibinfo {author} {\bibfnamefont {Y.}~\bibnamefont {{Bao}}},
  \bibinfo {author} {\bibfnamefont {E.}~\bibnamefont {{Altman}}},\ and\
  \bibinfo {author} {\bibfnamefont {A.}~\bibnamefont {{Vishwanath}}},\
  }\bibfield  {title} {\bibinfo {title} {{Diagnostics of mixed-state
  topological order and breakdown of quantum memory}},\ }\bibfield  {journal}
  {\bibinfo  {journal} {arXiv e-prints}\ }\href
  {https://doi.org/10.48550/arXiv.2301.05689} {10.48550/arXiv.2301.05689}
  (\bibinfo {year} {2023}),\ \Eprint {https://arxiv.org/abs/2301.05689}
  {arXiv:2301.05689 [quant-ph]} \BibitemShut {NoStop}%
\bibitem [{\citenamefont {Lee}\ \emph {et~al.}(2023)\citenamefont {Lee},
  \citenamefont {Jian},\ and\ \citenamefont {Xu}}]{lee2023decohere}%
  \BibitemOpen
  \bibfield  {author} {\bibinfo {author} {\bibfnamefont {J.~Y.}\ \bibnamefont
  {Lee}}, \bibinfo {author} {\bibfnamefont {C.-M.}\ \bibnamefont {Jian}},\ and\
  \bibinfo {author} {\bibfnamefont {C.}~\bibnamefont {Xu}},\ }\bibfield
  {title} {\bibinfo {title} {Quantum criticality under decoherence or weak
  measurement},\ }\href {https://doi.org/10.1103/PRXQuantum.4.030317}
  {\bibfield  {journal} {\bibinfo  {journal} {PRX Quantum}\ }\textbf {\bibinfo
  {volume} {4}},\ \bibinfo {pages} {030317} (\bibinfo {year}
  {2023})}\BibitemShut {NoStop}%
\bibitem [{\citenamefont {{Chen}}\ and\ \citenamefont
  {{Grover}}(2023{\natexlab{a}})}]{chen2023separable}%
  \BibitemOpen
  \bibfield  {author} {\bibinfo {author} {\bibfnamefont {Y.-H.}\ \bibnamefont
  {{Chen}}}\ and\ \bibinfo {author} {\bibfnamefont {T.}~\bibnamefont
  {{Grover}}},\ }\bibfield  {title} {\bibinfo {title} {{Symmetry-enforced
  many-body separability transitions}},\ }\bibfield  {journal} {\bibinfo
  {journal} {arXiv e-prints}\ }\href
  {https://doi.org/10.48550/arXiv.2310.07286} {10.48550/arXiv.2310.07286}
  (\bibinfo {year} {2023}{\natexlab{a}}),\ \Eprint
  {https://arxiv.org/abs/2310.07286} {arXiv:2310.07286 [quant-ph]} \BibitemShut
  {NoStop}%
\bibitem [{\citenamefont {{Chen}}\ and\ \citenamefont
  {{Grover}}(2023{\natexlab{b}})}]{chen2023separable2}%
  \BibitemOpen
  \bibfield  {author} {\bibinfo {author} {\bibfnamefont {Y.-H.}\ \bibnamefont
  {{Chen}}}\ and\ \bibinfo {author} {\bibfnamefont {T.}~\bibnamefont
  {{Grover}}},\ }\bibfield  {title} {\bibinfo {title} {{Separability
  transitions in topological states induced by local decoherence}},\ }\bibfield
   {journal} {\bibinfo  {journal} {arXiv e-prints}\ }\href
  {https://doi.org/10.48550/arXiv.2309.11879} {10.48550/arXiv.2309.11879}
  (\bibinfo {year} {2023}{\natexlab{b}}),\ \Eprint
  {https://arxiv.org/abs/2309.11879} {arXiv:2309.11879 [quant-ph]} \BibitemShut
  {NoStop}%
\bibitem [{\citenamefont {{Ma}}\ \emph {et~al.}(2023)\citenamefont {{Ma}},
  \citenamefont {{Zhang}}, \citenamefont {{Bi}}, \citenamefont {{Cheng}},\ and\
  \citenamefont {{Wang}}}]{ma2023avg}%
  \BibitemOpen
  \bibfield  {author} {\bibinfo {author} {\bibfnamefont {R.}~\bibnamefont
  {{Ma}}}, \bibinfo {author} {\bibfnamefont {J.-H.}\ \bibnamefont {{Zhang}}},
  \bibinfo {author} {\bibfnamefont {Z.}~\bibnamefont {{Bi}}}, \bibinfo {author}
  {\bibfnamefont {M.}~\bibnamefont {{Cheng}}},\ and\ \bibinfo {author}
  {\bibfnamefont {C.}~\bibnamefont {{Wang}}},\ }\bibfield  {title} {\bibinfo
  {title} {{Topological Phases with Average Symmetries: the Decohered, the
  Disordered, and the Intrinsic}},\ }\bibfield  {journal} {\bibinfo  {journal}
  {arXiv e-prints}\ }\href {https://doi.org/10.48550/arXiv.2305.16399}
  {10.48550/arXiv.2305.16399} (\bibinfo {year} {2023}),\ \Eprint
  {https://arxiv.org/abs/2305.16399} {arXiv:2305.16399 [cond-mat.str-el]}
  \BibitemShut {NoStop}%
\bibitem [{\citenamefont {{Sang}}\ \emph {et~al.}(2023)\citenamefont {{Sang}},
  \citenamefont {{Zou}},\ and\ \citenamefont {{Hsieh}}}]{sang2023mixed}%
  \BibitemOpen
  \bibfield  {author} {\bibinfo {author} {\bibfnamefont {S.}~\bibnamefont
  {{Sang}}}, \bibinfo {author} {\bibfnamefont {Y.}~\bibnamefont {{Zou}}},\ and\
  \bibinfo {author} {\bibfnamefont {T.~H.}\ \bibnamefont {{Hsieh}}},\
  }\bibfield  {title} {\bibinfo {title} {{Mixed-state Quantum Phases:
  Renormalization and Quantum Error Correction}},\ }\bibfield  {journal}
  {\bibinfo  {journal} {arXiv e-prints}\ }\href
  {https://doi.org/10.48550/arXiv.2310.08639} {10.48550/arXiv.2310.08639}
  (\bibinfo {year} {2023}),\ \Eprint {https://arxiv.org/abs/2310.08639}
  {arXiv:2310.08639 [quant-ph]} \BibitemShut {NoStop}%
\bibitem [{\citenamefont {{Dai}}\ \emph {et~al.}(2023)\citenamefont {{Dai}},
  \citenamefont {{Wang}}, \citenamefont {{Wang}},\ and\ \citenamefont
  {{Wang}}}]{dai2023ssto}%
  \BibitemOpen
  \bibfield  {author} {\bibinfo {author} {\bibfnamefont {X.-D.}\ \bibnamefont
  {{Dai}}}, \bibinfo {author} {\bibfnamefont {Z.}~\bibnamefont {{Wang}}},
  \bibinfo {author} {\bibfnamefont {H.-R.}\ \bibnamefont {{Wang}}},\ and\
  \bibinfo {author} {\bibfnamefont {Z.}~\bibnamefont {{Wang}}},\ }\bibfield
  {title} {\bibinfo {title} {{Steady-state topological order}},\ }\bibfield
  {journal} {\bibinfo  {journal} {arXiv e-prints}\ }\href
  {https://doi.org/10.48550/arXiv.2310.17612} {10.48550/arXiv.2310.17612}
  (\bibinfo {year} {2023}),\ \Eprint {https://arxiv.org/abs/2310.17612}
  {arXiv:2310.17612 [quant-ph]} \BibitemShut {NoStop}%
\bibitem [{\citenamefont {{Li}}\ and\ \citenamefont
  {{Mong}}(2024)}]{li2024replica}%
  \BibitemOpen
  \bibfield  {author} {\bibinfo {author} {\bibfnamefont {Z.}~\bibnamefont
  {{Li}}}\ and\ \bibinfo {author} {\bibfnamefont {R.~S.~K.}\ \bibnamefont
  {{Mong}}},\ }\bibfield  {title} {\bibinfo {title} {{Replica topological order
  in quantum mixed states and quantum error correction}},\ }\href
  {https://doi.org/10.48550/arXiv.2402.09516} {\bibfield  {journal} {\bibinfo
  {journal} {arXiv e-prints}\ ,\ \bibinfo {eid} {arXiv:2402.09516}} (\bibinfo
  {year} {2024})},\ \Eprint {https://arxiv.org/abs/2402.09516}
  {arXiv:2402.09516 [quant-ph]} \BibitemShut {NoStop}%
\bibitem [{\citenamefont {{Lavasani}}\ and\ \citenamefont
  {{Vijay}}(2024)}]{lavasani2024qec}%
  \BibitemOpen
  \bibfield  {author} {\bibinfo {author} {\bibfnamefont {A.}~\bibnamefont
  {{Lavasani}}}\ and\ \bibinfo {author} {\bibfnamefont {S.}~\bibnamefont
  {{Vijay}}},\ }\bibfield  {title} {\bibinfo {title} {{The Stability of Gapped
  Quantum Matter and Error-Correction with Adiabatic Noise}},\ }\bibfield
  {journal} {\bibinfo  {journal} {arXiv e-prints}\ }\href
  {https://doi.org/10.48550/arXiv.2402.14906} {10.48550/arXiv.2402.14906}
  (\bibinfo {year} {2024}),\ \Eprint {https://arxiv.org/abs/2402.14906}
  {arXiv:2402.14906 [cond-mat.str-el]} \BibitemShut {NoStop}%
\bibitem [{\citenamefont {{de Groot}}\ \emph {et~al.}(2022)\citenamefont {{de
  Groot}}, \citenamefont {{Turzillo}},\ and\ \citenamefont
  {{Schuch}}}]{degroot2022og}%
  \BibitemOpen
  \bibfield  {author} {\bibinfo {author} {\bibfnamefont {C.}~\bibnamefont {{de
  Groot}}}, \bibinfo {author} {\bibfnamefont {A.}~\bibnamefont {{Turzillo}}},\
  and\ \bibinfo {author} {\bibfnamefont {N.}~\bibnamefont {{Schuch}}},\
  }\bibfield  {title} {\bibinfo {title} {{Symmetry Protected Topological Order
  in Open Quantum Systems}},\ }\href
  {https://doi.org/10.22331/q-2022-11-10-856} {\bibfield  {journal} {\bibinfo
  {journal} {Quantum}\ }\textbf {\bibinfo {volume} {6}},\ \bibinfo {pages}
  {856} (\bibinfo {year} {2022})}\BibitemShut {NoStop}%
\bibitem [{\citenamefont {Ma}\ and\ \citenamefont {Wang}(2023)}]{ma2023prx}%
  \BibitemOpen
  \bibfield  {author} {\bibinfo {author} {\bibfnamefont {R.}~\bibnamefont
  {Ma}}\ and\ \bibinfo {author} {\bibfnamefont {C.}~\bibnamefont {Wang}},\
  }\bibfield  {title} {\bibinfo {title} {Average symmetry-protected topological
  phases},\ }\href {https://doi.org/10.1103/PhysRevX.13.031016} {\bibfield
  {journal} {\bibinfo  {journal} {Phys. Rev. X}\ }\textbf {\bibinfo {volume}
  {13}},\ \bibinfo {pages} {031016} (\bibinfo {year} {2023})}\BibitemShut
  {NoStop}%
\bibitem [{\citenamefont {{Zhang}}\ \emph {et~al.}(2022)\citenamefont
  {{Zhang}}, \citenamefont {{Qi}},\ and\ \citenamefont
  {{Bi}}}]{zhang2022strange}%
  \BibitemOpen
  \bibfield  {author} {\bibinfo {author} {\bibfnamefont {J.-H.}\ \bibnamefont
  {{Zhang}}}, \bibinfo {author} {\bibfnamefont {Y.}~\bibnamefont {{Qi}}},\ and\
  \bibinfo {author} {\bibfnamefont {Z.}~\bibnamefont {{Bi}}},\ }\bibfield
  {title} {\bibinfo {title} {{Strange Correlation Function for Average
  Symmetry-Protected Topological Phases}},\ }\bibfield  {journal} {\bibinfo
  {journal} {arXiv e-prints}\ }\href
  {https://doi.org/10.48550/arXiv.2210.17485} {10.48550/arXiv.2210.17485}
  (\bibinfo {year} {2022}),\ \Eprint {https://arxiv.org/abs/2210.17485}
  {arXiv:2210.17485 [cond-mat.str-el]} \BibitemShut {NoStop}%
\bibitem [{\citenamefont {{Lee}}\ \emph {et~al.}(2022)\citenamefont {{Lee}},
  \citenamefont {{You}},\ and\ \citenamefont {{Xu}}}]{lee2022aspt}%
  \BibitemOpen
  \bibfield  {author} {\bibinfo {author} {\bibfnamefont {J.~Y.}\ \bibnamefont
  {{Lee}}}, \bibinfo {author} {\bibfnamefont {Y.-Z.}\ \bibnamefont {{You}}},\
  and\ \bibinfo {author} {\bibfnamefont {C.}~\bibnamefont {{Xu}}},\ }\bibfield
  {title} {\bibinfo {title} {{Symmetry protected topological phases under
  decoherence}},\ }\bibfield  {journal} {\bibinfo  {journal} {arXiv e-prints}\
  }\href {https://doi.org/10.48550/arXiv.2210.16323}
  {10.48550/arXiv.2210.16323} (\bibinfo {year} {2022}),\ \Eprint
  {https://arxiv.org/abs/2210.16323} {arXiv:2210.16323 [cond-mat.str-el]}
  \BibitemShut {NoStop}%
\bibitem [{\citenamefont {{Chirame}}\ \emph {et~al.}(2024)\citenamefont
  {{Chirame}}, \citenamefont {{Burnell}}, \citenamefont {{Gopalakrishnan}},\
  and\ \citenamefont {{Prem}}}]{chirame2024}%
  \BibitemOpen
  \bibfield  {author} {\bibinfo {author} {\bibfnamefont {S.}~\bibnamefont
  {{Chirame}}}, \bibinfo {author} {\bibfnamefont {F.~J.}\ \bibnamefont
  {{Burnell}}}, \bibinfo {author} {\bibfnamefont {S.}~\bibnamefont
  {{Gopalakrishnan}}},\ and\ \bibinfo {author} {\bibfnamefont {A.}~\bibnamefont
  {{Prem}}},\ }\bibfield  {title} {\bibinfo {title} {{Stable Symmetry-Protected
  Topological Phases in Systems with Heralded Noise}},\ }\bibfield  {journal}
  {\bibinfo  {journal} {arXiv e-prints}\ }\href
  {https://doi.org/10.48550/arXiv.2404.16962} {10.48550/arXiv.2404.16962}
  (\bibinfo {year} {2024}),\ \Eprint {https://arxiv.org/abs/2404.16962}
  {arXiv:2404.16962 [quant-ph]} \BibitemShut {NoStop}%
\bibitem [{\citenamefont {{Guo}}\ \emph {et~al.}(2024)\citenamefont {{Guo}},
  \citenamefont {{Ding}},\ and\ \citenamefont {{Yang}}}]{guo2024}%
  \BibitemOpen
  \bibfield  {author} {\bibinfo {author} {\bibfnamefont {Y.}~\bibnamefont
  {{Guo}}}, \bibinfo {author} {\bibfnamefont {K.}~\bibnamefont {{Ding}}},\ and\
  \bibinfo {author} {\bibfnamefont {S.}~\bibnamefont {{Yang}}},\ }\bibfield
  {title} {\bibinfo {title} {{A New Framework for Quantum Phases in Open
  Systems: Steady State of Imaginary-Time Lindbladian Evolution}},\ }\bibfield
  {journal} {\bibinfo  {journal} {arXiv e-prints}\ }\href
  {https://doi.org/10.48550/arXiv.2408.03239} {10.48550/arXiv.2408.03239}
  (\bibinfo {year} {2024}),\ \Eprint {https://arxiv.org/abs/2408.03239}
  {arXiv:2408.03239 [quant-ph]} \BibitemShut {NoStop}%
\bibitem [{\citenamefont {{Wang}}\ \emph {et~al.}(2023)\citenamefont {{Wang}},
  \citenamefont {{Wu}},\ and\ \citenamefont {{Wang}}}]{wang2023mixed}%
  \BibitemOpen
  \bibfield  {author} {\bibinfo {author} {\bibfnamefont {Z.}~\bibnamefont
  {{Wang}}}, \bibinfo {author} {\bibfnamefont {Z.}~\bibnamefont {{Wu}}},\ and\
  \bibinfo {author} {\bibfnamefont {Z.}~\bibnamefont {{Wang}}},\ }\bibfield
  {title} {\bibinfo {title} {{Intrinsic Mixed-state Topological Order Without
  Quantum Memory}},\ }\href {https://doi.org/10.48550/arXiv.2307.13758}
  {\bibfield  {journal} {\bibinfo  {journal} {arXiv e-prints}\ ,\ \bibinfo
  {eid} {arXiv:2307.13758}} (\bibinfo {year} {2023})},\ \Eprint
  {https://arxiv.org/abs/2307.13758} {arXiv:2307.13758 [quant-ph]} \BibitemShut
  {NoStop}%
\bibitem [{\citenamefont {Bombin}\ \emph {et~al.}(2009)\citenamefont {Bombin},
  \citenamefont {Kargarian},\ and\ \citenamefont
  {Martin-Delgado}}]{bombin2009}%
  \BibitemOpen
  \bibfield  {author} {\bibinfo {author} {\bibfnamefont {H.}~\bibnamefont
  {Bombin}}, \bibinfo {author} {\bibfnamefont {M.}~\bibnamefont {Kargarian}},\
  and\ \bibinfo {author} {\bibfnamefont {M.~A.}\ \bibnamefont
  {Martin-Delgado}},\ }\bibfield  {title} {\bibinfo {title} {Interacting
  anyonic fermions in a two-body color code model},\ }\href
  {https://doi.org/10.1103/PhysRevB.80.075111} {\bibfield  {journal} {\bibinfo
  {journal} {Phys. Rev. B}\ }\textbf {\bibinfo {volume} {80}},\ \bibinfo
  {pages} {075111} (\bibinfo {year} {2009})}\BibitemShut {NoStop}%
\bibitem [{\citenamefont {Bombin}(2010)}]{bombin2010}%
  \BibitemOpen
  \bibfield  {author} {\bibinfo {author} {\bibfnamefont {H.}~\bibnamefont
  {Bombin}},\ }\bibfield  {title} {\bibinfo {title} {Topological subsystem
  codes},\ }\href {https://doi.org/10.1103/PhysRevA.81.032301} {\bibfield
  {journal} {\bibinfo  {journal} {Phys. Rev. A}\ }\textbf {\bibinfo {volume}
  {81}},\ \bibinfo {pages} {032301} (\bibinfo {year} {2010})}\BibitemShut
  {NoStop}%
\bibitem [{\citenamefont {{Bombin}}\ \emph {et~al.}(2012)\citenamefont
  {{Bombin}}, \citenamefont {{Duclos-Cianci}},\ and\ \citenamefont
  {{Poulin}}}]{bombin2012}%
  \BibitemOpen
  \bibfield  {author} {\bibinfo {author} {\bibfnamefont {H.}~\bibnamefont
  {{Bombin}}}, \bibinfo {author} {\bibfnamefont {G.}~\bibnamefont
  {{Duclos-Cianci}}},\ and\ \bibinfo {author} {\bibfnamefont {D.}~\bibnamefont
  {{Poulin}}},\ }\bibfield  {title} {\bibinfo {title} {{Universal topological
  phase of two-dimensional stabilizer codes}},\ }\href
  {https://doi.org/10.1088/1367-2630/14/7/073048} {\bibfield  {journal}
  {\bibinfo  {journal} {New Journal of Physics}\ }\textbf {\bibinfo {volume}
  {14}},\ \bibinfo {eid} {073048} (\bibinfo {year} {2012})}\BibitemShut
  {NoStop}%
\bibitem [{\citenamefont {{Bomb{\'\i}n}}(2014)}]{bombin2014}%
  \BibitemOpen
  \bibfield  {author} {\bibinfo {author} {\bibfnamefont {H.}~\bibnamefont
  {{Bomb{\'\i}n}}},\ }\bibfield  {title} {\bibinfo {title} {{Structure of 2D
  Topological Stabilizer Codes}},\ }\href
  {https://doi.org/10.1007/s00220-014-1893-4} {\bibfield  {journal} {\bibinfo
  {journal} {Communications in Mathematical Physics}\ }\textbf {\bibinfo
  {volume} {327}},\ \bibinfo {pages} {387} (\bibinfo {year}
  {2014})}\BibitemShut {NoStop}%
\bibitem [{\citenamefont {Ellison}\ \emph {et~al.}(2022)\citenamefont
  {Ellison}, \citenamefont {Chen}, \citenamefont {Dua}, \citenamefont
  {Shirley}, \citenamefont {Tantivasadakarn},\ and\ \citenamefont
  {Williamson}}]{ellison2022}%
  \BibitemOpen
  \bibfield  {author} {\bibinfo {author} {\bibfnamefont {T.~D.}\ \bibnamefont
  {Ellison}}, \bibinfo {author} {\bibfnamefont {Y.-A.}\ \bibnamefont {Chen}},
  \bibinfo {author} {\bibfnamefont {A.}~\bibnamefont {Dua}}, \bibinfo {author}
  {\bibfnamefont {W.}~\bibnamefont {Shirley}}, \bibinfo {author} {\bibfnamefont
  {N.}~\bibnamefont {Tantivasadakarn}},\ and\ \bibinfo {author} {\bibfnamefont
  {D.~J.}\ \bibnamefont {Williamson}},\ }\bibfield  {title} {\bibinfo {title}
  {Pauli stabilizer models of twisted quantum doubles},\ }\href
  {https://doi.org/10.1103/PRXQuantum.3.010353} {\bibfield  {journal} {\bibinfo
   {journal} {PRX Quantum}\ }\textbf {\bibinfo {volume} {3}},\ \bibinfo {pages}
  {010353} (\bibinfo {year} {2022})}\BibitemShut {NoStop}%
\bibitem [{\citenamefont {Piroli}\ and\ \citenamefont
  {Cirac}(2020)}]{piroli2020}%
  \BibitemOpen
  \bibfield  {author} {\bibinfo {author} {\bibfnamefont {L.}~\bibnamefont
  {Piroli}}\ and\ \bibinfo {author} {\bibfnamefont {J.~I.}\ \bibnamefont
  {Cirac}},\ }\bibfield  {title} {\bibinfo {title} {Quantum cellular automata,
  tensor networks, and area laws},\ }\href
  {https://doi.org/10.1103/PhysRevLett.125.190402} {\bibfield  {journal}
  {\bibinfo  {journal} {Phys. Rev. Lett.}\ }\textbf {\bibinfo {volume} {125}},\
  \bibinfo {pages} {190402} (\bibinfo {year} {2020})}\BibitemShut {NoStop}%
\bibitem [{\citenamefont {Shirley}\ \emph {et~al.}(2022)\citenamefont
  {Shirley}, \citenamefont {Chen}, \citenamefont {Dua}, \citenamefont
  {Ellison}, \citenamefont {Tantivasadakarn},\ and\ \citenamefont
  {Williamson}}]{shirley2022qca}%
  \BibitemOpen
  \bibfield  {author} {\bibinfo {author} {\bibfnamefont {W.}~\bibnamefont
  {Shirley}}, \bibinfo {author} {\bibfnamefont {Y.-A.}\ \bibnamefont {Chen}},
  \bibinfo {author} {\bibfnamefont {A.}~\bibnamefont {Dua}}, \bibinfo {author}
  {\bibfnamefont {T.~D.}\ \bibnamefont {Ellison}}, \bibinfo {author}
  {\bibfnamefont {N.}~\bibnamefont {Tantivasadakarn}},\ and\ \bibinfo {author}
  {\bibfnamefont {D.~J.}\ \bibnamefont {Williamson}},\ }\bibfield  {title}
  {\bibinfo {title} {Three-dimensional quantum cellular automata from chiral
  semion surface topological order and beyond},\ }\href
  {https://doi.org/10.1103/PRXQuantum.3.030326} {\bibfield  {journal} {\bibinfo
   {journal} {PRX Quantum}\ }\textbf {\bibinfo {volume} {3}},\ \bibinfo {pages}
  {030326} (\bibinfo {year} {2022})}\BibitemShut {NoStop}%
\bibitem [{\citenamefont {{Chen}}\ \emph {et~al.}(2024)\citenamefont {{Chen}},
  \citenamefont {{Hermele}},\ and\ \citenamefont
  {{Stephen}}}]{chen2024sequence}%
  \BibitemOpen
  \bibfield  {author} {\bibinfo {author} {\bibfnamefont {X.}~\bibnamefont
  {{Chen}}}, \bibinfo {author} {\bibfnamefont {M.}~\bibnamefont {{Hermele}}},\
  and\ \bibinfo {author} {\bibfnamefont {D.~T.}\ \bibnamefont {{Stephen}}},\
  }\bibfield  {title} {\bibinfo {title} {{Sequential Adiabatic Generation of
  Chiral Topological States}},\ }\bibfield  {journal} {\bibinfo  {journal}
  {arXiv e-prints}\ }\href {https://doi.org/10.48550/arXiv.2402.03433}
  {10.48550/arXiv.2402.03433} (\bibinfo {year} {2024}),\ \Eprint
  {https://arxiv.org/abs/2402.03433} {arXiv:2402.03433 [cond-mat.str-el]}
  \BibitemShut {NoStop}%
\bibitem [{\citenamefont {Ellison}\ \emph {et~al.}(2023)\citenamefont
  {Ellison}, \citenamefont {Chen}, \citenamefont {Dua}, \citenamefont
  {Shirley}, \citenamefont {Tantivasadakarn},\ and\ \citenamefont
  {Williamson}}]{ellison2023}%
  \BibitemOpen
  \bibfield  {author} {\bibinfo {author} {\bibfnamefont {T.~D.}\ \bibnamefont
  {Ellison}}, \bibinfo {author} {\bibfnamefont {Y.-A.}\ \bibnamefont {Chen}},
  \bibinfo {author} {\bibfnamefont {A.}~\bibnamefont {Dua}}, \bibinfo {author}
  {\bibfnamefont {W.}~\bibnamefont {Shirley}}, \bibinfo {author} {\bibfnamefont
  {N.}~\bibnamefont {Tantivasadakarn}},\ and\ \bibinfo {author} {\bibfnamefont
  {D.~J.}\ \bibnamefont {Williamson}},\ }\bibfield  {title} {\bibinfo {title}
  {Pauli topological subsystem codes from abelian anyon theories},\ }\href
  {https://doi.org/10.22331/q-2023-10-12-1137} {\bibfield  {journal} {\bibinfo
  {journal} {Quantum}\ }\textbf {\bibinfo {volume} {7}},\ \bibinfo {pages}
  {1137} (\bibinfo {year} {2023})}\BibitemShut {NoStop}%
\bibitem [{\citenamefont {{Burnell}}(2018)}]{burnellreview}%
  \BibitemOpen
  \bibfield  {author} {\bibinfo {author} {\bibfnamefont {F.~J.}\ \bibnamefont
  {{Burnell}}},\ }\bibfield  {title} {\bibinfo {title} {{Anyon Condensation and
  Its Applications}},\ }\href
  {https://doi.org/10.1146/annurev-conmatphys-033117-054154} {\bibfield
  {journal} {\bibinfo  {journal} {Annual Review of Condensed Matter Physics}\
  }\textbf {\bibinfo {volume} {9}},\ \bibinfo {pages} {307} (\bibinfo {year}
  {2018})}\BibitemShut {NoStop}%
\bibitem [{\citenamefont {{Walker}}\ and\ \citenamefont
  {{Wang}}(2012)}]{walkerwang}%
  \BibitemOpen
  \bibfield  {author} {\bibinfo {author} {\bibfnamefont {K.}~\bibnamefont
  {{Walker}}}\ and\ \bibinfo {author} {\bibfnamefont {Z.}~\bibnamefont
  {{Wang}}},\ }\bibfield  {title} {\bibinfo {title} {{(3+1)-TQFTs and
  topological insulators}},\ }\href {https://doi.org/10.1007/s11467-011-0194-z}
  {\bibfield  {journal} {\bibinfo  {journal} {Frontiers of Physics}\ }\textbf
  {\bibinfo {volume} {7}},\ \bibinfo {pages} {150} (\bibinfo {year}
  {2012})}\BibitemShut {NoStop}%
\bibitem [{\citenamefont {Hsin}\ \emph {et~al.}(2019)\citenamefont {Hsin},
  \citenamefont {Lam},\ and\ \citenamefont {Seiberg}}]{hsin2019oneform}%
  \BibitemOpen
  \bibfield  {author} {\bibinfo {author} {\bibfnamefont {P.-S.}\ \bibnamefont
  {Hsin}}, \bibinfo {author} {\bibfnamefont {H.~T.}\ \bibnamefont {Lam}},\ and\
  \bibinfo {author} {\bibfnamefont {N.}~\bibnamefont {Seiberg}},\ }\bibfield
  {title} {\bibinfo {title} {{Comments on one-form global symmetries and their
  gauging in 3d and 4d}},\ }\href
  {https://doi.org/10.21468/SciPostPhys.6.3.039} {\bibfield  {journal}
  {\bibinfo  {journal} {SciPost Phys.}\ }\textbf {\bibinfo {volume} {6}},\
  \bibinfo {pages} {039} (\bibinfo {year} {2019})}\BibitemShut {NoStop}%
\bibitem [{\citenamefont {Moy}\ \emph {et~al.}(2023)\citenamefont {Moy},
  \citenamefont {Goldman}, \citenamefont {Sohal},\ and\ \citenamefont
  {Fradkin}}]{moy2023oblique}%
  \BibitemOpen
  \bibfield  {author} {\bibinfo {author} {\bibfnamefont {B.}~\bibnamefont
  {Moy}}, \bibinfo {author} {\bibfnamefont {H.}~\bibnamefont {Goldman}},
  \bibinfo {author} {\bibfnamefont {R.}~\bibnamefont {Sohal}},\ and\ \bibinfo
  {author} {\bibfnamefont {E.}~\bibnamefont {Fradkin}},\ }\bibfield  {title}
  {\bibinfo {title} {{Theory of oblique topological insulators}},\ }\href
  {https://doi.org/10.21468/SciPostPhys.14.2.023} {\bibfield  {journal}
  {\bibinfo  {journal} {SciPost Phys.}\ }\textbf {\bibinfo {volume} {14}},\
  \bibinfo {pages} {023} (\bibinfo {year} {2023})}\BibitemShut {NoStop}%
\bibitem [{\citenamefont {{Kapustin}}\ and\ \citenamefont
  {{Saulina}}(2011)}]{kapustin2011}%
  \BibitemOpen
  \bibfield  {author} {\bibinfo {author} {\bibfnamefont {A.}~\bibnamefont
  {{Kapustin}}}\ and\ \bibinfo {author} {\bibfnamefont {N.}~\bibnamefont
  {{Saulina}}},\ }\bibfield  {title} {\bibinfo {title} {{Topological boundary
  conditions in abelian Chern-Simons theory}},\ }\href
  {https://doi.org/10.1016/j.nuclphysb.2010.12.017} {\bibfield  {journal}
  {\bibinfo  {journal} {Nuclear Physics B}\ }\textbf {\bibinfo {volume}
  {845}},\ \bibinfo {pages} {393} (\bibinfo {year} {2011})},\ \Eprint
  {https://arxiv.org/abs/1008.0654} {arXiv:1008.0654 [hep-th]} \BibitemShut
  {NoStop}%
\bibitem [{\citenamefont {{Jamio{\l}kowski}}(1972)}]{jamio1972}%
  \BibitemOpen
  \bibfield  {author} {\bibinfo {author} {\bibfnamefont {A.}~\bibnamefont
  {{Jamio{\l}kowski}}},\ }\bibfield  {title} {\bibinfo {title} {{Linear
  transformations which preserve trace and positive semidefiniteness of
  operators}},\ }\href {https://doi.org/10.1016/0034-4877(72)90011-0}
  {\bibfield  {journal} {\bibinfo  {journal} {Reports on Mathematical Physics}\
  }\textbf {\bibinfo {volume} {3}},\ \bibinfo {pages} {275} (\bibinfo {year}
  {1972})}\BibitemShut {NoStop}%
\bibitem [{\citenamefont {Choi}(1975)}]{choi1975}%
  \BibitemOpen
  \bibfield  {author} {\bibinfo {author} {\bibfnamefont {M.-D.}\ \bibnamefont
  {Choi}},\ }\bibfield  {title} {\bibinfo {title} {Completely positive linear
  maps on complex matrices},\ }\href
  {https://doi.org/https://doi.org/10.1016/0024-3795(75)90075-0} {\bibfield
  {journal} {\bibinfo  {journal} {Linear Algebra and its Applications}\
  }\textbf {\bibinfo {volume} {10}},\ \bibinfo {pages} {285} (\bibinfo {year}
  {1975})}\BibitemShut {NoStop}%
\bibitem [{\citenamefont {{Yoshida}}(2011)}]{yoshida2011}%
  \BibitemOpen
  \bibfield  {author} {\bibinfo {author} {\bibfnamefont {B.}~\bibnamefont
  {{Yoshida}}},\ }\bibfield  {title} {\bibinfo {title} {{Feasibility of
  self-correcting quantum memory and thermal stability of topological order}},\
  }\href {https://doi.org/10.1016/j.aop.2011.06.001} {\bibfield  {journal}
  {\bibinfo  {journal} {Annals of Physics}\ }\textbf {\bibinfo {volume}
  {326}},\ \bibinfo {pages} {2566} (\bibinfo {year} {2011})}\BibitemShut
  {NoStop}%
\bibitem [{\citenamefont {Poulin}\ \emph {et~al.}(2019)\citenamefont {Poulin},
  \citenamefont {Melko},\ and\ \citenamefont {Hastings}}]{poulin2019}%
  \BibitemOpen
  \bibfield  {author} {\bibinfo {author} {\bibfnamefont {D.}~\bibnamefont
  {Poulin}}, \bibinfo {author} {\bibfnamefont {R.~G.}\ \bibnamefont {Melko}},\
  and\ \bibinfo {author} {\bibfnamefont {M.~B.}\ \bibnamefont {Hastings}},\
  }\bibfield  {title} {\bibinfo {title} {Self-correction in wegner's
  three-dimensional ising lattice gauge theory},\ }\href
  {https://doi.org/10.1103/PhysRevB.99.094103} {\bibfield  {journal} {\bibinfo
  {journal} {Phys. Rev. B}\ }\textbf {\bibinfo {volume} {99}},\ \bibinfo
  {pages} {094103} (\bibinfo {year} {2019})}\BibitemShut {NoStop}%
\bibitem [{\citenamefont {{Kitaev}}(2006)}]{kitaev2006}%
  \BibitemOpen
  \bibfield  {author} {\bibinfo {author} {\bibfnamefont {A.}~\bibnamefont
  {{Kitaev}}},\ }\bibfield  {title} {\bibinfo {title} {{Anyons in an exactly
  solved model and beyond}},\ }\href
  {https://doi.org/10.1016/j.aop.2005.10.005} {\bibfield  {journal} {\bibinfo
  {journal} {Annals of Physics}\ }\textbf {\bibinfo {volume} {321}},\ \bibinfo
  {pages} {2} (\bibinfo {year} {2006})}\BibitemShut {NoStop}%
\bibitem [{\citenamefont {Tantivasadakarn}\ \emph {et~al.}(2023)\citenamefont
  {Tantivasadakarn}, \citenamefont {Vishwanath},\ and\ \citenamefont
  {Verresen}}]{feedforward2023}%
  \BibitemOpen
  \bibfield  {author} {\bibinfo {author} {\bibfnamefont {N.}~\bibnamefont
  {Tantivasadakarn}}, \bibinfo {author} {\bibfnamefont {A.}~\bibnamefont
  {Vishwanath}},\ and\ \bibinfo {author} {\bibfnamefont {R.}~\bibnamefont
  {Verresen}},\ }\bibfield  {title} {\bibinfo {title} {Hierarchy of topological
  order from finite-depth unitaries, measurement, and feedforward},\ }\href
  {https://doi.org/10.1103/PRXQuantum.4.020339} {\bibfield  {journal} {\bibinfo
   {journal} {PRX Quantum}\ }\textbf {\bibinfo {volume} {4}},\ \bibinfo {pages}
  {020339} (\bibinfo {year} {2023})}\BibitemShut {NoStop}%
\bibitem [{\citenamefont {Zanardi}\ and\ \citenamefont
  {Rasetti}(1997)}]{zanardi1997}%
  \BibitemOpen
  \bibfield  {author} {\bibinfo {author} {\bibfnamefont {P.}~\bibnamefont
  {Zanardi}}\ and\ \bibinfo {author} {\bibfnamefont {M.}~\bibnamefont
  {Rasetti}},\ }\bibfield  {title} {\bibinfo {title} {Noiseless quantum
  codes},\ }\href {https://doi.org/10.1103/PhysRevLett.79.3306} {\bibfield
  {journal} {\bibinfo  {journal} {Phys. Rev. Lett.}\ }\textbf {\bibinfo
  {volume} {79}},\ \bibinfo {pages} {3306} (\bibinfo {year}
  {1997})}\BibitemShut {NoStop}%
\bibitem [{\citenamefont {Zanardi}(1998)}]{zanardi1998}%
  \BibitemOpen
  \bibfield  {author} {\bibinfo {author} {\bibfnamefont {P.}~\bibnamefont
  {Zanardi}},\ }\bibfield  {title} {\bibinfo {title} {Dissipation and
  decoherence in a quantum register},\ }\href
  {https://doi.org/10.1103/PhysRevA.57.3276} {\bibfield  {journal} {\bibinfo
  {journal} {Phys. Rev. A}\ }\textbf {\bibinfo {volume} {57}},\ \bibinfo
  {pages} {3276} (\bibinfo {year} {1998})}\BibitemShut {NoStop}%
\bibitem [{\citenamefont {{Lidar}}\ \emph {et~al.}(1998)\citenamefont
  {{Lidar}}, \citenamefont {{Chuang}},\ and\ \citenamefont
  {{Whaley}}}]{lidar1998}%
  \BibitemOpen
  \bibfield  {author} {\bibinfo {author} {\bibfnamefont {D.~A.}\ \bibnamefont
  {{Lidar}}}, \bibinfo {author} {\bibfnamefont {I.~L.}\ \bibnamefont
  {{Chuang}}},\ and\ \bibinfo {author} {\bibfnamefont {K.~B.}\ \bibnamefont
  {{Whaley}}},\ }\bibfield  {title} {\bibinfo {title} {{Decoherence-Free
  Subspaces for Quantum Computation}},\ }\href
  {https://doi.org/10.1103/PhysRevLett.81.2594} {\bibfield  {journal} {\bibinfo
   {journal} {\prl}\ }\textbf {\bibinfo {volume} {81}},\ \bibinfo {pages}
  {2594} (\bibinfo {year} {1998})}\BibitemShut {NoStop}%
\bibitem [{\citenamefont {{Zanardi}}\ and\ \citenamefont
  {{Lloyd}}(2003)}]{zanardi2003}%
  \BibitemOpen
  \bibfield  {author} {\bibinfo {author} {\bibfnamefont {P.}~\bibnamefont
  {{Zanardi}}}\ and\ \bibinfo {author} {\bibfnamefont {S.}~\bibnamefont
  {{Lloyd}}},\ }\bibfield  {title} {\bibinfo {title} {{Topological Protection
  and Quantum Noiseless Subsystems}},\ }\href
  {https://doi.org/10.1103/PhysRevLett.90.067902} {\bibfield  {journal}
  {\bibinfo  {journal} {\prl}\ }\textbf {\bibinfo {volume} {90}},\ \bibinfo
  {eid} {067902} (\bibinfo {year} {2003})}\BibitemShut {NoStop}%
\bibitem [{\citenamefont {Bonderson}(2012)}]{bonderson2012thesis}%
  \BibitemOpen
  \bibfield  {author} {\bibinfo {author} {\bibfnamefont {P.~H.}\ \bibnamefont
  {Bonderson}},\ }\href@noop {} {\emph {\bibinfo {title} {Non-Abelian anyons
  and interferometry}}}\ (\bibinfo  {publisher} {California Institute of
  Technology},\ \bibinfo {year} {2012})\BibitemShut {NoStop}%
\bibitem [{\citenamefont {Brown}\ \emph {et~al.}(2016)\citenamefont {Brown},
  \citenamefont {Loss}, \citenamefont {Pachos}, \citenamefont {Self},\ and\
  \citenamefont {Wootton}}]{brownreview}%
  \BibitemOpen
  \bibfield  {author} {\bibinfo {author} {\bibfnamefont {B.~J.}\ \bibnamefont
  {Brown}}, \bibinfo {author} {\bibfnamefont {D.}~\bibnamefont {Loss}},
  \bibinfo {author} {\bibfnamefont {J.~K.}\ \bibnamefont {Pachos}}, \bibinfo
  {author} {\bibfnamefont {C.~N.}\ \bibnamefont {Self}},\ and\ \bibinfo
  {author} {\bibfnamefont {J.~R.}\ \bibnamefont {Wootton}},\ }\bibfield
  {title} {\bibinfo {title} {Quantum memories at finite temperature},\ }\href
  {https://doi.org/10.1103/RevModPhys.88.045005} {\bibfield  {journal}
  {\bibinfo  {journal} {Rev. Mod. Phys.}\ }\textbf {\bibinfo {volume} {88}},\
  \bibinfo {pages} {045005} (\bibinfo {year} {2016})}\BibitemShut {NoStop}%
\bibitem [{\citenamefont {Castelnovo}\ and\ \citenamefont
  {Chamon}(2007)}]{castelnovo2007classical}%
  \BibitemOpen
  \bibfield  {author} {\bibinfo {author} {\bibfnamefont {C.}~\bibnamefont
  {Castelnovo}}\ and\ \bibinfo {author} {\bibfnamefont {C.}~\bibnamefont
  {Chamon}},\ }\bibfield  {title} {\bibinfo {title} {Topological order and
  topological entropy in classical systems},\ }\href
  {https://doi.org/10.1103/PhysRevB.76.174416} {\bibfield  {journal} {\bibinfo
  {journal} {Phys. Rev. B}\ }\textbf {\bibinfo {volume} {76}},\ \bibinfo
  {pages} {174416} (\bibinfo {year} {2007})}\BibitemShut {NoStop}%
\bibitem [{\citenamefont {Wildeboer}\ \emph {et~al.}(2022)\citenamefont
  {Wildeboer}, \citenamefont {Iadecola},\ and\ \citenamefont
  {Williamson}}]{domtom}%
  \BibitemOpen
  \bibfield  {author} {\bibinfo {author} {\bibfnamefont {J.}~\bibnamefont
  {Wildeboer}}, \bibinfo {author} {\bibfnamefont {T.}~\bibnamefont
  {Iadecola}},\ and\ \bibinfo {author} {\bibfnamefont {D.~J.}\ \bibnamefont
  {Williamson}},\ }\bibfield  {title} {\bibinfo {title} {Symmetry-protected
  infinite-temperature quantum memory from subsystem codes},\ }\href
  {https://doi.org/10.1103/PRXQuantum.3.020330} {\bibfield  {journal} {\bibinfo
   {journal} {PRX Quantum}\ }\textbf {\bibinfo {volume} {3}},\ \bibinfo {pages}
  {020330} (\bibinfo {year} {2022})}\BibitemShut {NoStop}%
\bibitem [{\citenamefont {Levin}\ and\ \citenamefont {Wen}(2005)}]{levinwen}%
  \BibitemOpen
  \bibfield  {author} {\bibinfo {author} {\bibfnamefont {M.~A.}\ \bibnamefont
  {Levin}}\ and\ \bibinfo {author} {\bibfnamefont {X.-G.}\ \bibnamefont
  {Wen}},\ }\bibfield  {title} {\bibinfo {title} {String-net condensation: A
  physical mechanism for topological phases},\ }\href
  {https://doi.org/10.1103/PhysRevB.71.045110} {\bibfield  {journal} {\bibinfo
  {journal} {Phys. Rev. B}\ }\textbf {\bibinfo {volume} {71}},\ \bibinfo
  {pages} {045110} (\bibinfo {year} {2005})}\BibitemShut {NoStop}%
\bibitem [{\citenamefont {Burnell}\ \emph {et~al.}(2011)\citenamefont
  {Burnell}, \citenamefont {Simon},\ and\ \citenamefont
  {Slingerland}}]{burnell2011}%
  \BibitemOpen
  \bibfield  {author} {\bibinfo {author} {\bibfnamefont {F.~J.}\ \bibnamefont
  {Burnell}}, \bibinfo {author} {\bibfnamefont {S.~H.}\ \bibnamefont {Simon}},\
  and\ \bibinfo {author} {\bibfnamefont {J.~K.}\ \bibnamefont {Slingerland}},\
  }\bibfield  {title} {\bibinfo {title} {Condensation of achiral simple
  currents in topological lattice models: Hamiltonian study of topological
  symmetry breaking},\ }\href {https://doi.org/10.1103/PhysRevB.84.125434}
  {\bibfield  {journal} {\bibinfo  {journal} {Phys. Rev. B}\ }\textbf {\bibinfo
  {volume} {84}},\ \bibinfo {pages} {125434} (\bibinfo {year}
  {2011})}\BibitemShut {NoStop}%
\bibitem [{\citenamefont {{Burnell}}\ \emph {et~al.}(2012)\citenamefont
  {{Burnell}}, \citenamefont {{Simon}},\ and\ \citenamefont
  {{Slingerland}}}]{burnell2012}%
  \BibitemOpen
  \bibfield  {author} {\bibinfo {author} {\bibfnamefont {F.~J.}\ \bibnamefont
  {{Burnell}}}, \bibinfo {author} {\bibfnamefont {S.~H.}\ \bibnamefont
  {{Simon}}},\ and\ \bibinfo {author} {\bibfnamefont {J.~K.}\ \bibnamefont
  {{Slingerland}}},\ }\bibfield  {title} {\bibinfo {title} {{Phase transitions
  in topological lattice models via topological symmetry breaking}},\ }\href
  {https://doi.org/10.1088/1367-2630/14/1/015004} {\bibfield  {journal}
  {\bibinfo  {journal} {New Journal of Physics}\ }\textbf {\bibinfo {volume}
  {14}},\ \bibinfo {eid} {015004} (\bibinfo {year} {2012})}\BibitemShut
  {NoStop}%
\bibitem [{\citenamefont {Prem}\ \emph {et~al.}(2019)\citenamefont {Prem},
  \citenamefont {Huang}, \citenamefont {Song},\ and\ \citenamefont
  {Hermele}}]{cagenet}%
  \BibitemOpen
  \bibfield  {author} {\bibinfo {author} {\bibfnamefont {A.}~\bibnamefont
  {Prem}}, \bibinfo {author} {\bibfnamefont {S.-J.}\ \bibnamefont {Huang}},
  \bibinfo {author} {\bibfnamefont {H.}~\bibnamefont {Song}},\ and\ \bibinfo
  {author} {\bibfnamefont {M.}~\bibnamefont {Hermele}},\ }\bibfield  {title}
  {\bibinfo {title} {Cage-net fracton models},\ }\href
  {https://doi.org/10.1103/PhysRevX.9.021010} {\bibfield  {journal} {\bibinfo
  {journal} {Phys. Rev. X}\ }\textbf {\bibinfo {volume} {9}},\ \bibinfo {pages}
  {021010} (\bibinfo {year} {2019})}\BibitemShut {NoStop}%
\bibitem [{\citenamefont {{Mueger}}(2002)}]{mueger2002}%
  \BibitemOpen
  \bibfield  {author} {\bibinfo {author} {\bibfnamefont {M.}~\bibnamefont
  {{Mueger}}},\ }\bibfield  {title} {\bibinfo {title} {{On the Structure of
  Modular Categories}},\ }\href {https://doi.org/10.48550/arXiv.math/0201017}
  {\bibfield  {journal} {\bibinfo  {journal} {arXiv Mathematics e-prints}\ ,\
  \bibinfo {eid} {math/0201017}} (\bibinfo {year} {2002})},\ \Eprint
  {https://arxiv.org/abs/math/0201017} {arXiv:math/0201017 [math.CT]}
  \BibitemShut {NoStop}%
\bibitem [{\citenamefont {{Shokrian Zini}}\ and\ \citenamefont
  {{Wang}}(2021)}]{wangmixedtqft}%
  \BibitemOpen
  \bibfield  {author} {\bibinfo {author} {\bibfnamefont {M.}~\bibnamefont
  {{Shokrian Zini}}}\ and\ \bibinfo {author} {\bibfnamefont {Z.}~\bibnamefont
  {{Wang}}},\ }\bibfield  {title} {\bibinfo {title} {{Mixed-state TQFTs}},\
  }\href {https://doi.org/10.48550/arXiv.2110.13946} {\bibfield  {journal}
  {\bibinfo  {journal} {arXiv e-prints}\ ,\ \bibinfo {eid} {arXiv:2110.13946}}
  (\bibinfo {year} {2021})},\ \Eprint {https://arxiv.org/abs/2110.13946}
  {arXiv:2110.13946 [math.QA]} \BibitemShut {NoStop}%
\bibitem [{\citenamefont {Ma}\ \emph {et~al.}(2017)\citenamefont {Ma},
  \citenamefont {Lake}, \citenamefont {Chen},\ and\ \citenamefont
  {Hermele}}]{ma2017pstring}%
  \BibitemOpen
  \bibfield  {author} {\bibinfo {author} {\bibfnamefont {H.}~\bibnamefont
  {Ma}}, \bibinfo {author} {\bibfnamefont {E.}~\bibnamefont {Lake}}, \bibinfo
  {author} {\bibfnamefont {X.}~\bibnamefont {Chen}},\ and\ \bibinfo {author}
  {\bibfnamefont {M.}~\bibnamefont {Hermele}},\ }\bibfield  {title} {\bibinfo
  {title} {Fracton topological order via coupled layers},\ }\href
  {https://doi.org/10.1103/PhysRevB.95.245126} {\bibfield  {journal} {\bibinfo
  {journal} {Phys. Rev. B}\ }\textbf {\bibinfo {volume} {95}},\ \bibinfo
  {pages} {245126} (\bibinfo {year} {2017})}\BibitemShut {NoStop}%
\bibitem [{\citenamefont {{Sang}}\ and\ \citenamefont
  {{Hsieh}}(2024)}]{sang2024stable}%
  \BibitemOpen
  \bibfield  {author} {\bibinfo {author} {\bibfnamefont {S.}~\bibnamefont
  {{Sang}}}\ and\ \bibinfo {author} {\bibfnamefont {T.~H.}\ \bibnamefont
  {{Hsieh}}},\ }\bibfield  {title} {\bibinfo {title} {{Stability of mixed-state
  quantum phases via finite Markov length}},\ }\bibfield  {journal} {\bibinfo
  {journal} {arXiv e-prints}\ }\href
  {https://doi.org/10.48550/arXiv.2404.07251} {10.48550/arXiv.2404.07251}
  (\bibinfo {year} {2024}),\ \Eprint {https://arxiv.org/abs/2404.07251}
  {arXiv:2404.07251 [quant-ph]} \BibitemShut {NoStop}%
\bibitem [{\citenamefont {Gaiotto}\ \emph {et~al.}(2015)\citenamefont
  {Gaiotto}, \citenamefont {Kapustin}, \citenamefont {Seiberg},\ and\
  \citenamefont {Willett}}]{gaiotto2015generalized}%
  \BibitemOpen
  \bibfield  {author} {\bibinfo {author} {\bibfnamefont {D.}~\bibnamefont
  {Gaiotto}}, \bibinfo {author} {\bibfnamefont {A.}~\bibnamefont {Kapustin}},
  \bibinfo {author} {\bibfnamefont {N.}~\bibnamefont {Seiberg}},\ and\ \bibinfo
  {author} {\bibfnamefont {B.}~\bibnamefont {Willett}},\ }\bibfield  {title}
  {\bibinfo {title} {Generalized global symmetries},\ }\bibfield  {journal}
  {\bibinfo  {journal} {Journal of High Energy Physics}\ }\textbf {\bibinfo
  {volume} {2015}},\ \href {https://doi.org/10.1007/jhep02(2015)172}
  {10.1007/jhep02(2015)172} (\bibinfo {year} {2015})\BibitemShut {NoStop}%
\bibitem [{\citenamefont {McGreevy}(2023)}]{mcgreevy2023generalized}%
  \BibitemOpen
  \bibfield  {author} {\bibinfo {author} {\bibfnamefont {J.}~\bibnamefont
  {McGreevy}},\ }\bibfield  {title} {\bibinfo {title} {Generalized symmetries
  in condensed matter},\ }\href
  {https://doi.org/10.1146/annurev-conmatphys-040721-021029} {\bibfield
  {journal} {\bibinfo  {journal} {Annual Review of Condensed Matter Physics}\
  }\textbf {\bibinfo {volume} {14}},\ \bibinfo {pages} {57–82} (\bibinfo
  {year} {2023})}\BibitemShut {NoStop}%
\bibitem [{\citenamefont {Buča}\ and\ \citenamefont
  {Prosen}(2012)}]{buca2012}%
  \BibitemOpen
  \bibfield  {author} {\bibinfo {author} {\bibfnamefont {B.}~\bibnamefont
  {Buča}}\ and\ \bibinfo {author} {\bibfnamefont {T.}~\bibnamefont {Prosen}},\
  }\bibfield  {title} {\bibinfo {title} {A note on symmetry reductions of the
  lindblad equation: transport in constrained open spin chains},\ }\href
  {https://doi.org/10.1088/1367-2630/14/7/073007} {\bibfield  {journal}
  {\bibinfo  {journal} {New Journal of Physics}\ }\textbf {\bibinfo {volume}
  {14}},\ \bibinfo {pages} {073007} (\bibinfo {year} {2012})}\BibitemShut
  {NoStop}%
\bibitem [{\citenamefont {Albert}\ and\ \citenamefont
  {Jiang}(2014)}]{albert2014sym}%
  \BibitemOpen
  \bibfield  {author} {\bibinfo {author} {\bibfnamefont {V.~V.}\ \bibnamefont
  {Albert}}\ and\ \bibinfo {author} {\bibfnamefont {L.}~\bibnamefont {Jiang}},\
  }\bibfield  {title} {\bibinfo {title} {Symmetries and conserved quantities in
  lindblad master equations},\ }\href
  {https://doi.org/10.1103/PhysRevA.89.022118} {\bibfield  {journal} {\bibinfo
  {journal} {Phys. Rev. A}\ }\textbf {\bibinfo {volume} {89}},\ \bibinfo
  {pages} {022118} (\bibinfo {year} {2014})}\BibitemShut {NoStop}%
\bibitem [{\citenamefont {{Rakovszky}}\ \emph {et~al.}(2023)\citenamefont
  {{Rakovszky}}, \citenamefont {{Gopalakrishnan}},\ and\ \citenamefont {{von
  Keyserlingk}}}]{rakovszky2023stable}%
  \BibitemOpen
  \bibfield  {author} {\bibinfo {author} {\bibfnamefont {T.}~\bibnamefont
  {{Rakovszky}}}, \bibinfo {author} {\bibfnamefont {S.}~\bibnamefont
  {{Gopalakrishnan}}},\ and\ \bibinfo {author} {\bibfnamefont {C.}~\bibnamefont
  {{von Keyserlingk}}},\ }\bibfield  {title} {\bibinfo {title} {{Defining
  stable phases of open quantum systems}},\ }\bibfield  {journal} {\bibinfo
  {journal} {arXiv e-prints}\ }\href
  {https://doi.org/10.48550/arXiv.2308.15495} {10.48550/arXiv.2308.15495}
  (\bibinfo {year} {2023}),\ \Eprint {https://arxiv.org/abs/2308.15495}
  {arXiv:2308.15495 [quant-ph]} \BibitemShut {NoStop}%
\bibitem [{\citenamefont {Ma}\ and\ \citenamefont
  {Turzillo}(2024)}]{ma2024symmetryprotectedtopologicalphases}%
  \BibitemOpen
  \bibfield  {author} {\bibinfo {author} {\bibfnamefont {R.}~\bibnamefont
  {Ma}}\ and\ \bibinfo {author} {\bibfnamefont {A.}~\bibnamefont {Turzillo}},\
  }\href {https://arxiv.org/abs/2403.13280} {\bibinfo {title} {Symmetry
  protected topological phases of mixed states in the doubled space}} (\bibinfo
  {year} {2024}),\ \Eprint {https://arxiv.org/abs/2403.13280} {arXiv:2403.13280
  [quant-ph]} \BibitemShut {NoStop}%
\bibitem [{\citenamefont {Hermanns}\ and\ \citenamefont
  {Trebst}(2014)}]{hermanns2014classical}%
  \BibitemOpen
  \bibfield  {author} {\bibinfo {author} {\bibfnamefont {M.}~\bibnamefont
  {Hermanns}}\ and\ \bibinfo {author} {\bibfnamefont {S.}~\bibnamefont
  {Trebst}},\ }\bibfield  {title} {\bibinfo {title} {Renyi entropies for
  classical string-net models},\ }\href
  {https://doi.org/10.1103/PhysRevB.89.205107} {\bibfield  {journal} {\bibinfo
  {journal} {Phys. Rev. B}\ }\textbf {\bibinfo {volume} {89}},\ \bibinfo
  {pages} {205107} (\bibinfo {year} {2014})}\BibitemShut {NoStop}%
\bibitem [{\citenamefont {Peres}(1996)}]{peres1996}%
  \BibitemOpen
  \bibfield  {author} {\bibinfo {author} {\bibfnamefont {A.}~\bibnamefont
  {Peres}},\ }\bibfield  {title} {\bibinfo {title} {{Separability criterion for
  density matrices}},\ }\href@noop {} {\bibfield  {journal} {\bibinfo
  {journal} {Phys. Rev. Lett.}\ }\textbf {\bibinfo {volume} {77}},\ \bibinfo
  {pages} {1413} (\bibinfo {year} {1996})},\ \Eprint
  {https://arxiv.org/abs/quant-ph/9604005} {arXiv:quant-ph/9604005}
  \BibitemShut {NoStop}%
\bibitem [{\citenamefont {Zyczkowski}\ \emph {et~al.}(1998)\citenamefont
  {Zyczkowski}, \citenamefont {Horodecki}, \citenamefont {Sanpera},\ and\
  \citenamefont {Lewenstein}}]{zyczkowski1998}%
  \BibitemOpen
  \bibfield  {author} {\bibinfo {author} {\bibfnamefont {K.}~\bibnamefont
  {Zyczkowski}}, \bibinfo {author} {\bibfnamefont {P.}~\bibnamefont
  {Horodecki}}, \bibinfo {author} {\bibfnamefont {A.}~\bibnamefont {Sanpera}},\
  and\ \bibinfo {author} {\bibfnamefont {M.}~\bibnamefont {Lewenstein}},\
  }\bibfield  {title} {\bibinfo {title} {{On the volume of the set of mixed
  entangled states}},\ }\href@noop {} {\bibfield  {journal} {\bibinfo
  {journal} {Phys. Rev. A}\ }\textbf {\bibinfo {volume} {58}},\ \bibinfo
  {pages} {883} (\bibinfo {year} {1998})},\ \Eprint
  {https://arxiv.org/abs/quant-ph/9804024} {arXiv:quant-ph/9804024}
  \BibitemShut {NoStop}%
\bibitem [{\citenamefont {Eisert}\ and\ \citenamefont
  {Plenio}(1999)}]{eisert1999}%
  \BibitemOpen
  \bibfield  {author} {\bibinfo {author} {\bibfnamefont {J.}~\bibnamefont
  {Eisert}}\ and\ \bibinfo {author} {\bibfnamefont {M.~B.}\ \bibnamefont
  {Plenio}},\ }\bibfield  {title} {\bibinfo {title} {A comparison of
  entanglement measures},\ }\href {https://doi.org/10.1080/09500349908231260}
  {\bibfield  {journal} {\bibinfo  {journal} {Journal of Modern Optics}\
  }\textbf {\bibinfo {volume} {46}},\ \bibinfo {pages} {145} (\bibinfo {year}
  {1999})}\BibitemShut {NoStop}%
\bibitem [{\citenamefont {Vidal}\ and\ \citenamefont
  {Werner}(2002)}]{vidal2002}%
  \BibitemOpen
  \bibfield  {author} {\bibinfo {author} {\bibfnamefont {G.}~\bibnamefont
  {Vidal}}\ and\ \bibinfo {author} {\bibfnamefont {R.~F.}\ \bibnamefont
  {Werner}},\ }\bibfield  {title} {\bibinfo {title} {Computable measure of
  entanglement},\ }\href {https://doi.org/10.1103/PhysRevA.65.032314}
  {\bibfield  {journal} {\bibinfo  {journal} {Phys. Rev. A}\ }\textbf {\bibinfo
  {volume} {65}},\ \bibinfo {pages} {032314} (\bibinfo {year}
  {2002})}\BibitemShut {NoStop}%
\bibitem [{\citenamefont {Plenio}(2005)}]{plenio2005}%
  \BibitemOpen
  \bibfield  {author} {\bibinfo {author} {\bibfnamefont {M.~B.}\ \bibnamefont
  {Plenio}},\ }\bibfield  {title} {\bibinfo {title} {Logarithmic negativity: A
  full entanglement monotone that is not convex},\ }\href
  {https://doi.org/10.1103/PhysRevLett.95.090503} {\bibfield  {journal}
  {\bibinfo  {journal} {Phys. Rev. Lett.}\ }\textbf {\bibinfo {volume} {95}},\
  \bibinfo {pages} {090503} (\bibinfo {year} {2005})}\BibitemShut {NoStop}%
\bibitem [{\citenamefont {Kitaev}\ and\ \citenamefont
  {Preskill}(2006)}]{kitaev2006tee}%
  \BibitemOpen
  \bibfield  {author} {\bibinfo {author} {\bibfnamefont {A.}~\bibnamefont
  {Kitaev}}\ and\ \bibinfo {author} {\bibfnamefont {J.}~\bibnamefont
  {Preskill}},\ }\bibfield  {title} {\bibinfo {title} {Topological entanglement
  entropy},\ }\bibfield  {journal} {\bibinfo  {journal} {Physical Review
  Letters}\ }\textbf {\bibinfo {volume} {96}},\ \href
  {https://doi.org/10.1103/physrevlett.96.110404}
  {10.1103/physrevlett.96.110404} (\bibinfo {year} {2006})\BibitemShut
  {NoStop}%
\bibitem [{\citenamefont {Levin}\ and\ \citenamefont
  {Wen}(2006)}]{levin2006tee}%
  \BibitemOpen
  \bibfield  {author} {\bibinfo {author} {\bibfnamefont {M.}~\bibnamefont
  {Levin}}\ and\ \bibinfo {author} {\bibfnamefont {X.-G.}\ \bibnamefont
  {Wen}},\ }\bibfield  {title} {\bibinfo {title} {Detecting topological order
  in a ground state wave function},\ }\href
  {https://doi.org/10.1103/PhysRevLett.96.110405} {\bibfield  {journal}
  {\bibinfo  {journal} {Phys. Rev. Lett.}\ }\textbf {\bibinfo {volume} {96}},\
  \bibinfo {pages} {110405} (\bibinfo {year} {2006})}\BibitemShut {NoStop}%
\bibitem [{\citenamefont {Lee}\ and\ \citenamefont
  {Vidal}(2013)}]{lee2013negativity}%
  \BibitemOpen
  \bibfield  {author} {\bibinfo {author} {\bibfnamefont {Y.~A.}\ \bibnamefont
  {Lee}}\ and\ \bibinfo {author} {\bibfnamefont {G.}~\bibnamefont {Vidal}},\
  }\bibfield  {title} {\bibinfo {title} {Entanglement negativity and
  topological order},\ }\href {https://doi.org/10.1103/PhysRevA.88.042318}
  {\bibfield  {journal} {\bibinfo  {journal} {Phys. Rev. A}\ }\textbf {\bibinfo
  {volume} {88}},\ \bibinfo {pages} {042318} (\bibinfo {year}
  {2013})}\BibitemShut {NoStop}%
\bibitem [{\citenamefont {Castelnovo}(2013)}]{castelnovo2013negativity}%
  \BibitemOpen
  \bibfield  {author} {\bibinfo {author} {\bibfnamefont {C.}~\bibnamefont
  {Castelnovo}},\ }\bibfield  {title} {\bibinfo {title} {Negativity and
  topological order in the toric code},\ }\href
  {https://doi.org/10.1103/PhysRevA.88.042319} {\bibfield  {journal} {\bibinfo
  {journal} {Phys. Rev. A}\ }\textbf {\bibinfo {volume} {88}},\ \bibinfo
  {pages} {042319} (\bibinfo {year} {2013})}\BibitemShut {NoStop}%
\bibitem [{\citenamefont {Wen}\ \emph {et~al.}(2016{\natexlab{a}})\citenamefont
  {Wen}, \citenamefont {Matsuura},\ and\ \citenamefont {Ryu}}]{wen2016edge}%
  \BibitemOpen
  \bibfield  {author} {\bibinfo {author} {\bibfnamefont {X.}~\bibnamefont
  {Wen}}, \bibinfo {author} {\bibfnamefont {S.}~\bibnamefont {Matsuura}},\ and\
  \bibinfo {author} {\bibfnamefont {S.}~\bibnamefont {Ryu}},\ }\bibfield
  {title} {\bibinfo {title} {Edge theory approach to topological entanglement
  entropy, mutual information, and entanglement negativity in chern-simons
  theories},\ }\href {https://doi.org/10.1103/PhysRevB.93.245140} {\bibfield
  {journal} {\bibinfo  {journal} {Phys. Rev. B}\ }\textbf {\bibinfo {volume}
  {93}},\ \bibinfo {pages} {245140} (\bibinfo {year}
  {2016}{\natexlab{a}})}\BibitemShut {NoStop}%
\bibitem [{\citenamefont {Wen}\ \emph {et~al.}(2016{\natexlab{b}})\citenamefont
  {Wen}, \citenamefont {Chang},\ and\ \citenamefont {Ryu}}]{wen2016surgery}%
  \BibitemOpen
  \bibfield  {author} {\bibinfo {author} {\bibfnamefont {X.}~\bibnamefont
  {Wen}}, \bibinfo {author} {\bibfnamefont {P.-Y.}\ \bibnamefont {Chang}},\
  and\ \bibinfo {author} {\bibfnamefont {S.}~\bibnamefont {Ryu}},\ }\bibfield
  {title} {\bibinfo {title} {Topological entanglement negativity in
  chern-simons theories},\ }\href {https://doi.org/10.1007/JHEP09(2016)012}
  {\bibfield  {journal} {\bibinfo  {journal} {Journal of High Energy Physics}\
  }\textbf {\bibinfo {volume} {2016}} (\bibinfo {year}
  {2016}{\natexlab{b}})}\BibitemShut {NoStop}%
\bibitem [{\citenamefont {Hart}\ and\ \citenamefont
  {Castelnovo}(2018)}]{hart2018negativity}%
  \BibitemOpen
  \bibfield  {author} {\bibinfo {author} {\bibfnamefont {O.}~\bibnamefont
  {Hart}}\ and\ \bibinfo {author} {\bibfnamefont {C.}~\bibnamefont
  {Castelnovo}},\ }\bibfield  {title} {\bibinfo {title} {Entanglement
  negativity and sudden death in the toric code at finite temperature},\ }\href
  {https://doi.org/10.1103/PhysRevB.97.144410} {\bibfield  {journal} {\bibinfo
  {journal} {Phys. Rev. B}\ }\textbf {\bibinfo {volume} {97}},\ \bibinfo
  {pages} {144410} (\bibinfo {year} {2018})}\BibitemShut {NoStop}%
\bibitem [{\citenamefont {{Eckstein}}\ \emph {et~al.}(2024)\citenamefont
  {{Eckstein}}, \citenamefont {{Han}}, \citenamefont {{Trebst}},\ and\
  \citenamefont {{Zhu}}}]{eckstein2024}%
  \BibitemOpen
  \bibfield  {author} {\bibinfo {author} {\bibfnamefont {F.}~\bibnamefont
  {{Eckstein}}}, \bibinfo {author} {\bibfnamefont {B.}~\bibnamefont {{Han}}},
  \bibinfo {author} {\bibfnamefont {S.}~\bibnamefont {{Trebst}}},\ and\
  \bibinfo {author} {\bibfnamefont {G.-Y.}\ \bibnamefont {{Zhu}}},\ }\bibfield
  {title} {\bibinfo {title} {{Robust teleportation of a surface code and
  cascade of topological quantum phase transitions}},\ }\href
  {https://doi.org/10.48550/arXiv.2403.04767} {\bibfield  {journal} {\bibinfo
  {journal} {arXiv e-prints}\ ,\ \bibinfo {eid} {arXiv:2403.04767}} (\bibinfo
  {year} {2024})}\BibitemShut {NoStop}%
\bibitem [{\citenamefont {Chen}\ and\ \citenamefont
  {Grover}(2024)}]{chen2024unconventional}%
  \BibitemOpen
  \bibfield  {author} {\bibinfo {author} {\bibfnamefont {Y.-H.}\ \bibnamefont
  {Chen}}\ and\ \bibinfo {author} {\bibfnamefont {T.}~\bibnamefont {Grover}},\
  }\href@noop {} {\bibinfo {title} {Unconventional topological mixed-state
  transition and critical phase induced by self-dual coherent errors}}
  (\bibinfo {year} {2024}),\ \Eprint {https://arxiv.org/abs/2403.06553}
  {arXiv:2403.06553 [quant-ph]} \BibitemShut {NoStop}%
\bibitem [{\citenamefont {Siva}\ \emph {et~al.}(2022)\citenamefont {Siva},
  \citenamefont {Zou}, \citenamefont {Soejima}, \citenamefont {Mong},\ and\
  \citenamefont {Zaletel}}]{siva2021}%
  \BibitemOpen
  \bibfield  {author} {\bibinfo {author} {\bibfnamefont {K.}~\bibnamefont
  {Siva}}, \bibinfo {author} {\bibfnamefont {Y.}~\bibnamefont {Zou}}, \bibinfo
  {author} {\bibfnamefont {T.}~\bibnamefont {Soejima}}, \bibinfo {author}
  {\bibfnamefont {R.~S.~K.}\ \bibnamefont {Mong}},\ and\ \bibinfo {author}
  {\bibfnamefont {M.~P.}\ \bibnamefont {Zaletel}},\ }\bibfield  {title}
  {\bibinfo {title} {Universal tripartite entanglement signature of ungappable
  edge states},\ }\href {https://doi.org/10.1103/PhysRevB.106.L041107}
  {\bibfield  {journal} {\bibinfo  {journal} {Phys. Rev. B}\ }\textbf {\bibinfo
  {volume} {106}},\ \bibinfo {pages} {L041107} (\bibinfo {year}
  {2022})}\BibitemShut {NoStop}%
\bibitem [{\citenamefont {Liu}\ \emph {et~al.}(2022)\citenamefont {Liu},
  \citenamefont {Sohal}, \citenamefont {Kudler-Flam},\ and\ \citenamefont
  {Ryu}}]{liu2021}%
  \BibitemOpen
  \bibfield  {author} {\bibinfo {author} {\bibfnamefont {Y.}~\bibnamefont
  {Liu}}, \bibinfo {author} {\bibfnamefont {R.}~\bibnamefont {Sohal}}, \bibinfo
  {author} {\bibfnamefont {J.}~\bibnamefont {Kudler-Flam}},\ and\ \bibinfo
  {author} {\bibfnamefont {S.}~\bibnamefont {Ryu}},\ }\bibfield  {title}
  {\bibinfo {title} {Multipartitioning topological phases by vertex states and
  quantum entanglement},\ }\href {https://doi.org/10.1103/PhysRevB.105.115107}
  {\bibfield  {journal} {\bibinfo  {journal} {Phys. Rev. B}\ }\textbf {\bibinfo
  {volume} {105}},\ \bibinfo {pages} {115107} (\bibinfo {year}
  {2022})}\BibitemShut {NoStop}%
\bibitem [{\citenamefont {Zou}\ \emph {et~al.}(2022)\citenamefont {Zou},
  \citenamefont {Shi}, \citenamefont {Sorce}, \citenamefont {Lim},\ and\
  \citenamefont {Kim}}]{zou2022modular}%
  \BibitemOpen
  \bibfield  {author} {\bibinfo {author} {\bibfnamefont {Y.}~\bibnamefont
  {Zou}}, \bibinfo {author} {\bibfnamefont {B.}~\bibnamefont {Shi}}, \bibinfo
  {author} {\bibfnamefont {J.}~\bibnamefont {Sorce}}, \bibinfo {author}
  {\bibfnamefont {I.~T.}\ \bibnamefont {Lim}},\ and\ \bibinfo {author}
  {\bibfnamefont {I.~H.}\ \bibnamefont {Kim}},\ }\bibfield  {title} {\bibinfo
  {title} {Modular commutators in conformal field theory},\ }\href
  {https://doi.org/10.1103/PhysRevLett.129.260402} {\bibfield  {journal}
  {\bibinfo  {journal} {Phys. Rev. Lett.}\ }\textbf {\bibinfo {volume} {129}},\
  \bibinfo {pages} {260402} (\bibinfo {year} {2022})}\BibitemShut {NoStop}%
\bibitem [{\citenamefont {Kim}\ \emph {et~al.}(2022{\natexlab{a}})\citenamefont
  {Kim}, \citenamefont {Shi}, \citenamefont {Kato},\ and\ \citenamefont
  {Albert}}]{kim2022modulara}%
  \BibitemOpen
  \bibfield  {author} {\bibinfo {author} {\bibfnamefont {I.~H.}\ \bibnamefont
  {Kim}}, \bibinfo {author} {\bibfnamefont {B.}~\bibnamefont {Shi}}, \bibinfo
  {author} {\bibfnamefont {K.}~\bibnamefont {Kato}},\ and\ \bibinfo {author}
  {\bibfnamefont {V.~V.}\ \bibnamefont {Albert}},\ }\bibfield  {title}
  {\bibinfo {title} {Chiral central charge from a single bulk wave function},\
  }\href {https://doi.org/10.1103/PhysRevLett.128.176402} {\bibfield  {journal}
  {\bibinfo  {journal} {Phys. Rev. Lett.}\ }\textbf {\bibinfo {volume} {128}},\
  \bibinfo {pages} {176402} (\bibinfo {year} {2022}{\natexlab{a}})}\BibitemShut
  {NoStop}%
\bibitem [{\citenamefont {Kim}\ \emph {et~al.}(2022{\natexlab{b}})\citenamefont
  {Kim}, \citenamefont {Shi}, \citenamefont {Kato},\ and\ \citenamefont
  {Albert}}]{kim2022modularb}%
  \BibitemOpen
  \bibfield  {author} {\bibinfo {author} {\bibfnamefont {I.~H.}\ \bibnamefont
  {Kim}}, \bibinfo {author} {\bibfnamefont {B.}~\bibnamefont {Shi}}, \bibinfo
  {author} {\bibfnamefont {K.}~\bibnamefont {Kato}},\ and\ \bibinfo {author}
  {\bibfnamefont {V.~V.}\ \bibnamefont {Albert}},\ }\bibfield  {title}
  {\bibinfo {title} {Modular commutator in gapped quantum many-body systems},\
  }\href {https://doi.org/10.1103/PhysRevB.106.075147} {\bibfield  {journal}
  {\bibinfo  {journal} {Phys. Rev. B}\ }\textbf {\bibinfo {volume} {106}},\
  \bibinfo {pages} {075147} (\bibinfo {year} {2022}{\natexlab{b}})}\BibitemShut
  {NoStop}%
\bibitem [{\citenamefont {Fan}(2022)}]{fan2022modular}%
  \BibitemOpen
  \bibfield  {author} {\bibinfo {author} {\bibfnamefont {R.}~\bibnamefont
  {Fan}},\ }\bibfield  {title} {\bibinfo {title} {From entanglement generated
  dynamics to the gravitational anomaly and chiral central charge},\ }\href
  {https://doi.org/10.1103/PhysRevLett.129.260403} {\bibfield  {journal}
  {\bibinfo  {journal} {Phys. Rev. Lett.}\ }\textbf {\bibinfo {volume} {129}},\
  \bibinfo {pages} {260403} (\bibinfo {year} {2022})}\BibitemShut {NoStop}%
\bibitem [{\citenamefont {Sohal}\ and\ \citenamefont
  {Ryu}(2023)}]{sohal2023tripartite}%
  \BibitemOpen
  \bibfield  {author} {\bibinfo {author} {\bibfnamefont {R.}~\bibnamefont
  {Sohal}}\ and\ \bibinfo {author} {\bibfnamefont {S.}~\bibnamefont {Ryu}},\
  }\bibfield  {title} {\bibinfo {title} {Entanglement in tripartitions of
  topological orders: A diagrammatic approach},\ }\href
  {https://doi.org/10.1103/PhysRevB.108.045104} {\bibfield  {journal} {\bibinfo
   {journal} {Phys. Rev. B}\ }\textbf {\bibinfo {volume} {108}},\ \bibinfo
  {pages} {045104} (\bibinfo {year} {2023})}\BibitemShut {NoStop}%
\bibitem [{\citenamefont {Liu}\ \emph {et~al.}(2024)\citenamefont {Liu},
  \citenamefont {Kusuki}, \citenamefont {Kudler-Flam}, \citenamefont {Sohal},\
  and\ \citenamefont {Ryu}}]{liu2024tripartite}%
  \BibitemOpen
  \bibfield  {author} {\bibinfo {author} {\bibfnamefont {Y.}~\bibnamefont
  {Liu}}, \bibinfo {author} {\bibfnamefont {Y.}~\bibnamefont {Kusuki}},
  \bibinfo {author} {\bibfnamefont {J.}~\bibnamefont {Kudler-Flam}}, \bibinfo
  {author} {\bibfnamefont {R.}~\bibnamefont {Sohal}},\ and\ \bibinfo {author}
  {\bibfnamefont {S.}~\bibnamefont {Ryu}},\ }\bibfield  {title} {\bibinfo
  {title} {Multipartite entanglement in two-dimensional chiral topological
  liquids},\ }\href {https://doi.org/10.1103/PhysRevB.109.085108} {\bibfield
  {journal} {\bibinfo  {journal} {Phys. Rev. B}\ }\textbf {\bibinfo {volume}
  {109}},\ \bibinfo {pages} {085108} (\bibinfo {year} {2024})}\BibitemShut
  {NoStop}%
\bibitem [{\citenamefont {{Lessa}}\ \emph {et~al.}(2024)\citenamefont
  {{Lessa}}, \citenamefont {{Cheng}},\ and\ \citenamefont
  {{Wang}}}]{lessa2024anomaly}%
  \BibitemOpen
  \bibfield  {author} {\bibinfo {author} {\bibfnamefont {L.~A.}\ \bibnamefont
  {{Lessa}}}, \bibinfo {author} {\bibfnamefont {M.}~\bibnamefont {{Cheng}}},\
  and\ \bibinfo {author} {\bibfnamefont {C.}~\bibnamefont {{Wang}}},\
  }\bibfield  {title} {\bibinfo {title} {{Mixed-state quantum anomaly and
  multipartite entanglement}},\ }\bibfield  {journal} {\bibinfo  {journal}
  {arXiv e-prints}\ }\href {https://doi.org/10.48550/arXiv.2401.17357}
  {10.48550/arXiv.2401.17357} (\bibinfo {year} {2024}),\ \Eprint
  {https://arxiv.org/abs/2401.17357} {arXiv:2401.17357 [cond-mat.str-el]}
  \BibitemShut {NoStop}%
\bibitem [{\citenamefont {Li}\ \emph {et~al.}(2024)\citenamefont {Li},
  \citenamefont {Lee},\ and\ \citenamefont
  {Yoshida}}]{li2024entanglementneededemergentanyons}%
  \BibitemOpen
  \bibfield  {author} {\bibinfo {author} {\bibfnamefont {Z.}~\bibnamefont
  {Li}}, \bibinfo {author} {\bibfnamefont {D.}~\bibnamefont {Lee}},\ and\
  \bibinfo {author} {\bibfnamefont {B.}~\bibnamefont {Yoshida}},\ }\href
  {https://arxiv.org/abs/2405.07970} {\bibinfo {title} {How much entanglement
  is needed for emergent anyons and fermions?}} (\bibinfo {year} {2024}),\
  \Eprint {https://arxiv.org/abs/2405.07970} {arXiv:2405.07970 [quant-ph]}
  \BibitemShut {NoStop}%
\bibitem [{\citenamefont {Fidkowski}\ \emph {et~al.}(2022)\citenamefont
  {Fidkowski}, \citenamefont {Haah},\ and\ \citenamefont
  {Hastings}}]{fidkowski2022-3d}%
  \BibitemOpen
  \bibfield  {author} {\bibinfo {author} {\bibfnamefont {L.}~\bibnamefont
  {Fidkowski}}, \bibinfo {author} {\bibfnamefont {J.}~\bibnamefont {Haah}},\
  and\ \bibinfo {author} {\bibfnamefont {M.~B.}\ \bibnamefont {Hastings}},\
  }\bibfield  {title} {\bibinfo {title} {Gravitational anomaly of
  $(3+1)$-dimensional ${\mathbb{z}}_{2}$ toric code with fermionic charges and
  fermionic loop self-statistics},\ }\href
  {https://doi.org/10.1103/PhysRevB.106.165135} {\bibfield  {journal} {\bibinfo
   {journal} {Phys. Rev. B}\ }\textbf {\bibinfo {volume} {106}},\ \bibinfo
  {pages} {165135} (\bibinfo {year} {2022})}\BibitemShut {NoStop}%
\bibitem [{\citenamefont {Chen}\ \emph {et~al.}(2023)\citenamefont {Chen},
  \citenamefont {Dua}, \citenamefont {Hsin}, \citenamefont {Jian},
  \citenamefont {Shirley},\ and\ \citenamefont {Xu}}]{chen2023-4d}%
  \BibitemOpen
  \bibfield  {author} {\bibinfo {author} {\bibfnamefont {X.}~\bibnamefont
  {Chen}}, \bibinfo {author} {\bibfnamefont {A.}~\bibnamefont {Dua}}, \bibinfo
  {author} {\bibfnamefont {P.-S.}\ \bibnamefont {Hsin}}, \bibinfo {author}
  {\bibfnamefont {C.-M.}\ \bibnamefont {Jian}}, \bibinfo {author}
  {\bibfnamefont {W.}~\bibnamefont {Shirley}},\ and\ \bibinfo {author}
  {\bibfnamefont {C.}~\bibnamefont {Xu}},\ }\bibfield  {title} {\bibinfo
  {title} {{Loops in 4+1d topological phases}},\ }\href
  {https://doi.org/10.21468/SciPostPhys.15.1.001} {\bibfield  {journal}
  {\bibinfo  {journal} {SciPost Phys.}\ }\textbf {\bibinfo {volume} {15}},\
  \bibinfo {pages} {001} (\bibinfo {year} {2023})}\BibitemShut {NoStop}%
\bibitem [{\citenamefont {Chen}\ and\ \citenamefont
  {Hsin}(2023)}]{chen2023-4dlattice}%
  \BibitemOpen
  \bibfield  {author} {\bibinfo {author} {\bibfnamefont {Y.-A.}\ \bibnamefont
  {Chen}}\ and\ \bibinfo {author} {\bibfnamefont {P.-S.}\ \bibnamefont
  {Hsin}},\ }\bibfield  {title} {\bibinfo {title} {{Exactly solvable lattice
  Hamiltonians and gravitational anomalies}},\ }\href
  {https://doi.org/10.21468/SciPostPhys.14.5.089} {\bibfield  {journal}
  {\bibinfo  {journal} {SciPost Phys.}\ }\textbf {\bibinfo {volume} {14}},\
  \bibinfo {pages} {089} (\bibinfo {year} {2023})}\BibitemShut {NoStop}%
\bibitem [{\citenamefont {{Ellison}}\ and\ \citenamefont
  {{Cheng}}(2024)}]{ellison2024}%
  \BibitemOpen
  \bibfield  {author} {\bibinfo {author} {\bibfnamefont {T.}~\bibnamefont
  {{Ellison}}}\ and\ \bibinfo {author} {\bibfnamefont {M.}~\bibnamefont
  {{Cheng}}},\ }\bibfield  {title} {\bibinfo {title} {{Towards a classification
  of mixed-state topological orders in two dimensions}},\ }\bibfield  {journal}
  {\bibinfo  {journal} {arXiv e-prints}\ }\href
  {https://doi.org/10.48550/arXiv.2405.02390} {10.48550/arXiv.2405.02390}
  (\bibinfo {year} {2024}),\ \Eprint {https://arxiv.org/abs/2405.02390}
  {arXiv:2405.02390 [cond-mat.str-el]} \BibitemShut {NoStop}%
\end{thebibliography}%


\end{document}